\documentclass[useAMS,usenatbib, usegraphicx]{mn2e}
\pdfoutput=1
\usepackage{fancyhdr}
\usepackage{makeidx}
\usepackage{lscape}
\usepackage{hyperref}
\usepackage{wasysym}
\usepackage{longtable}
\usepackage{aas_macros}
\usepackage{txfonts}
\usepackage{times}
\usepackage{graphicx}
\usepackage{color}
\usepackage{ulem,hyperref}
\def\limepy{{\sc limepy}}
\def\spes{{\sc spes}}

\def\rt{r_{\rm t}}
\def\rj{r_{\rm J}}
\def\fpe{f_{\rm PE}}

\begin{document}

\title[Globular cluster number density profiles using {\it Gaia} DR2]{Globular cluster number density profiles using {\it Gaia} DR2}
\author[T.J.L. de Boer et al.]{T.J.L. de Boer$^{1,2}$\thanks{E-mail:
tdeboer@hawaii.edu}, M. Gieles$^{1,3,4}$, E. Balbinot$^{1,5}$,V. H\'{e}nault-Brunet$^{6}$, A. Sollima$^{7}$,  \newauthor L.L. Watkins$^{8}$ and I. Claydon$^{1}$\\
$^{1}$ Department of Physics, University of Surrey, Guildford, GU2 7XH, UK\\
$^2$ Institute for Astronomy, University of Hawai`i, 2680 Woodlawn Drive, Honolulu, HI 96822, USA\\
$^3$ Institut de Ci\`{e}ncies del Cosmos (ICCUB), Universitat de Barcelona, Mart\'{i} i Franqu\`{e}s 1, E08028 Barcelona, Spain\\
$^4$ ICREA, Pg. Lluis Companys 23, 08010 Barcelona, Spain.\\
$^5$ Kapteyn Astronomical Institute, University of Groningen, Postbus 800, 9700AV Groningen, the Netherlands \\
$^{6}$ National Research Council, Herzberg Astronomy \& Astrophysics, 5071 West Saanich Road, Victoria, BC, V9E 2E7, Canada\\
$^{7}$ INAF Osservatorio Astronomico di Bologna, via Ranzani 1, 40127 Bologna, Italy\\
$^{8}$ European Southern Observatory, Karl-Schwarzschild-Str. 2, 85748 Garching, Germany\\
}
\date{Received ...; accepted ...}

%\pagerange{\pageref{firstpage}--\pageref{lastpage}} \pubyear{2017}

\maketitle

\begin{abstract}
Using data from {\it Gaia} DR2, we study the radial number density profiles of the Galactic globular cluster sample. Proper motions are used for accurate membership selection, especially crucial in the cluster outskirts. Due to the severe crowding in the centres, the {\it Gaia} data is supplemented by literature data from {\it HST} and surface brightness measurements, where available. This results in 81 clusters with a complete density profile covering the full tidal radius (and beyond) for each cluster.
We model the density profiles using a set of single-mass models ranging from King and Wilson models to generalised lowered isothermal \limepy\ models and the recently introduced \spes\ models, which allow for the inclusion of potential escapers. We find that both King and Wilson models are too simple to fully reproduce the density profiles, with King (Wilson) models on average underestimating(overestimating) the radial extent of the clusters.
The truncation radii derived from the \limepy\ models are similar to estimates for the Jacobi radii based on the cluster masses and their orbits. We show clear correlations between structural and environmental parameters, as a function of Galactocentric radius and integrated luminosity. Notably, the recovered fraction of potential escapers correlates with cluster pericentre radius, luminosity and cluster concentration. The ratio of half mass over Jacobi radius also correlates with both truncation parameter and PE fraction, showing the effect of Roche lobe filling.
\end{abstract}

\begin{keywords}
methods: numerical --- galaxies: star cluster --- globular clusters: general --- stars: kinematics and dynamics
\end{keywords}

\label{firstpage}

\section{Introduction}
\label{introduction}
Globular clusters (GCs) are amongst the oldest known structures in the Universe, believed to have been formed between redshifts of $z\sim5-10$ \citep[e.g.][]{KravtsovGnedin05}. They have long been used as the principal stellar population calibration source against which to compare other systems, or as simple tracer particles to probe the gravitational potential of the systems they inhabit. Through their use, they have contributed to invaluable progress in e.g. early Universe cosmology \citep{Peebles68}, the formation and evolution of the Milky Way (MW) disc \citep{Freeman02} and halo \citep{SearleZinn78}, and external galaxies \citep{Brodie06}. The present day spatial distribution and motions of GCs provide a dynamical probe of the MW dark matter (DM) potential, the hierarchical assembly of the Milky Way \citep{Moore06} and a constraint on the re-ionisation of the Universe \citep{Couchman86,Spitler12}.

During the last two decades, the field of GC formation has been reinvigorated due to the discovery that GCs are not simple, spherical, non-rotating stellar systems. An ever increasing number of studies has shown that their stellar populations are anything but simple, with clear evidence for multiple populations due to light element abundance variations and discrete sequences in colour-magnitude space \citep[e.g.][]{Carretta091,Gratton12,Bastian18}. Dynamical studies of GCs have shown the presence of kinematic signatures, concluding that rotation is common in these systems \citep[e.g.][]{Mackey13,Fabricius14,Ferraro18,Kamann18,Bianchini18}. Studies of the dynamical mass-to-light ratios conclude there is no signature of DM in the inner parts of GCs \citep[e.g.][]{Watkins15b,Kimmig15,Baumgardt17a}, with the discovery of tidal tails around GCs further arguing against significant fractions of DM in at least some GCs \citep{Moore96,Odenkirchen01,DES_streams}. 

Nonetheless, the mechanism of GC formation in a DM halo is by no means ruled out, since collisional relaxation pushes the DM to the peripheries where tidal interaction with the MW can effectively strip the entire DM content \citep{Mashchenko05,Baumgardt08}. Furthermore, the discovery of extended, spherical stellar halos around some GCs \citep{Carballo-Bello12,Kuzma18} are in good agreement with models of GC evolution within their own DM halo, in which stars are scattered to large radii and move on long radial orbits as their escape is prevented by their DM halo \citep{Penarrubia17}. This has highlighted the need for a comprehensive kinematic study of the outer regions of GCs, which remain largely unexplored. 

The spatial structure of GCs has been extensively studied within the Local Group, leading to the discovery of numerous scaling relations \citep{Trager95, Harris96} and the constraining of the GC fundamental plane \citep{Djorgovski94,McLaughlin00}. Traditionally, the density distribution of GCs has been analysed in the context of isotropic, isothermal sphere models, such as King models \citep{King66}. More recent studies found that the outer regions of GCs are more extended than allowed by King models \citep{Elson87,Larsen04} and models with a power law distribution provide a better fit to the outer parts of GCs due to their shallower density fall-off~\citep{McLaughlin05,Carballo-Bello12,Williams12, Kuzma18}. Once again, studying the outer regions of the GCs is the only way to distinguish between the different models. 

King models are isotropic, lowered isothermal models, which are described by a distribution function (DF): $f(E)\propto \exp(-E/s^2)-1$, for $E<0$ and $f(E)=0$ otherwise. Here $E$ is the specific energy, `lowered' by a truncation energy $\phi_{\rm t}$ (i.e. $E = 0.5v^2 + \phi(r) - \phi_{\rm t}$, where $\phi(r)$ is the specific potential at radius $r$) and $s$ is a velocity scale, which combined with the constant of proportionality in the DF sets the physical scales of the model. This model is fully specified by the dimensionless central potential $W_0$, which controls the central concentration (high $W_0$ implies more concentrated models). For concentrated models ($W_0\gtrsim5$), $s$ is approximately equal to the central 1-dimensional velocity dispersion. The DF of (isotropic and non-rotating) Wilson models is $f(E)\propto \exp(-E/s^2)-1 + E/s^2$, and has a more gradual decline in the density near the tidal radius.
\citet{Davoust77} showed that the King and Wilson models are members of a general family of models in which leading order terms of the exponential are subtracted from the isothemal model. \citet{GomezVelazquez14} showed that this can  be extended to non-integer terms, leading to a more general class of (isotropic) lowered isothermal model, which has an additional model parameter $g$ (with King and Wilson models recovered for $g=1$ and $g=2$, respectively). Because this additional parameter describes the sharpness of the truncation in energy, it affects mostly the mass and velocity profile at large distances. \citet{Gieles15} further expanded these models by introducing radial velocity anisotropy as in \citet{Eddington15} and \citet{Michie63}, multiple mass components as in \citet{DaCostaFreeman76} and \citet{GunnGriffin79}, and introduced the lowered isothermal model explorer in {\sc python} (\limepy)\footnote{\limepy\ is available from \href{https://github.com/mgieles/limepy}{https://github.com/mgieles/limepy}}. 

The \limepy\ models allow for a more elaborate description of stars near the escape energy, but do not include the effect of the Galactic tidal potential, unlike other models by \citep[e.g.][]{Heggie95,Varri09}. The tidal field makes the potential in which the stars move anisotropic and it slows down the escape of stars \citep{Fukushige00,Baumgardt2001}, because escape is limited to narrow apertures around the Lagrangian points. As a result, a GC builds up a population of so-called potential escapers (PEs) during its evolution. These are stars that are energetically unbound, but have not yet escaped because their orbits have not come near the Lagrangian points \citep[e.g.,][]{Daniel17}.
These PEs give rise to an elevation of the density  and velocity dispersion near the Jacobi radius \citep{Kupper10,Claydon17}. The fraction of PEs in a GC is dependent on the mass of the cluster (approximately) as $M^{1/4}$ \citep{Baumgardt2001} and the shape of the Jacobi surface \citep{Claydon17}, which in turns depends on the Galactic potential and GC orbit~\citep{Tanikawa10,Renaud15} and for GCs we expect typical fractions of a few per cent \citep{Claydon17}. The presence of PEs in GCs has been proposed as a way to explain peculiarities in GC outskirts not consistent with the expected behaviour of bound stars even in a generalised lowered isothermal model, such as unusual surface density profiles~\citep[e.g.][]{Cote02,Kupper11}, extended structures~\citep{Kuzma16} and stars with velocities above the escape speed~\citep{Meylan91,Lutzgendorf12}.

In this work, we will make use of data from {\it Gaia} DR2~\citep{Brown18} to study the outskirts of the sample of Galactic GCs presented in \citet[][2010 version]{Harris96}. The use of {\it Gaia} proper motions allows us to perform a membership selection which is far more accurate than any other study of GCs on this scale \citep[e.g.][]{Pancino17}. The density of stars in the outer regions will be combined with existing literature data to obtain a full sampling of GC densities covering the entire system. The resulting density profiles will be modelled using the different types of single-mass models described above to probe for the presence of tidal disturbances and PEs. Importantly, the density profiles will be constructed from a homogeneous dataset, while previous comprehensive works \citep[e.g.,][]{Trager95} have been based on a heterogeneous mix of star counts and integrated photometry, and other homogeneous works have been composed of only a few GCs \citep{Carballo-Bello12,Miocchi13}.

This paper is organised as follows: in Section~\ref{data} we discuss the use of {\it Gaia} data, adopted queries and initial processing. Following this, in Sections~\ref{membership} and~\ref{profiles} we determine the GC membership selection as well as the construction of density profiles extending from the centre out to $\sim$2 tidal radii. The profiles are then fit using a variety of different single-mass models in Section~\ref{model_fits}, followed by an analysis of the resulting parameters and their correlations (in Section~\ref{results}).  Finally, Section~\ref{conclusions} discusses the results and their implications for the study of initial conditions of GC formation.

\section{Data}\label{data}
To study the density profiles of GCs we will make use of data from the {\it Gaia} mission~\citep{GAIAmain1,GAIAmain2,Lindegren18}, which contains exquisite data for about 1.6 billion sources covering the full sky. In particular, we make use of the recently released Data Release 2 (DR2) data, which includes spectro-photometry in the $G$, $G_{\rm BP}$ and $G_{\rm RP}$ bands as well as accurate parallaxes and proper motions for stars down to $G=21$~\citep{Riello18, Evans18,Lindegren18}. Furthermore, for all bright stars ($G_{\rm RVS}<12$), {\it Gaia} measures radial velocities from the Gaia Radial Velocity Spectrometer (RVS) spectrograph~\citep{Cropper18,Sartoretti18}. The availability of proper motions on large spatial scales represents a key improvement for the study of GCs, allowing us to study the density in their heavily contaminated outskirts. The use of photometric membership selection followed by spectroscopic confirmation is very inefficient in these regions, leading to low (a few percent) success rates. This impedes a thorough study of GC outskirts, which is where many interesting dynamical processes linked to cluster formation and evolution can be constrained. 

We make use of the extensive catalog of GCs from \citet[][2010 version]{Harris96} for our input list of targets. To avoid regions of excessive crowding where {\it Gaia} measurements become less reliable, we limit our sample to $\vert b\vert>$5 deg, leaving 113 GCs. Each of these targets is queried in the {\it Gaia} data archive (\url{https://gea.esac.esa.int/archive/}) using a cone search out to a radius of 2.5 times the Jacobi radius ($\rj$) determined by \citet{Balbinot18}. The dataset is further processed to include tangent plane projection coordinates and extinction values using dust maps from~\citet{Schlegel98} with coefficients from \citet{Schlafly11}, on a star by star basis. In heavily extincted regions ($E(B-V)>0.3$) the~\citet{Schlegel98} maps become unreliable, and literature extinction values from \citet[][2010 version]{Harris96} are used instead. 

In the determination of cluster-centric coordinates, position angles and ellipticities are assumed to be zero. These parameters are available in the literature, but different studies find different mean values which vary with radius, and often do not probe the cluster outskirts \citep{Harris96,Chenchen10}. Therefore, we assume each cluster is perfectly spherical, and conduct a detailed study of GC shape in a future work. 
\begin{figure}
\centering
\includegraphics[angle=0, width=0.495\textwidth]{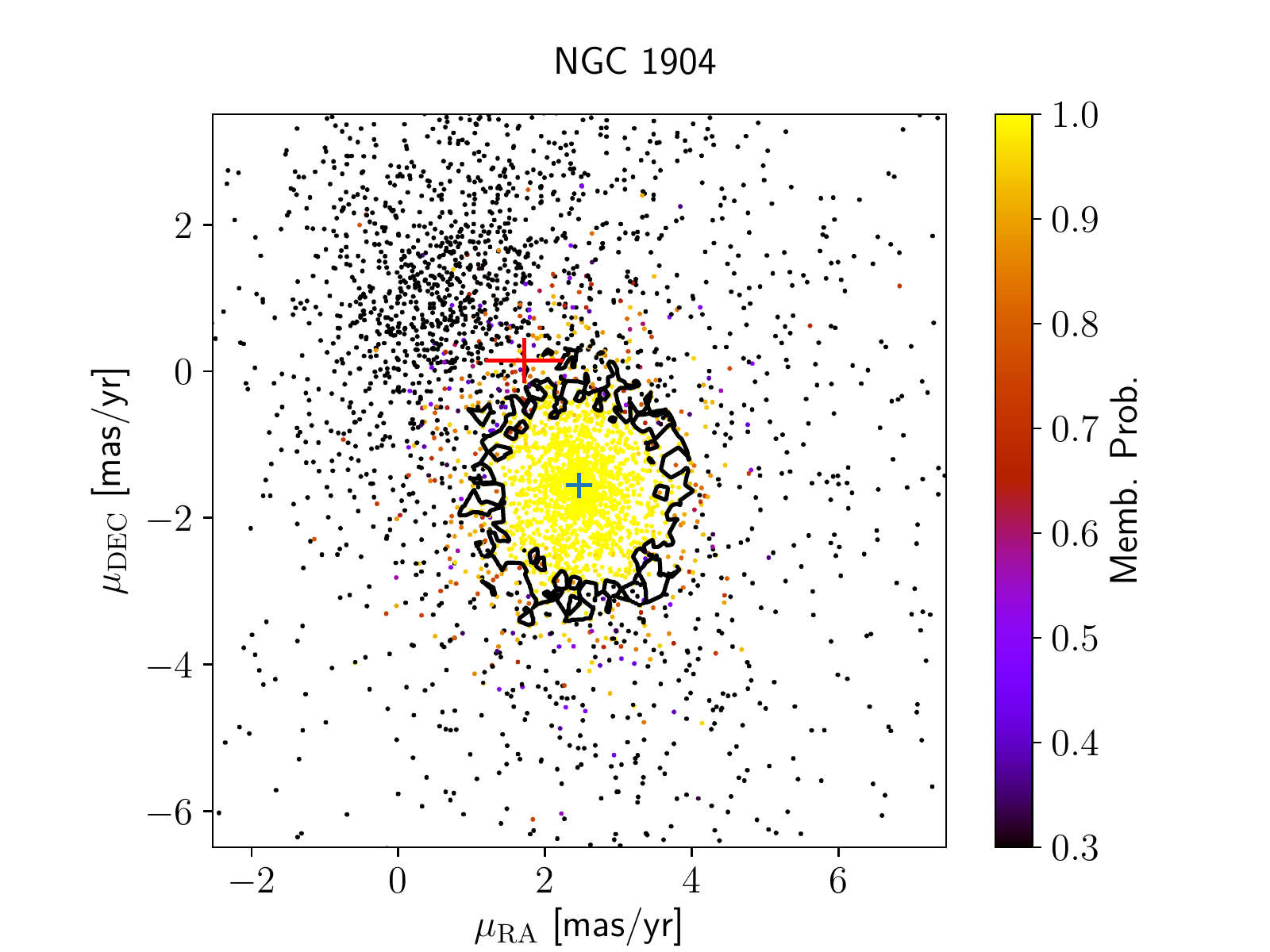}
\caption{The proper motion distribution of stars in our NGC1904 sample, coloured with the computed membership probability. The sample shown has already been cleaned using CMD isochrone cuts and parallax selections. The blue marker indicates the peak of the GC PM distribution, while the red marker indicates the peak of the background distribution. A contour is drawn for membership probability of 0.9 for reference. \label{ngc1904_PMs}}
\end{figure}

\section{Membership selection}\label{membership}
A crucial step in the study of GC density profiles is a reliable membership selection. In this work, we first employ a fixed parallax cut to remove nearby stars, followed by a selection in colour-magnitude space and proper motion space. To remove nearby stars we apply a cut to parallax $\vert\varpi-\varpi_{0}\vert < 2\times\delta\varpi$ with $\varpi_{0}$ the mean parallax of the GC and $\delta\varpi$ the parallax uncertainty. No attempt is made to fit the distribution of parallaxes due to the ongoing characterisation of parallax systematics \citep{Luri18}. 

Colour-magnitude filtering is performed by making use of isochrones with {\it Gaia} bandpasses from the Padova library \citep{Marigo17}, as queried from \url{http://stev.oapd.inaf.it/cmd}. For the stellar population parameters of the GCs we make use of metallicities and distances from \citet{Harris10} and ages taken from~\citep{Marin-Franch09,VandenBerg13}. If no age is available, a cluster is assumed to have an age of 13.5~Gyr. For each cluster, we selected member stars in a conservative region around the isochrone with $\vert(G_{\rm BP}-G_{\rm RP}) - (G_{\rm BP}-G_{\rm RP})_{0}\vert < 2\times\delta(G_{\rm BP}-G_{\rm RP})$ at each $G$ magnitude. For this procedure, a minimum colour error of 0.03 is adopted to avoid having an arbitrarily small selection window for bright stars with small photometric errors. We include only stars up to the tip of the Red Giant Branch (RGB) and forego selecting stars on horizontal branch (HB), to avoid including the potentially heavily contaminated regions corresponding to red HBs for metal-rich GCs. A magnitude limit of $G$=20 is adopted to avoid stars with proper motions of poor quality. Furthermore, we do not include a sample cleaning using the $\texttt{phot\_bp\_rp\_excess\_factor}$ variable as suggested in \citet{Evans18}. The cleaning of well-behaved single sources will make little difference in halo GCs with good PM separation, but reject a large fraction of sources in crowded regions like the Galactic bulge. Since this is expected to have a large impact on the radial density profiles, we choose to forego selections which are not homogeneous across the cluster field of view. 

Following these selections, we use the {\it Gaia} proper motions to compute the membership probability of each star. The proper motion cloud is fit using a Gaussian mixture model consisting of one Gaussian for the cluster distribution and another for the Milky Way foreground distribution. Initial guesses for the cluster Gaussian centres are taken from~\citet{Helmi18} where available and using a simple mean within half the Jacobi radius otherwise. Distributions are fit using the $\texttt{emcee}$ python MCMC package, after which membership probabilities for each star in our sample are computed~\citep{emcee}. The final adopted member samples are selected using a probability cut of 0.5, and made available at \url{https://github.com/tdboer/GC_profiles}. 
\begin{figure}
\centering
\includegraphics[angle=0, width=0.495\textwidth]{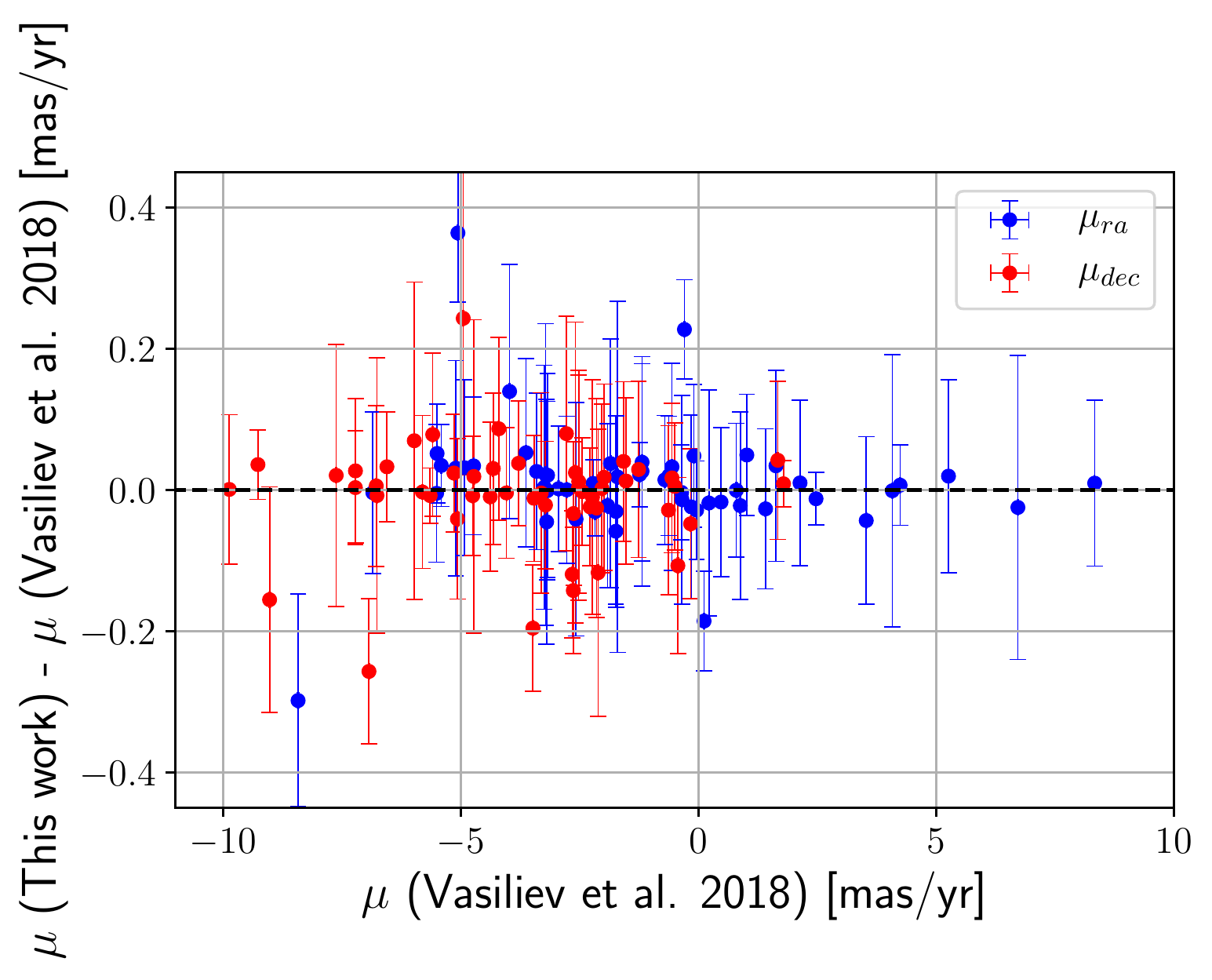}
\caption{Proper motion comparison between the GCs in common between this work and~\citet{Vasiliev18}. \label{Vasiliev_PMcomp}}
\end{figure}

Figure~\ref{ngc1904_PMs} shows the proper motion distribution for an example cluster, NGC1904. The best-fit GC PM peaks are shown as blue and red markers for GC and background sample respectively, while the contours shows the 0.9 member probability. The MCMC fit cleanly separates the cluster and foreground distributions, resulting in a secure sample of member stars with a cut at prob$>$0.5. The GC peak values of ($\mu_{\mathrm{ra}}$,$\mu_{\mathrm{dec}}$) = (2.51$\pm$0.08,-1.51$\pm$0.09) are consistent with values from~\citet{Helmi18}.

Figure~\ref{Vasiliev_PMcomp} compares the determined mean proper motions in RA and Dec for GCs in common with the sample from~\citet{Vasiliev18}. The errorbars display the uncertainties on the proper motion based on the 16th, 50th, and 84th percentiles from the MCMC runs. There is good agreement between both samples, with overall little scatter in both $\mu_{\mathrm{ra}}$ and $\mu_{\mathrm{dec}}$. Some GCs show large ($>$0.25 mas/yr) uncertainties in our sample, although the peak values are in good agreement with~\citet{Vasiliev18}. These are bulge GCs such as NGC6284 and NGC6388 which are low mass, but suffer from excessive ($>$75\%) foreground contamination. Given that we determine our PM values using the entire sample within 2.5 times the Jacobi radius, our uncertainties are naturally larger than the values in~\citet{Vasiliev18}, where a much smaller spatial area is utilised.

\begin{figure}
\centering
\includegraphics[angle=0, width=0.495\textwidth]{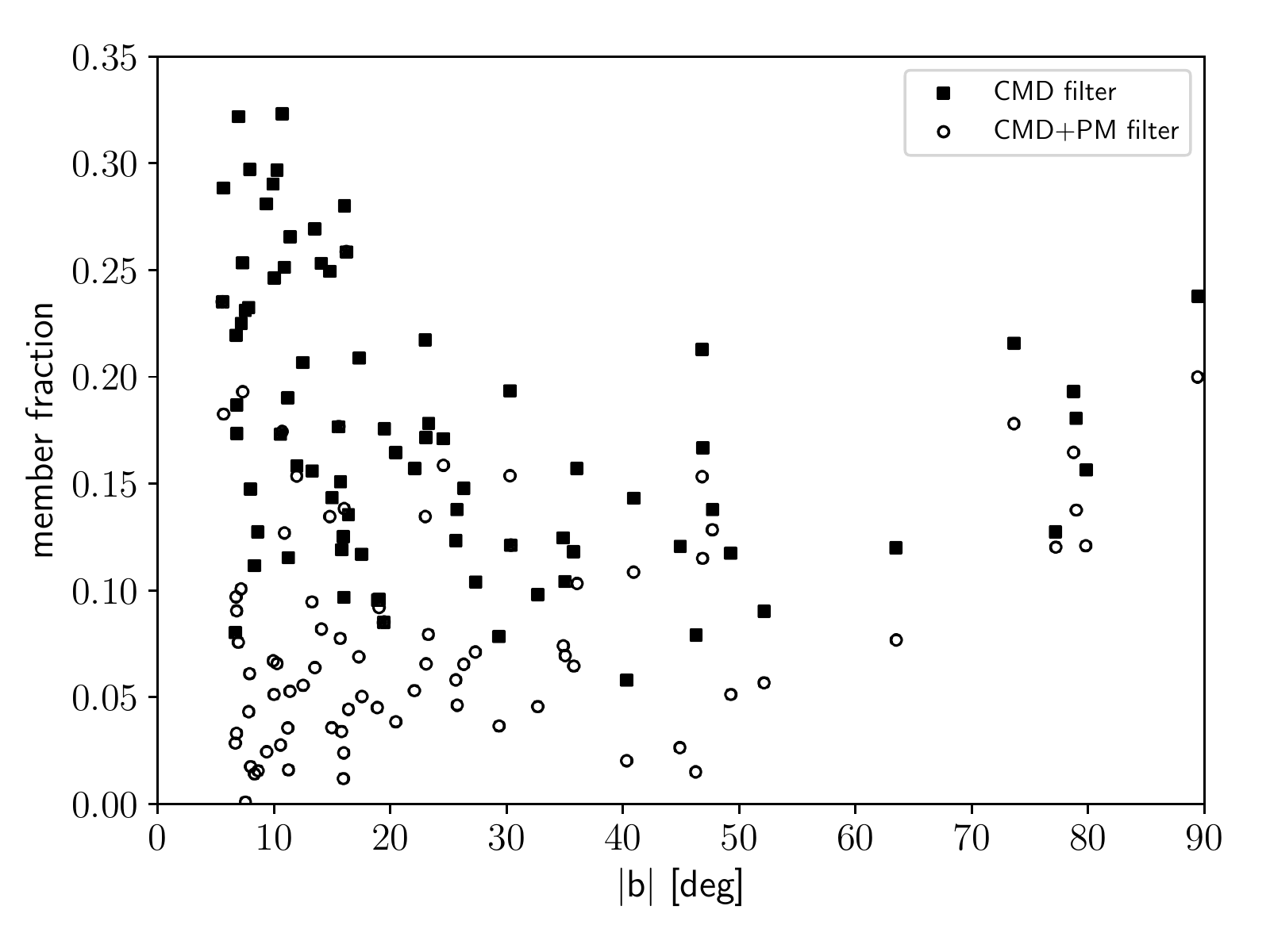}
\caption{The fraction of member stars in each cluster compared to the total number of stars within 2.5 Jacobi radii, as a function of absolute Galactic latitude $|$b$|$. Member fraction is shown after applying the colour-magnitude isochrone filter, as well as after applying the additional proper motion filter described in Section~\ref{membership}. The mean membership fraction after CMD filtering is 0.18 (a reduction of roughly a factor 5), while the mean membership fraction after CMD+PM filtering is 0.08 constituting a further factor 2 reduction. \label{GC_nrs}}
\end{figure}

Figure~\ref{GC_nrs} shows the fraction of remaining member stars for each cluster, after successive stages of membership cleaning, as a function of absolute Galactic latitude $|$b$|$. Filled squares show the membership fraction after applying the colour-magnitude filtering using isochrones, relative to the total number of sources within 2.5 times the Jacobi radius. Open circles show the membership fraction after applying the additional proper motion selection described above. The figure shows that the reduction in member stars using a simple CMD cut is roughly a factor of 5 (the mean fraction if 0.18$\pm$0.06), but that the cleaning is least efficient for clusters closest to the MW disk. The filtering using proper motions leads to a further reduction of a factor of two on average (the mean fraction is 0.08$\pm$0.06). However, the reduction is clearly larger for clusters close to the disk (with reductions of a factor $>$5), due to a better separation of cluster and disk stars in proper motion space. 

\section{Number density profiles}\label{profiles}
With membership probability for our GC sample in place, we construct the radial number density profiles by binning the radial data as a function of distance from the cluster centre. We adopt a fixed number of 50 radial bins, with an equal number of stars in each bin.  For ill-sampled or low density GCs, a fixed bin occupation of 10 stars per bin is used instead. We reiterate that sphericity is assumed when computing the radial distance from the cluster centre. 

\begin{figure}
\centering
\includegraphics[angle=0, width=0.495\textwidth]{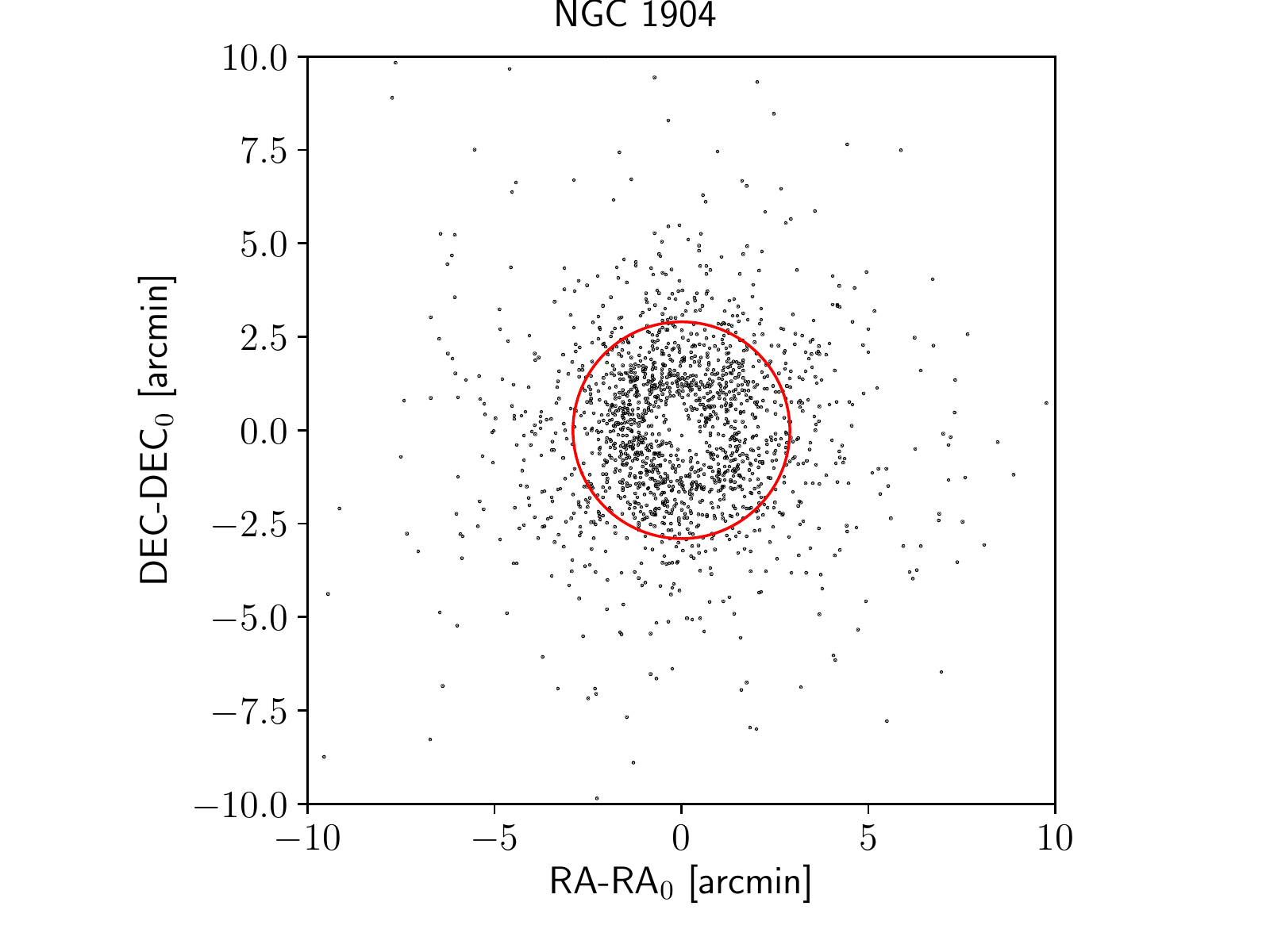}
\caption{Zoom of the inner regions of the spatial coverage of our NGC~1904 sample, after membership selection. A hole due to incompleteness is clearly visible in the cluster centre. The red circle indicates the innermost usable {\it Gaia} radius of 2.9$^{\prime}$, computed following the density prescription from~\citet{Arenou18}.\label{ngc1904_inrad}}
\end{figure}

The number density profiles constructed in this way provide a homogeneous coverage of the GC outskirts that is unmatched in other surveys. However, due to the increasing crowding toward the cluster centres, the inner parts of the profiles are incomplete for all but the lowest density clusters~\citep{Arenou18}. To obtain a complete profile for each GC, we complement the {\it Gaia} profiles with literature profiles from {\it the Hubble Space Telescope (HST)}~\citep{Miocchi13} and the compilation of ground-based surface brightness profile compilation of \citet{Trager95}. When both are available, \citet{Miocchi13} profiles are preferred over \citet{Trager95} profiles since they are more recent. These profiles are stitched to the inner regions of the {\it Gaia} profiles to provide a full coverage out and beyond the Jacobi radius. To stitch the profiles, we first need to determine out to which radius the {\it Gaia} data is reliable and complete. We make use of the comparison between {\it Gaia} and {\it HST} data for 26 clusters performed in~\citet{Arenou18}, which shows that densities of 10$^{5}$ stars/deg$^{2}$ are roughly 80\% complete at $G=20$~mag. Therefore, we assume the {\it Gaia} is free from radius dependent completeness effects outside this radius and adopt this density threshold as the cutoff for the {\it Gaia} profiles. Figure~\ref{ngc1904_inrad} displays a zoom of the spatial distribution of our NGC1904 sample after membership selection, clearly showing the incompleteness of the data in the central regions due to crowding. Using the density criterium from~\citet{Arenou18}, we compute an innermost usable radius of 2.9$^{\prime}$ for this cluster, which is shown in the figure as a red circle. Given that the completeness depends on more than just a simple function of local stellar density (e.g. scanning law coverage, extinction, foreground contamination), we adopt a default inner radius of 2 arcmin for GCs of low density, inside of which we will not use the {\it Gaia} data. The adopted innermost usable radii are presented in Table~\ref{GCpars} for each GC.  

\begin{figure}
\centering
\includegraphics[angle=0, width=0.495\textwidth]{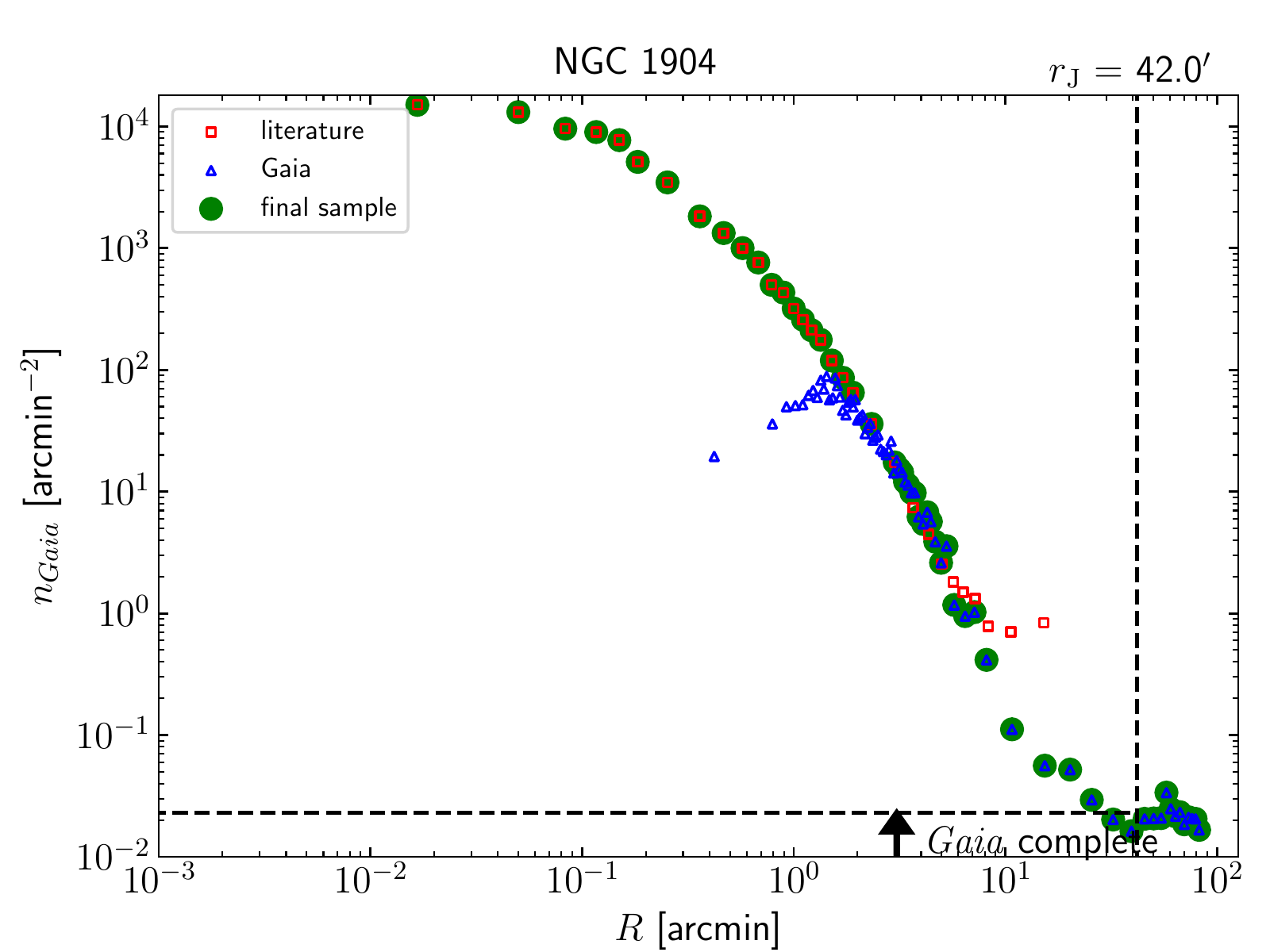}
\caption{The number density profile of NGC1904. Blue triangles show the profile as obtained from {\it Gaia} DR2 after the selections for parallax, proper motion and colours described in section~\ref{membership}. Red squares show the HST number density profile from \citep{Miocchi13}, scaled to the {\it Gaia} profile using all points in the overlapping region outside the {\it Gaia} inner usable radius (shown in Figure~\ref{ngc1904_inrad}, and indicated by the solid black arrow). Finally, the green circles shows the combination of both profiles, which will be used to fit mass models. For reference, the vertical dashed line shows the Jacobi radius from \citet{Balbinot18} while the dashed horizontal line shows the background level estimated using stars between 1.5 and 2 Jacobi radii. \label{ngc1904_dens}}
\end{figure}

Following this, the {\it Gaia} profiles are then tied together with literature profiles by using the overlapping region of both datasets (outside the inner usable Gaia radius) to calibrate the heterogeneous literature data to the homogeneous {\it Gaia} system. Within the overlap region, the literature profile data is interpolated to the same radial values as the {\it Gaia} profile, allowing us to compute a scaling fraction for each radial bin. The adopted scaling fraction is taken to be the average of all the individual fractions, after which the entire literature profile is scaled. Following this, the two profiles are combined, taking the scaled literature values within the innermost usable radius and the {\it Gaia} profile outside, taking care to rescale the number densities in overlapping bins straddling the adopted radius.  Figure~\ref{ngc1904_dens} shows the density profile of NGC1904 as determined from {\it Gaia} data (in blue triangles), along with the existing literature profile from \citet{Miocchi13} as red squares. The {\it Gaia} profile clearly becomes incomplete in the inner regions, as evidenced by the drop in density at a radius of $\sim1^{\prime}$. The green circles show the combined density profile adopted for the cluster, to which mass models will be fit. From Figure~\ref{ngc1904_dens} it is clear that using the {\it Gaia} membership allows us to make use of reliable stellar density data almost 1.5 order of magnitude below the background of the {\it HST} data, showing the added value of proper motion information.

In tying the two profiles together, we are making the implicit assumption that both profiles follow the same underlying number density distribution. While not necessarily true, we believe this to be a reasonable assumption, given that the {\it Gaia} profile is calculated from bright stars and the attached luminosity profiles are also dominated by bright stars. Furthermore, the effects of mass segregation should not be significant as the stars in both datasets have a small range of stellar mass. For these reasons, we believe the difference between the two profiles are small and our approach is justified. 

\section{Dynamical model fits}\label{model_fits}
We will consider different types of single-mass models to fit the number density profiles of our GC sample.  First off, we consider the  King and the isotropic and non-rotating Wilson models, which are often used to fit the spatial distributions of both GCs and dwarf galaxies~\citep{King66,Wilson75}. King and Wilson models provide a fairly simple description of GC morphology, with their shape entirely determined by the dimensionless central potential $W_0$ (high $W_0$ implies more concentrated models). For some GCs, Wilson models have been shown to fit the outer parts of GCs better than King models, due to their shallower density fall-off~\citep{McLaughlin05}. King models have been fitted to Galactic GCs by numerous previous works~\citep[e.g.][]{Djorgovski93}, while Wilson models have been fitted to the entire \citet{Trager95} data set by~\citet{McLaughlin05}. However, given the updated profiles for the GCs presented here, we have refit for the parameters of the King and Wilson models. We will also fit the isotropic, single-mass \limepy\ models to the data and simultaneously fit on $W_0$ and the  truncation parameter $g$.

%\footnote{The {\sc limepy} models are available from \url{https://github.com/mgieles/limepy}.} described in \citet{Gieles15}.

The second class of models we fit to the data are models with the inclusion of Potential Escapers (PEs), as recently presented in \citet{Claydon18}. These models allow for a more elaborate description of stars near the escape energy including the effect of marginally unbound stars. These spherical potential escapers stitched models (hereafter {\sc spes} models) have an energy truncation similar to the models discussed above, with the fundamental difference that the density of stars at the truncation energy can be non-zero. More importantly, the models include stars above the escape energy, with an isothermal DF that continuously and smoothly connects to the bound stars. Apart from $W_0$, the model has two additional parameters $B$ and $\eta$. The value of $B$ can be $0\le B\le1$, where for $B=1$ there are no PEs (i.e. the DF is the same as the King model) and for $0\le B<1$, the model contains PEs. The parameter $\eta$  is the ratio of the velocity dispersion of the PEs over the velocity scale $s$ (see above) and it can have values $0\le \eta \le1$. For $\eta=0$ there are no PEs, and (for fixed $B$) the fraction of PEs correlates with $\eta$. For a fixed $\eta$, the fraction of PEs anticorrelates with $B$ for $B$ close to 1. For smaller $B$, the fraction of PEs is approximately constant or correlates slightly with $B$ (for constant $\eta$). Finally, in the presence of PEs the \spes\ models are not continuous at $r_{\rm t}$, but the models have the ability to be solved (continuously and smoothly) beyond  $r_{\rm t}$ to mimic the effect of escaping stars \citep[see][for details]{Claydon18}. We solve the models out to 25 times the Jacobi radii determined by \citet{Balbinot18} to take into account the projected density in front of the cluster and allow a smooth transition between cluster and background counts. 

The models are fit to the combined number density profiles using the {\sc emcee} {\sc python} MCMC package~\citep{emcee}, fitting for the model parameters (one for the King/Wilson models, two for {\sc limepy} and three for {\sc spes} models), the radial scale (we use the tidal radius as a fitting parameter) and the vertical scaling of the profile. A constant contamination level is defined by taking the average stellar density between 1.5 and 2 Jacobi radii, where we expect the GC contribution to be negligible. Computed background levels are presented in Table~\ref{GCpars}. In the case of the {\sc spes} models, we also directly fit for the cluster tidal radius, without making any a priori assumption about the Jacobi radius. 
\begin{figure*}
\centering
\includegraphics[angle=0, width=0.95\textwidth]{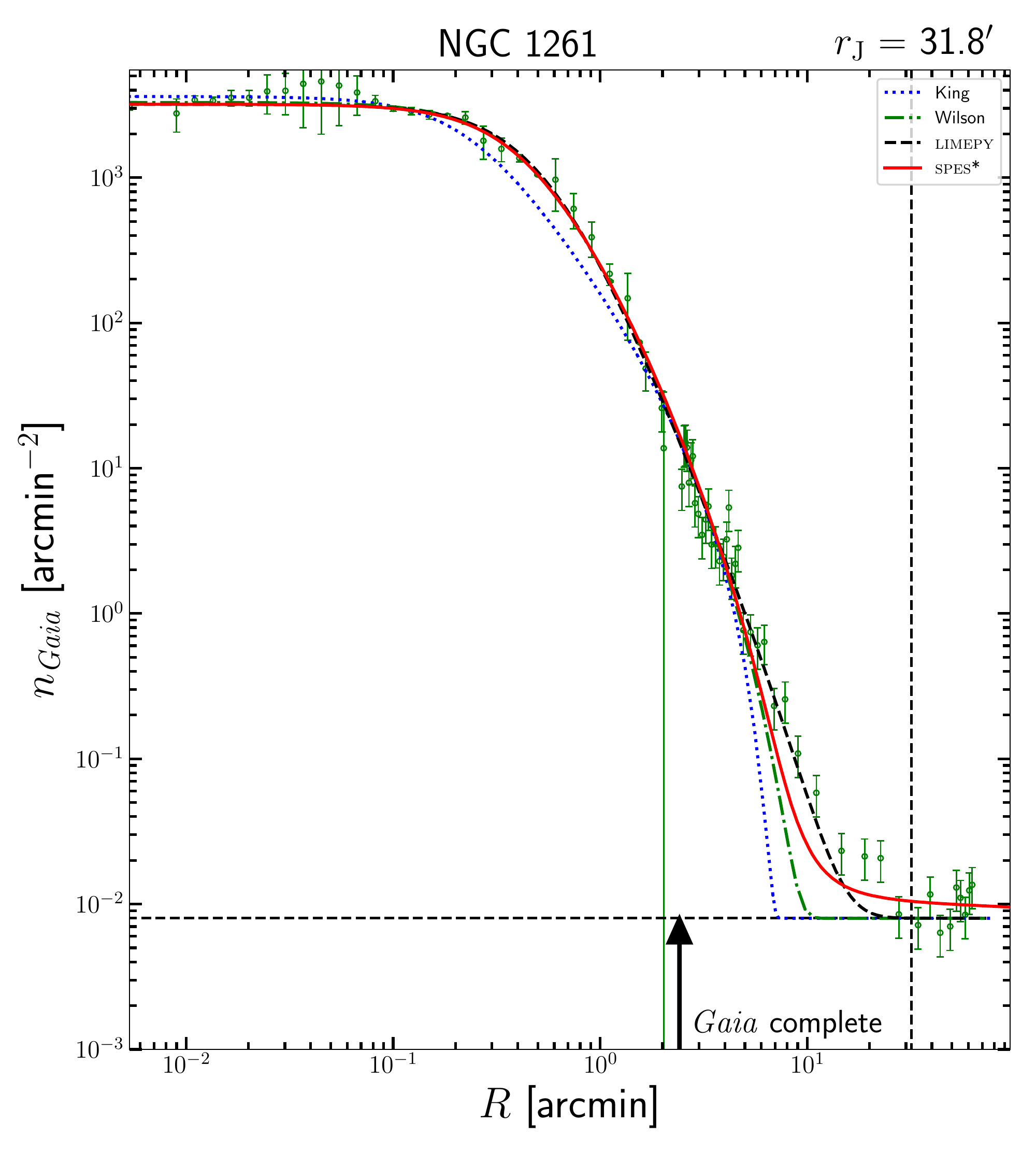}
\caption{The number density profile of NGC~1261 with best-fit dynamical models. Blue and green dashed lines indicate King and Wilson models respectively, with parameters taken from \citet{McLaughlin05}. The solid black line shows the best-fit {\sc limepy} model, while the red solid line shows the {\sc spes} model fit. The model indicated by an * is the one with the lowest reduced $\chi^{2}$ value. In this case, the best-fit Limepy model has a $W_0$ of $3.63\pm0.41$ while the Wilson model has $W_0$=5.09$\pm$0.03). The best-fit {\sc spes} model has $W_0= 4.99\pm0.10$, $\eta = 0.23\pm0.01$ and $\log_{10}(1-B)=-2.59\pm0.23$. This results in a PE fraction of 0.25$\pm$0.09 per cent of the total mass. Finally, the derived tidal radius is $\rt=51.51\pm4.52^{\prime}$, which is slightly larger than the estimated Jacobi radius of $\rj=31.80^{\prime}$ from \citet{Balbinot18}. \label{ngc1261_dens}}
\end{figure*}

Figure~\ref{ngc1261_dens} shows an example number density profile fit for NGC~1261 with best fitting models overlaid. The errorbars on individual data points are Poisson uncertainties for each radial bin. King and Wilson models with parameters taken from \citet{McLaughlin05} are shown as blue and green dashed lines respectively, while the {\sc limepy} model is shown as the solid black line. The red line shows the best-fit \spes\ model including PEs. It is clear from Figure~\ref{ngc1261_dens} that King and Wilson models do not manage to fit the outermost density profile, truncating at radii of $\approx$5 and 9$^{\prime}$ respectively, which falls far short of the 31.8$^{\prime}$ Jacobi radius from \citet{Balbinot18}. Even the {\sc limepy} model does not manage to reproduce the outer slope of the number density profile completely. However, the {\sc spes} model does provide a good fit of the GC profile, both in the very centre and in the outskirts. The best-fit parameters of the {\sc spes} model are $W_0= 4.99\pm$0.10, $\eta= 0.23\pm$0.01 and $\log_{10}(1-B)=-2.59\pm0.23$, resulting in a fraction of PEs of $0.25\pm0.09\%$ of the total mass. The derived tidal radius of the model is $\rt=51.51\pm4.52^{\prime}$, indicating this cluster is much more extended (factor of 5-10 larger $r_{\rm t}$) than can be inferred from simple single-mass models like King and Wilson. The number density profiles and model fits are shown for all GCs in Figure~\ref{GC_dens_plots} in Appendix~\ref{GC_numdens_all}.

\section{results}\label{results}
Our analysis of all GCs in the Harris catalogue with $\vert b\vert>5$ deg (113 clusters) resulted in PMs and number density profiles for 81 clusters. The remaining GCs are rejected from our final sample due to a variety of reasons, including being too distant to contain enough stars in {\it Gaia} DR2, suffering from poor scanning law coverage or sampling incompleteness resulting in profiles that could not be tied to literature values. The remaining GCs have been fit using single-mass models, with model fits shown in Figure~\ref{GC_dens_plots}. The best-fit parameters of the models are given in Table~\ref{GCpars} in Appendix~\ref{GC_numdens_fitpars}. 

\begin{figure}
\centering
\includegraphics[angle=0, width=0.495\textwidth]{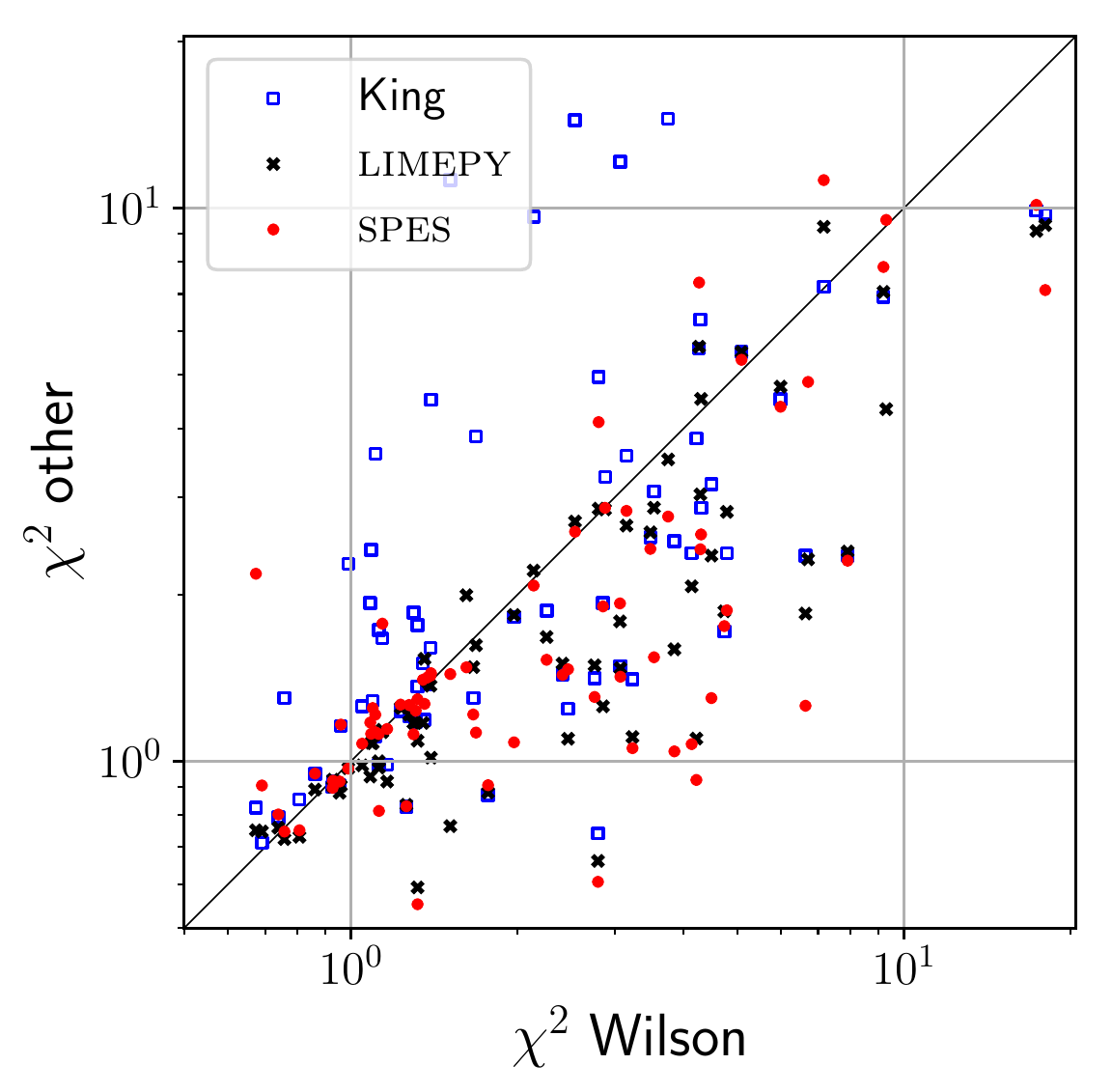}
\caption{Comparison between the reduced $\chi^{2}$ values of the Wilson fits and those for the other model fits. Points below the line indicate a fit better than Wilson, while points above indicate a fit worse than Wilson. For the majority of GCs, {\sc spes} models result in a better $\chi^{2}$ than {\sc limepy} models. \label{GC_chi2}}
\end{figure}

Analysis of the fits in Figure~\ref{GC_dens_plots} shows that King and Wilson models are typically not a good fit to our GC density profiles, especially in the outskirts. In almost all cases, a {\sc limepy} or {\sc spes} model results in a better or equally good fit. Nonetheless, there are some GCs for which a King or Wilson model results in the lowest  $\chi^{2}$ value (indicated by the * in the plot legend). In those cases, the profiles of all fitted models are very similar, but the simpler model is preferred due to the lower number of model parameters. Figure~\ref{GC_chi2} shows the reduced $\chi^{2}$ values for the different model fits as a function of the reduced $\chi^{2}$ value computed from the comparison between the Wilson model with \citet{McLaughlin05} parameters and our profile. The King models provide worse fits for the majority of GCs \citep[as found already by][]{McLaughlin05}, although a subsample of our clusters are fit much better by King than Wilson models. The fits for {\sc limepy} and {\sc spes} models result in fits better than Wilson profiles for all but 2 GCs. Furthermore, for the majority of GCs, a {\sc spes} model shows a smaller reduced $\chi^{2}$ value than a {\sc limepy} model, indicating the outer GC structure is better matched with the inclusion of PEs. Therefore, we can conclude that both King and Wilson models are too simplistic, and {\sc limepy} or {\sc spes} models are needed to explain the distribution of GC stars simultaneously in the inner and outer regions. 

\subsection{Model comparisons}\label{modelcomparison}
\begin{figure}
\centering
\includegraphics[angle=0, width=0.495\textwidth]{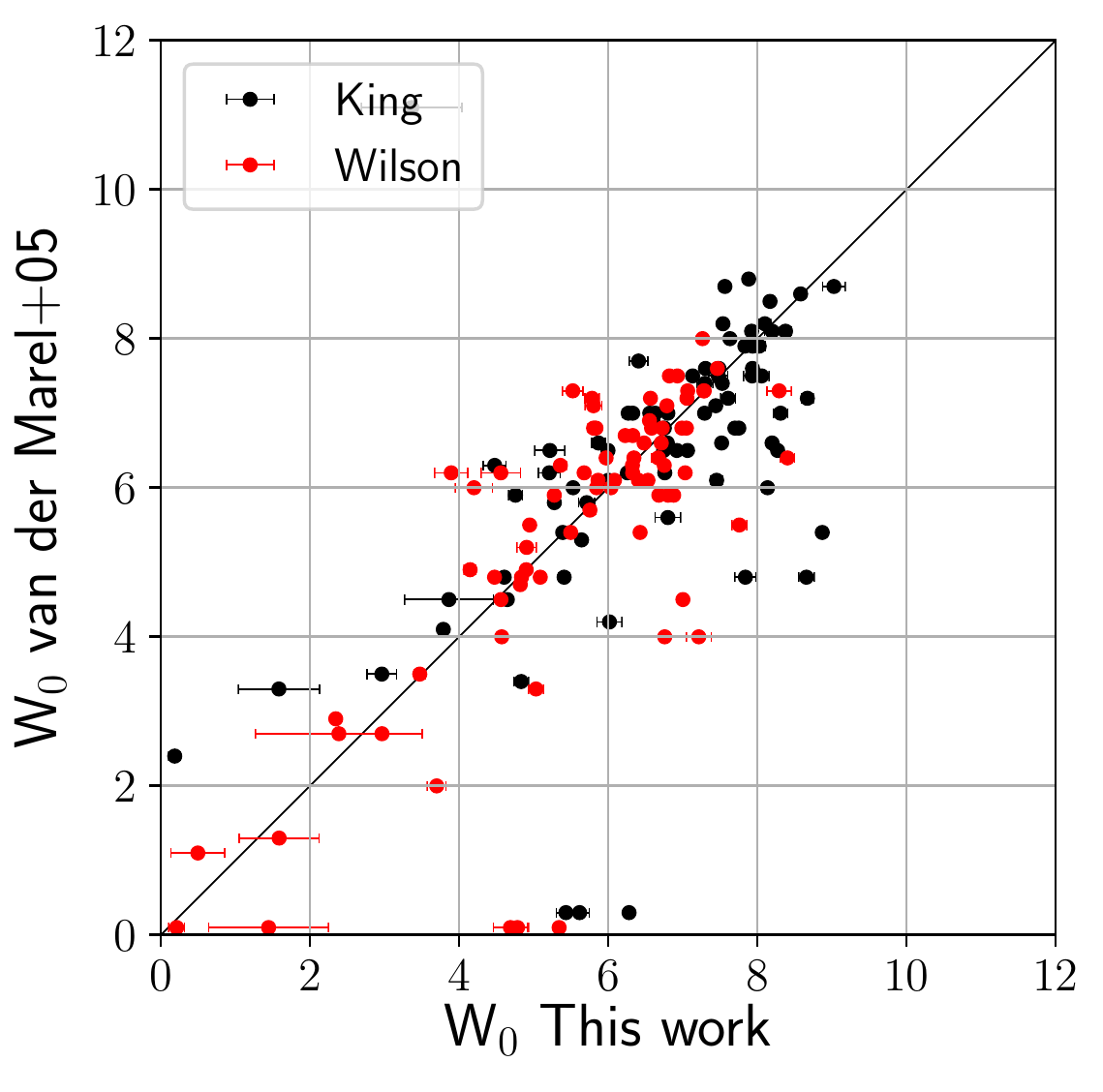}
\caption{Comparison between the $W_0$ parameter for King and Wilson as derived from our fits and the values from \citet{McLaughlin05}. The solid line indicates the one to one correlation.} \label{GC_KingWilson_W_comp}
\end{figure}

We can compare the structural parameters of the different model fits, and correlate them with literature values. First off, Figure~\ref{GC_KingWilson_W_comp} shows the comparison between the $W_0$ parameter as presented in Table~\ref{GCpars} as derived from the fits to our new profiles and the literature values from \citet{McLaughlin05}. The recovered values are in good agreement to the literature values for most of the GCs, with a few notable outliers at low $W_0$ such as NGC~6101 and NGC~6496 which were notably incompletely sampled in \citet{McLaughlin05}.

\begin{figure}
\centering
\includegraphics[angle=0, width=0.495\textwidth]{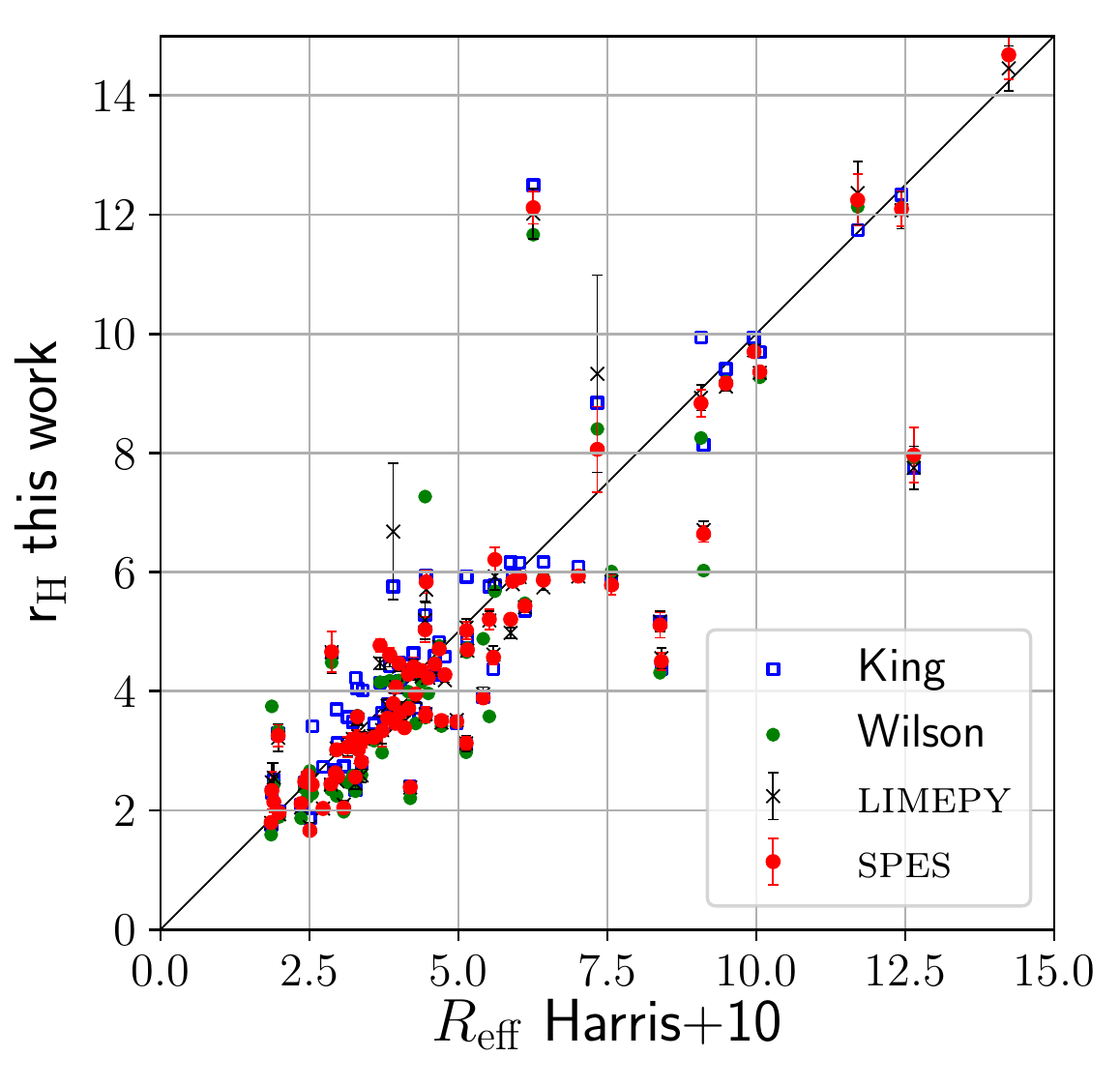}
\caption{Comparison between half mass radius R$_{\mathrm{half}}$ as derived from our fits and the values from \citet[][2010 version]{Harris96}. The solid line indicates the one to one correlation.} \label{GC_RH_comp}
\end{figure}

Next, Figure~\ref{GC_RH_comp} shows the values of the 3D half mass radius for each of the different model fits, in comparison to effective half mass radii from \citet[][2010 version]{Harris96} (which are mostly from \citet{McLaughlin05}), multiplied by a factor of 4/3 to correct for the radius projection. We note that we are neglecting any possible effect due to mass segregation. Our models all fall along the one-to-one correlation line, indicating good agreement between the literature and our models.

\begin{figure}
\centering
\includegraphics[angle=0, width=0.495\textwidth]{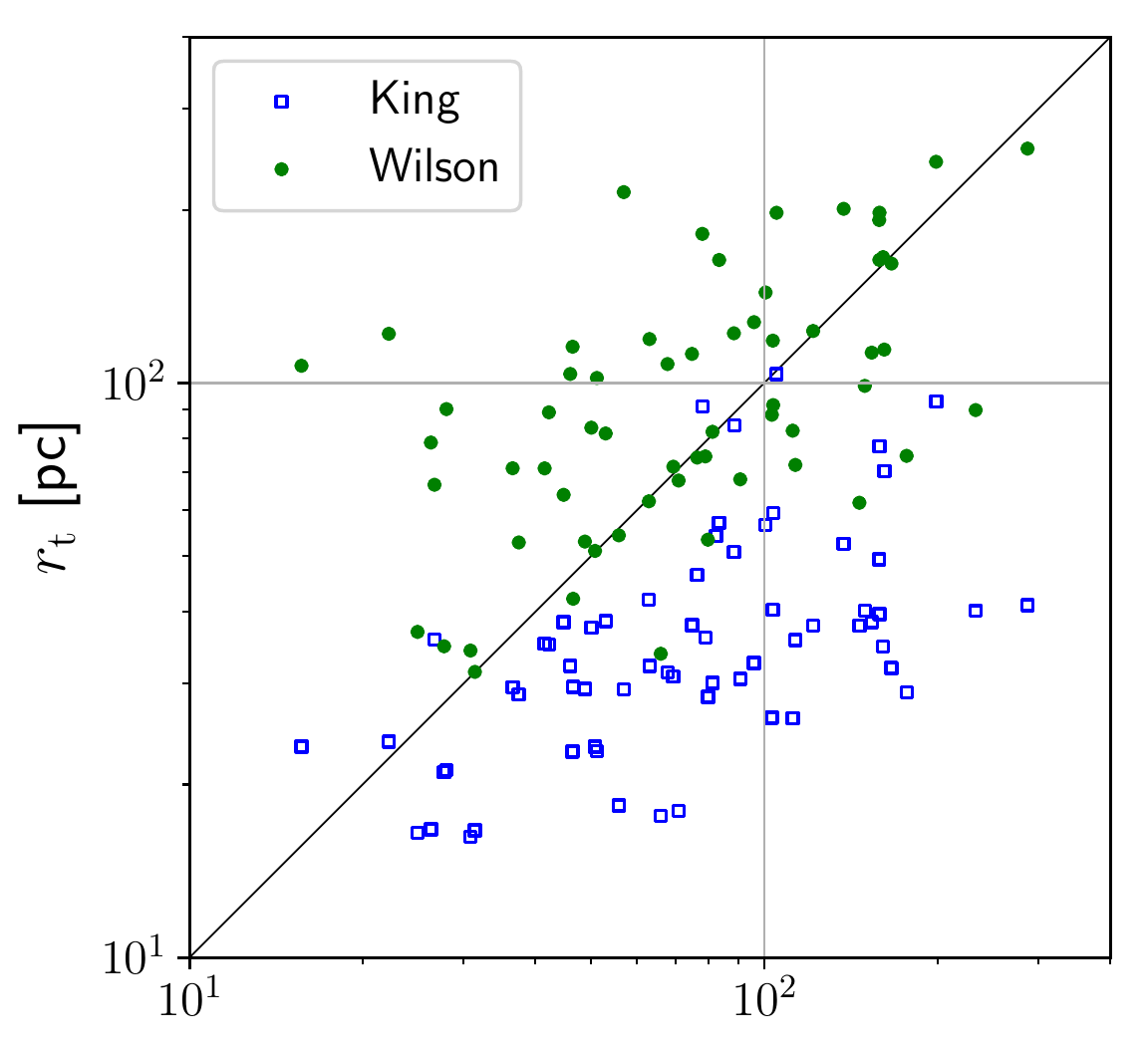}
\includegraphics[angle=0, width=0.495\textwidth]{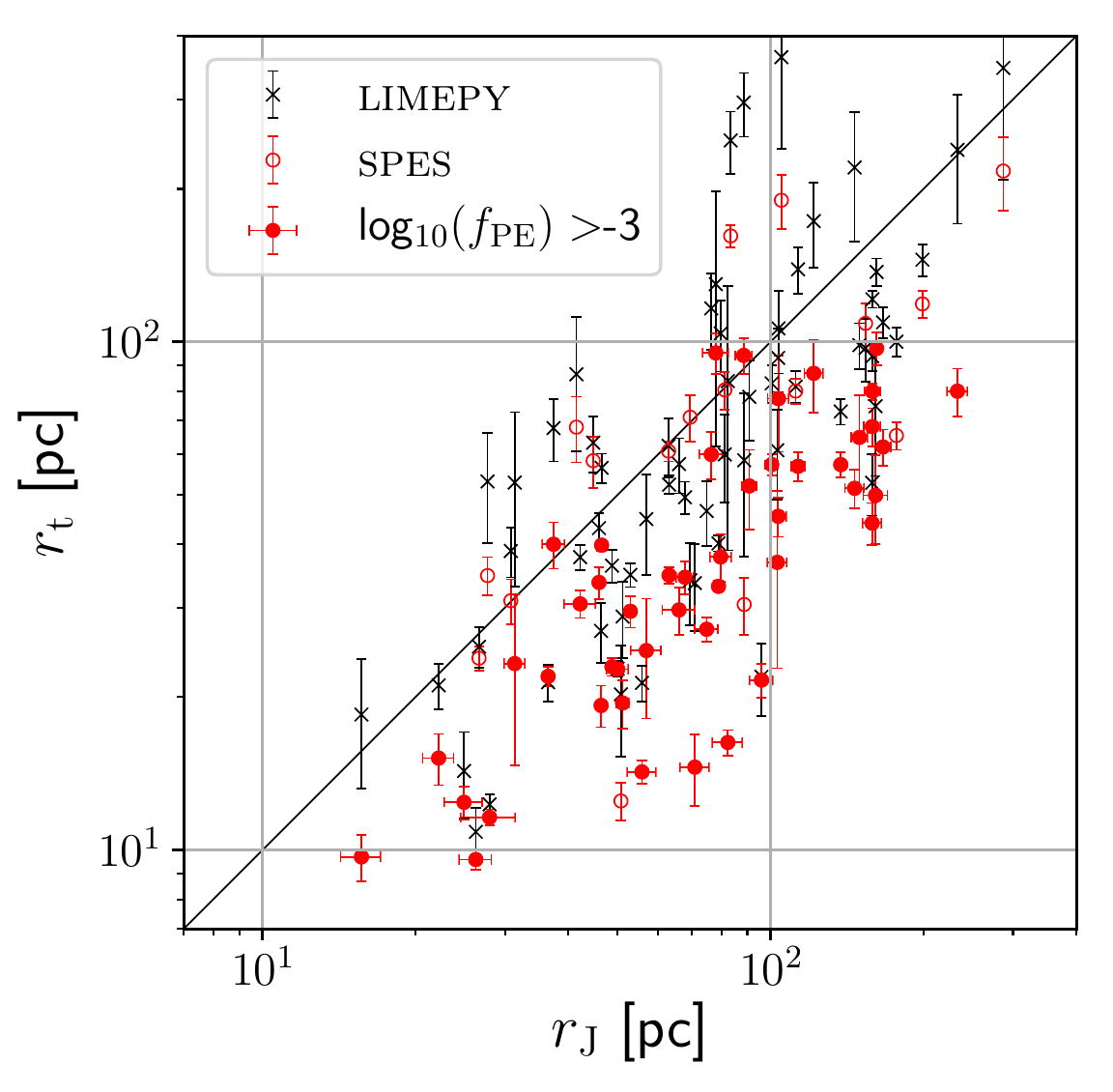}
\caption{Comparison between $r_{\rm t}$ as derived from our fits and the R$_{\mathrm{J}}$ values from \citet{Balbinot18}. GCs with large ($>$0.1 percent) fractions of PEs are shown as solid red symbols. We note that King and Wilson models are not fits, but simply use parameters as given in \citet{McLaughlin05}. \label{GC_RJ_comp}}
\end{figure}

Given the large radial extent of the {\it Gaia} DR2 data, it is insightful to look at the tidal radii as derived from our fits. In Figure~\ref{GC_RJ_comp} we show the tidal radius of each model in comparison to the values of the Jacobi radius as determined by \citet{Balbinot18}. For reference, the Jacobi radii are computed following R$_{\mathrm{J}}$ = [G M$_{cluster}$ / 2* 220$^{2}$]$^{1/3}$ R$_{GC}^{2/3}$, in which M$_{cluster}$ is the present day mass of the GC and R$_{GC}$ is the Galactocentric radius. The top panel of Figure~\ref{GC_RJ_comp} indicates the truncation radii of King fits is too small, owing to the intrinsic shape of the model. Values derived from Wilson fits are more diverse, with roughly half showing larger truncation radii than Jacobi radii. Comparison of model fits in Figure~\ref{GC_dens_plots} makes it clear that \citet{McLaughlin05} parameters are simply not a good representation of the outskirts of many of these GCs, such as NGC6121. 

The bottom panel of Figure~\ref{GC_RJ_comp} shows that tidal radii from {\sc limepy} fits are mostly in agreement with the Jacobi radii estimates from \citet{Balbinot18}. The {\sc spes} fits result in tidal radii which are mostly below the Jacobi radii but with a clear subset with values above or in agreement with the estimates based on the mass and orbit. The difference between the two groups is related to fraction of PEs ($\fpe$) recovered in the best-fit. As expected, a larger $\fpe$ leads to a decrease in the fitted tidal radius. This can be understood by considering that the PEs can have an energy greater than the binding energy and can therefore reside at distances greater than the tidal radius. Conversely, for {\sc limepy} fits, the tidal radius will be larger to model the PEs as if they were bound stars. Fitting the density of these stars as bound objects therefore leads to an overestimate of the tidal radius when using {\sc limepy} models. Models with $\log_{10}(\fpe)>  -3$ (i.e. more than 0.1 per cent) are shown as full red symbols, and consistently show tidal radii smaller than Jacobi radii. 

\subsection{Structural parameters}\label{strucpars}
We will now focus on the results from the {\sc limepy} and {\sc spes} fits, and analyse them further to look for trends of GC structural parameters as a function of environment or initial parameters. 

\begin{figure}
\centering
\includegraphics[angle=0, width=0.495\textwidth]{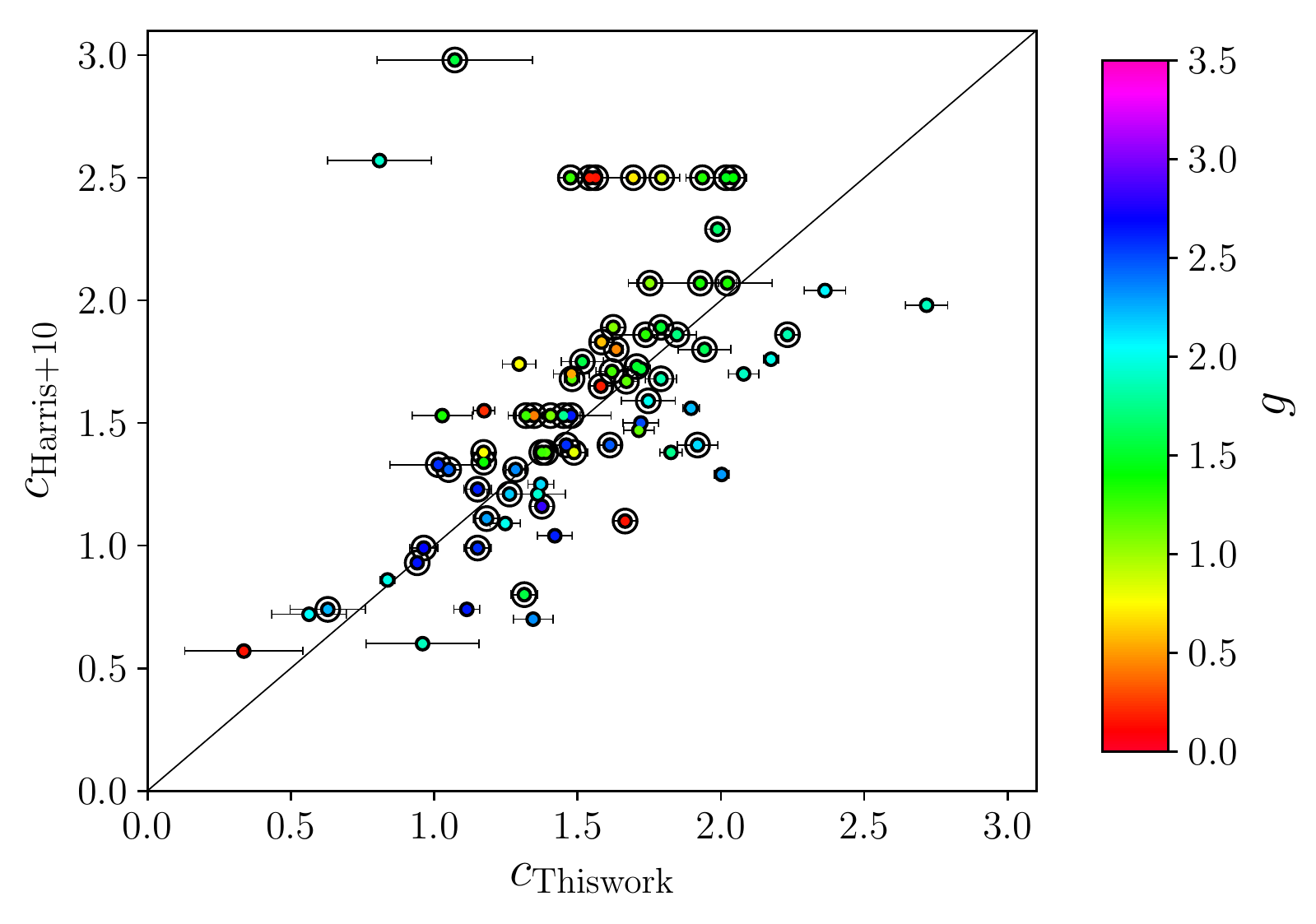}
\caption{Comparison between the concentration parameter $c$ as defined by log$_{10}$($r_{\rm t}$/$r_{\mathrm{core}}$) for our {\sc spes} models and those from \citet[][2010 version]{Harris96}. Higher values of $c$ indicate a higher concentration, with $c=2.5$ typically classified as core collapse GCs. Colours indicate the truncation parameter $g$ from {\sc limepy} fits. GCs with large ($>$0.1 percent) fractions of PEs are shown as points with an additional circle around them. \label{GC_c_comp}}
\end{figure}

First off, in Fig.~\ref{GC_c_comp} we compare the recovered concentration $c$ of our {\sc spes} models to those derived by \citet[][2010 version]{Harris96}. Concentration $c$ is defined as log$_{10}$($r_{\rm t}$/$r_{\mathrm{core}}$) with $r_{\mathrm{core}}$ being the core radius (the distance from the cluster centre at which the surface brightness drops by a factor of two from the central value). In the definition of $c$ employed in \citet{McLaughlin05} the King core radius is used, but the difference between the two quantities is negligible for all but the lowest $W_0$ GCs. There is good agreement between the two concentration parameters, indicating the concentration is largely consistent in between King and {\sc spes} models. Nonetheless, there is noticeable scatter around the 1:1 line due to the different tidal radii used, which are in some cases off by a factor of 2 or more. The colours of points in Fig.~\ref{GC_c_comp} represents the {\sc limepy} truncation parameter $g$, which is a measure of the extent of the cluster halo. The figure shows that more concentrated GCs typically show a lower value of $g$ (i.e. are for instance more King-like than Wilson-like), but there is clear variation in $c$ between GCs with the same truncation parameter $g$. This indicates that $g$ alone does not provide a unique measure of cluster concentration, but does anti-correlate with increased concentration. 

\begin{figure}
\centering
\includegraphics[angle=0, width=0.49\textwidth]{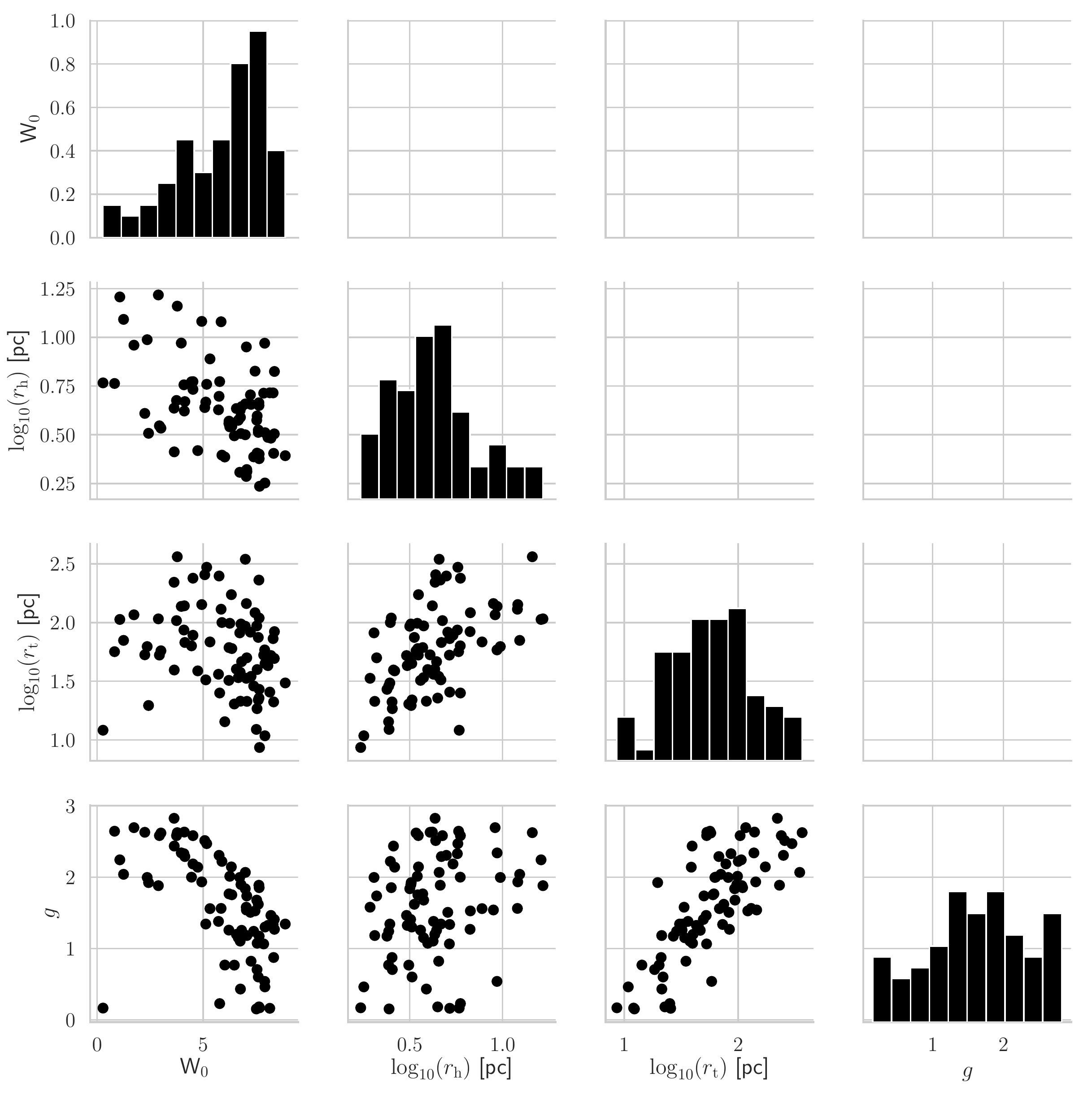}
\includegraphics[angle=0, width=0.49\textwidth]{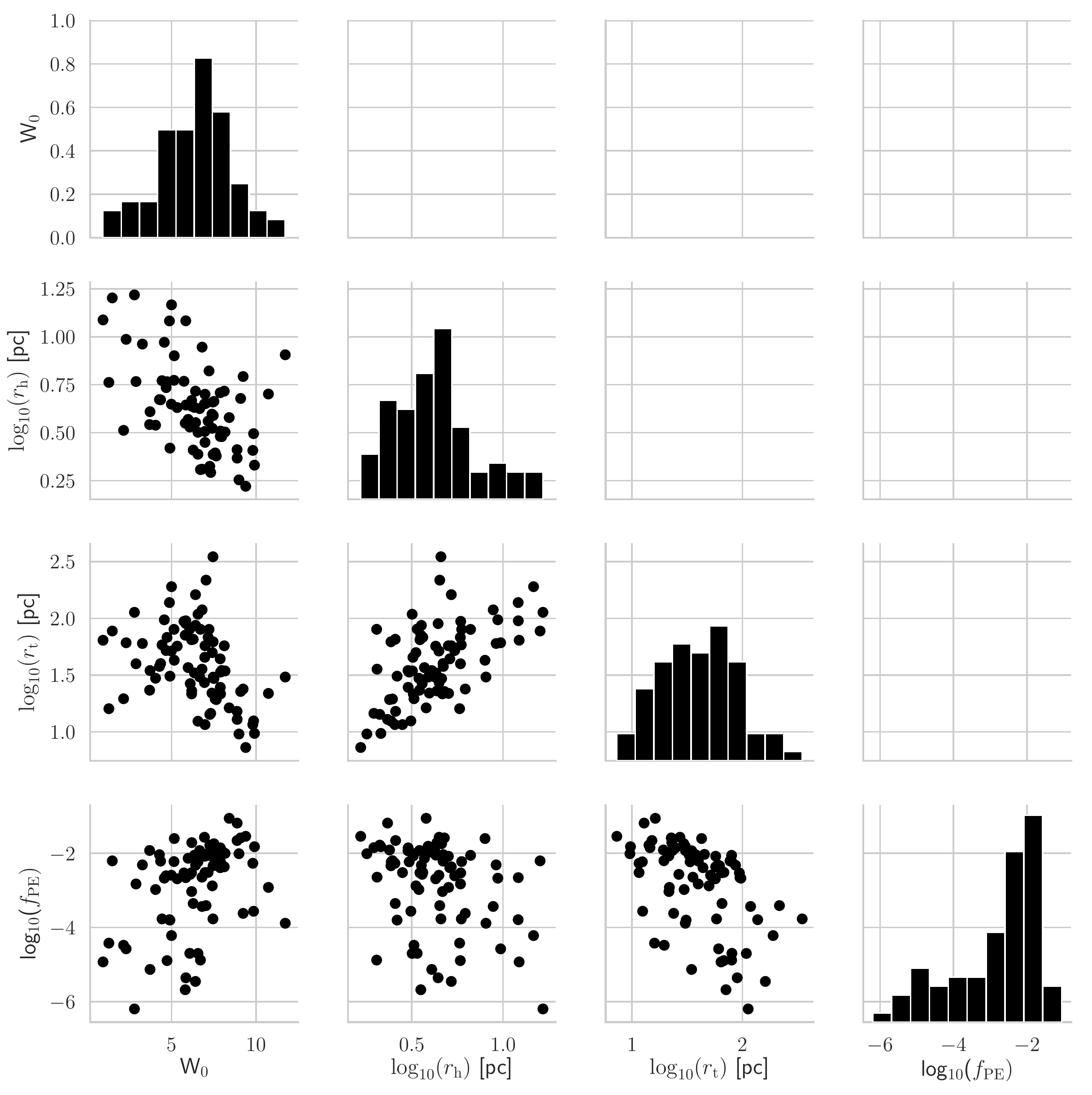}
\caption{Scatter matrix plots of {\sc limepy} and {\sc spes} fit parameters. \label{limepy_SPES_corner}}
\end{figure}

Next, we discuss the parameters derived for the {\sc limepy} and {\sc spes} model fits, as shown in Figure~\ref{limepy_SPES_corner} in both scatter plots and histograms. The truncation parameter $g$ correlates with the tidal radius, as expected, while $W_0$ weakly correlates with both half-mass and tidal radius. The best-fit {\sc limepy} fit parameters result in half mass radii which peak at $\approx$5 pc and tidal radii covering a range in between 30$-$130 pc. The GC sample from \citet[][2010 version]{Harris96} covers a variety of morphologies, with a wide range in both dimensionless potential $W_0$ and $g$. Strikingly, there is a clear correlation between the two parameters, with GCs with high $W_0$ having lower truncation parameter $g$ on average. The single GC showing both low $W_0$ and $g$ is Pal 11, for which the available data is low quality due to its distance and location close to the Galactic bulge.

\begin{figure*}
\centering
\includegraphics[angle=0, width=0.95\textwidth]{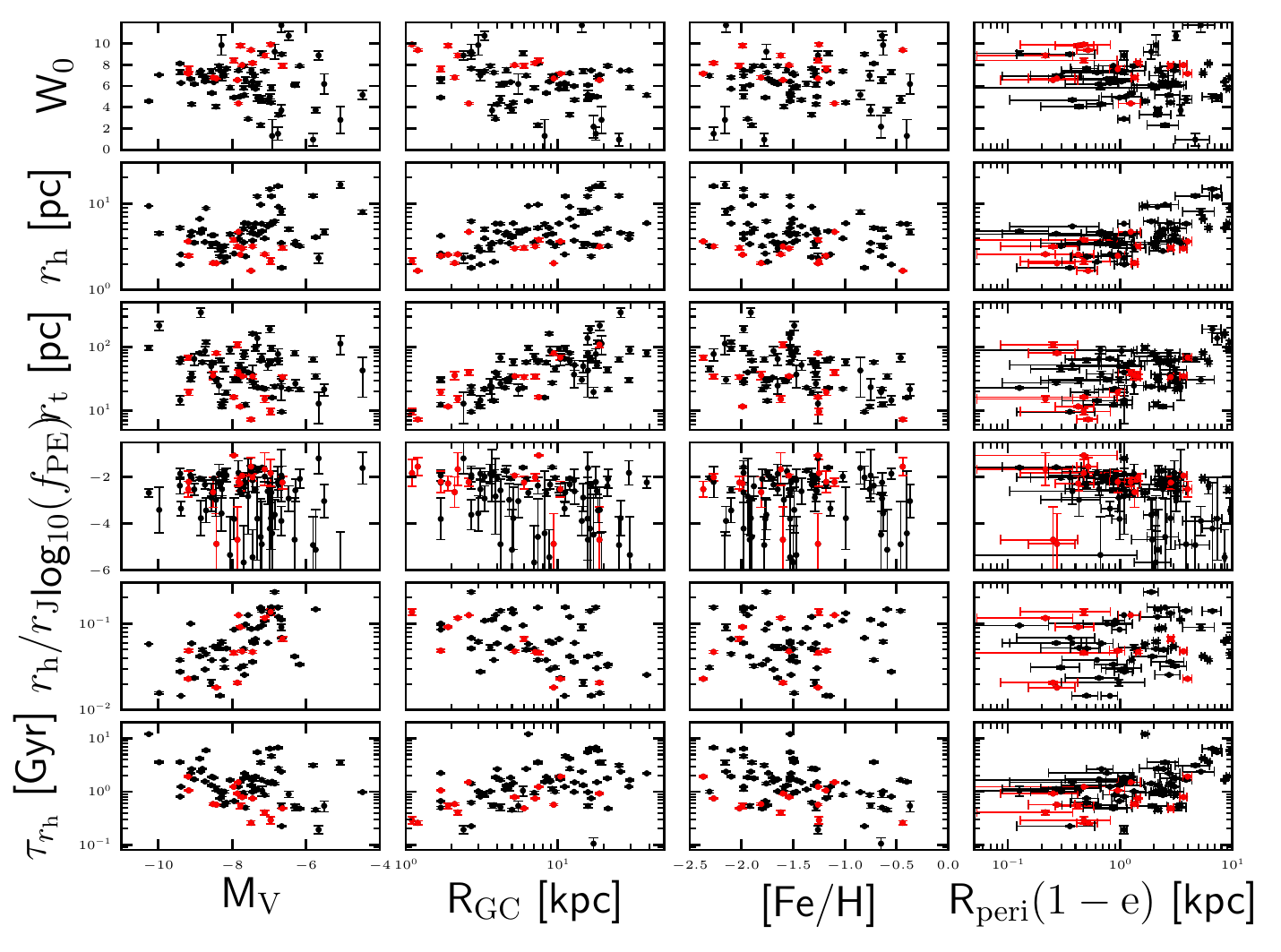}
\caption{Correlation plots showing structural values from the {\sc spes} fits ($W_0$, half mass radius $r_{\mathrm{h}}$, tidal radius, f$_{\mathrm{PE}}$, fractional $r_{\mathrm{h}}$ and half mass relaxation time $\tau_{\mathrm{r_{h}}}$) versus global values (integrated $V$-band luminosity, Galactocentric radius, metallicity and pericentre distance). Black points show all GCs, while red points show core-collapse clusters. \label{GC_corr_SPES}}
\end{figure*}

The {\sc spes} fit parameters are shown in the bottom panels of Figure~\ref{limepy_SPES_corner}. Once again, half mass radii peak around values of 5 pc, while tidal radii peak at values around 30-50 pc, consistent with results from Figures~\ref{GC_RH_comp} and \ref{GC_RJ_comp}. The fraction of PEs in the {\sc spes} fits shows a peak at $\log_{10}(\fpe)=-2$ with a long tail towards negligible PE fractions. The fraction of PEs does not strongly correlate with $W_0$ like the $g$ parameter of the {\sc limepy} fits, although higher values of $\fpe$ tend to be found for GCs with a higher value of $W_0$.
\begin{figure*}
\centering
\includegraphics[angle=0, width=0.95\textwidth]{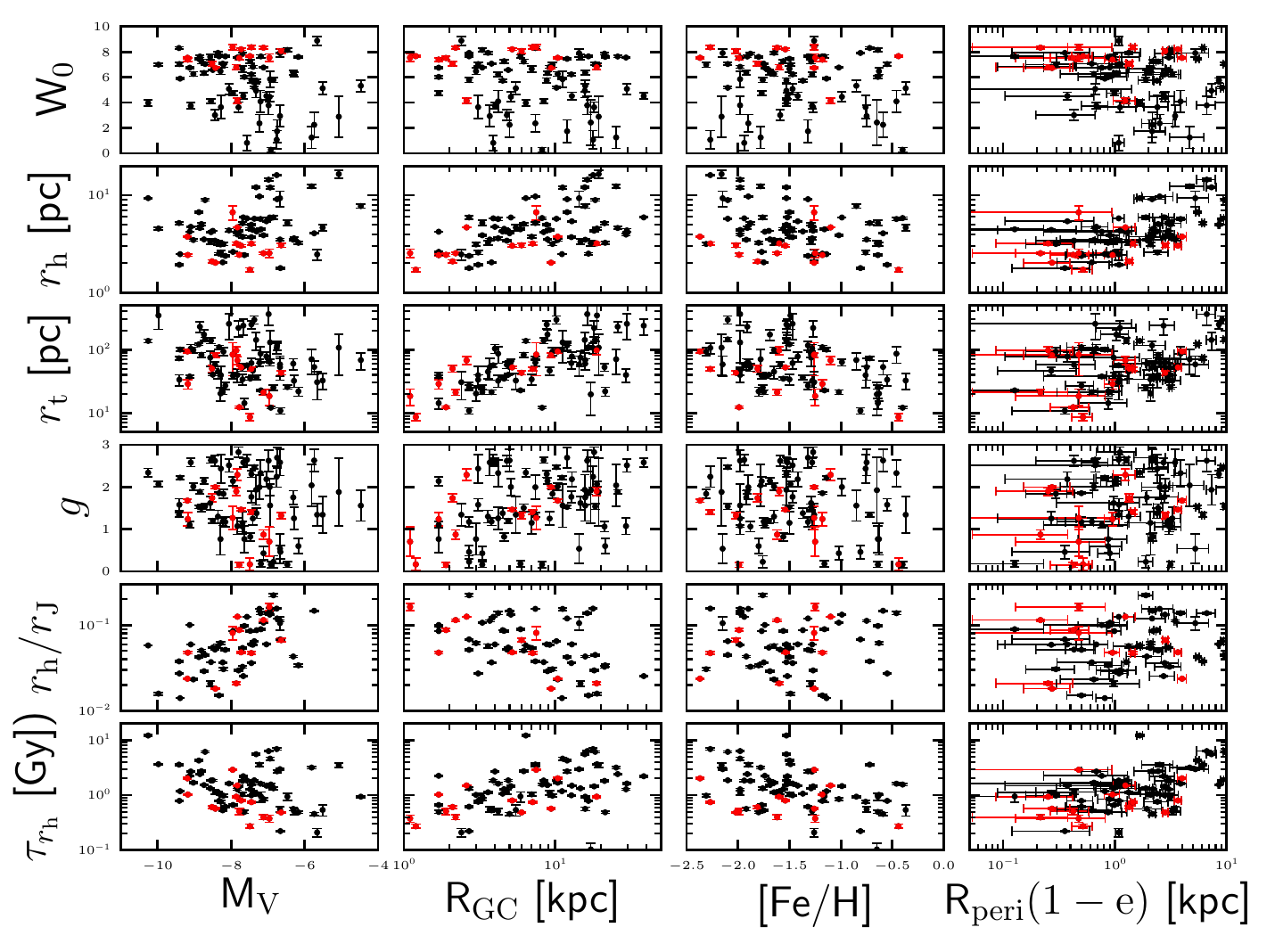}
\caption{Correlation plots showing structural values from the {\sc limepy} fits ($W_0$, half mass radius $r_{\mathrm{h}}$, tidal radius, $g$, fractional $r_{\mathrm{h}}$ and half mass relaxation time $\tau_{\mathrm{r_{h}}}$) versus global values (integrated $V$-band luminosity, Galactocentric radius, metallicity and pericentre distance). The red points show core-collapse clusters, according to \citet[][2010 version]{Harris96}. \label{GC_corr_limepy}}
\end{figure*}
Besides structural parameters, we can also compare the best-fit model values to environmental and global parameters. To that end, we have compiled a list of parameters from \citet[][2010 version]{Harris96} including integrated $V$-band luminosity, Galactocentric radius and metallicity. Furthermore, we also consider orbital information from \citet{Vasiliev18} and compute GC pericentre radii. Figure~\ref{GC_corr_limepy} displays the {\sc limepy} parameters as a function of the environmental parameters, while Figure~\ref{GC_corr_SPES} displays the {\sc spes} parameters. Besides basic structural parameters, we also included the half mass relaxation time, following the prescription by \citet{McLaughlin05} who in turn followed \citet{Binney87} ($\tau_{\mathrm{r_{h}}}/{\rm yr}$ = $\left[2.06\times10^6/\ln(0.4M_{\rm tot}/m_\star)\right]m_\star^{-1} M_{\rm tot}^{1/2} \mathrm{r_{h}}^{3/2}$ with m$_\star$ = 0.5M$_\odot$).

It is clear once again that the sample of 81 GCs displays a wide variety of morphologies and covers a range in both luminosity, metallicity and Galactocentric radius. There are a number of clear correlations in Figure~\ref{GC_corr_limepy}, some of which are obvious. For instance, $r_{\rm t}$ correlates with R$_{\mathrm{GC}}$ given the weaker tidal field at large Galactocentric radius\citep{vonHoerner57}. Additionally, we also see that metallicity correlates with half mass and tidal radii, which is likely a manifestation of the underlying correlation between Galactocentric radius and metallicity~\citep{vandenBergh11}. GCs with brighter integrated $V$-band luminosity typically display higher values of truncation parameter $g$ and lower values of $r_{\mathrm{h}}$/$r_{\mathrm{J}}$, due to their smaller half-mass radii leading to bright cores. 

There are several other correlating parameters among the best-fit {\sc limepy} parameters. The dimensionless potential $W_0$ correlates weakly with $V$-band luminosity, and GC pericentre radius. The concentration sensitive {\sc limepy} $g$ parameter clearly correlates with Galactocentric radius $R_{\mathrm{GC}}$ showing that outer MW GCs are less concentrated than those more inward, similar to results from~\citet{Djorgovski94,vandenBergh11}. The $r_{\mathrm{h}}$/$r_{\mathrm{J}}$ parameter also correlates with R$_{\mathrm{GC}}$, with lower values found at larger Galactocentric radius. Similar to \citet{Baumgardt10} we see a group of GCs with both a large Galactocentric radius and high $r_{\mathrm{h}}$/$r_{\mathrm{J}}$. The GCs found in this branch preferentially display lower $W_0$ and $g$ than the bulk of the clusters. Unfortunately, our sample does not include as many GCs in this group as in \citet{Baumgardt10} due to their large distance pushing them out of the observable window of {\it Gaia}. Finally, the {\sc limepy} truncation parameter $g$ also correlates with Galactocentric radius, with more King-like GCs found preferentially at smaller radii. 

Figure~\ref{GC_corr_SPES} shows that some of the same correlations are present in the best-fit {\sc spes} parameters. The correlations with tidal radius are more pronounced in the {\sc spes} fits, given the results of Figure~\ref{GC_RJ_comp}. The fraction of PEs correlates weakly with the $V$-band luminosity in the sense that higher luminosity GCs have less PEs. The fraction $f_{\mathrm{PE}}$ also correlates with both the Galactocentric radius and pericentre distance, with larger distance leading to a lower fraction of PEs, likely due to experiencing weaker gravitational fields. The pericentre distance also correlates with half-mass radius $r_{\mathrm{h}}$ and W$_0$, showing a higher W$_0$ and smaller $r_{\mathrm{h}}$ for GCs with small pericentres. Therefore, the Galactic tidal field exerts an influence not just on the very outskirts of GCs but also further into the cluster centre.
\begin{figure}
\centering
\includegraphics[angle=0, width=0.49\textwidth]{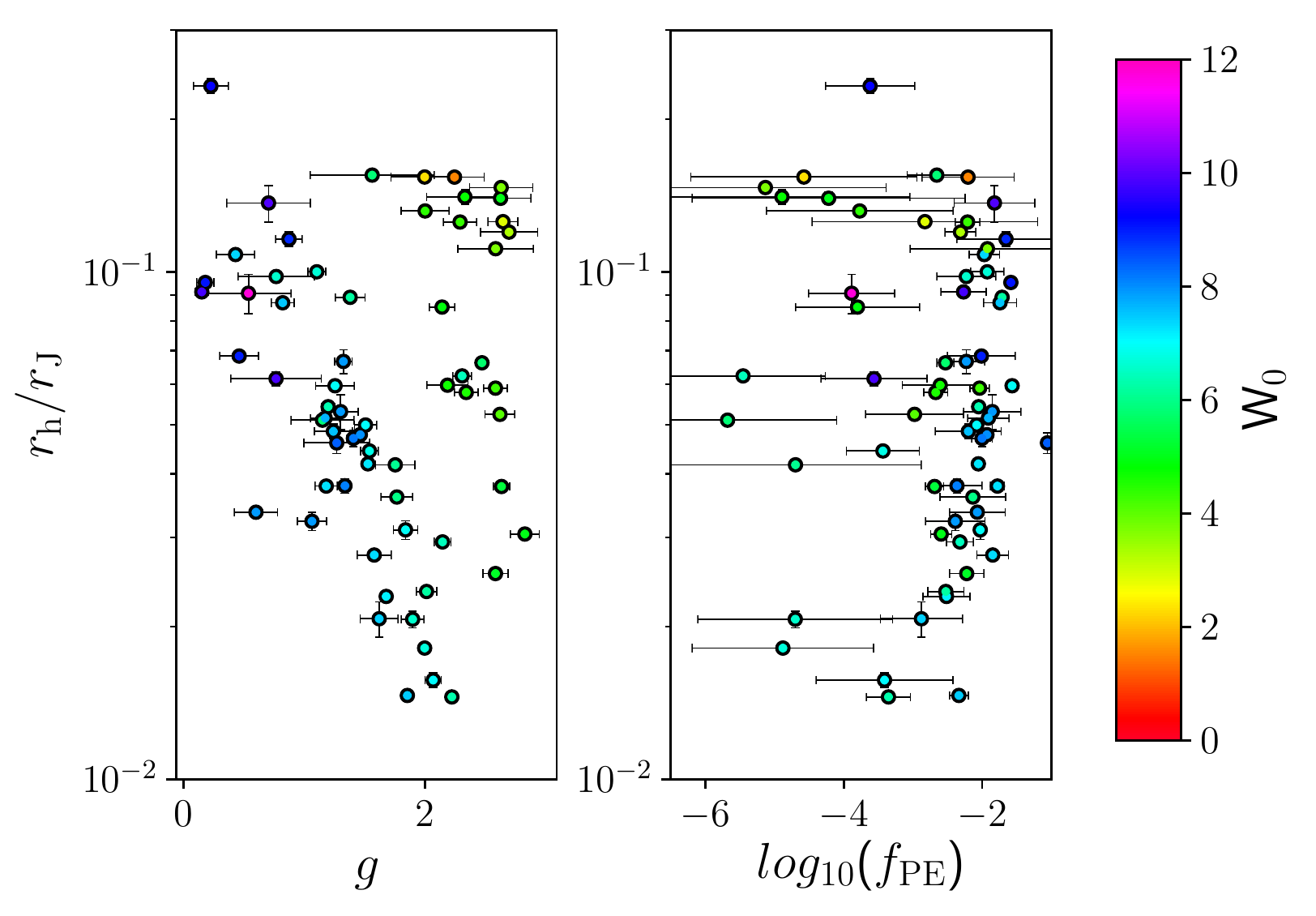}
\caption{Comparison between the ratio $r_{\mathrm{h}}$/$r_{\mathrm{J}}$ and $g$ or $f_{\mathrm{PE}}$ for the {\sc limepy} and {\sc spes} models respectively. The colour of individual points indicates the value of $W_0$. \label{GC_RHRJcomp}}
\end{figure}

To investigate the parameters in more detail, Figure~\ref{GC_RHRJcomp} shows $r_{\mathrm{h}}$/$r_{\mathrm{J}}$ as a function of the {\sc limepy} truncation parameter $g$ and the {\sc spes} fraction of PEs. In the figure, points are coloured according to the dimensionless potential $W_0$ for each model fit. The left panel shows that truncation parameter and $r_{\mathrm{h}}$/$r_{\mathrm{J}}$ are clearly correlated for GCs with similar $W_0$, with for instance a diagonal sequence for systems with $W_0=7-8$. There is also a correlation with $W_0$ at fixed truncation parameter. The right panel of Figure~\ref{GC_RHRJcomp} shows the correlation between $r_{\mathrm{h}}$/$r_{\mathrm{J}}$ and the fraction of PEs. Looking just at GCs with a fraction of PEs higher than 0.1\% we see a weak correlation with $r_{\mathrm{h}}$/$r_{\mathrm{J}}$. GCs with higher $r_{\mathrm{h}}$/$r_{\mathrm{J}}$ are more likely to be Roche filling, in which case a higher fraction of PEs is expected, and inferred.

\begin{figure}
\centering
\includegraphics[angle=0, width=0.495\textwidth]{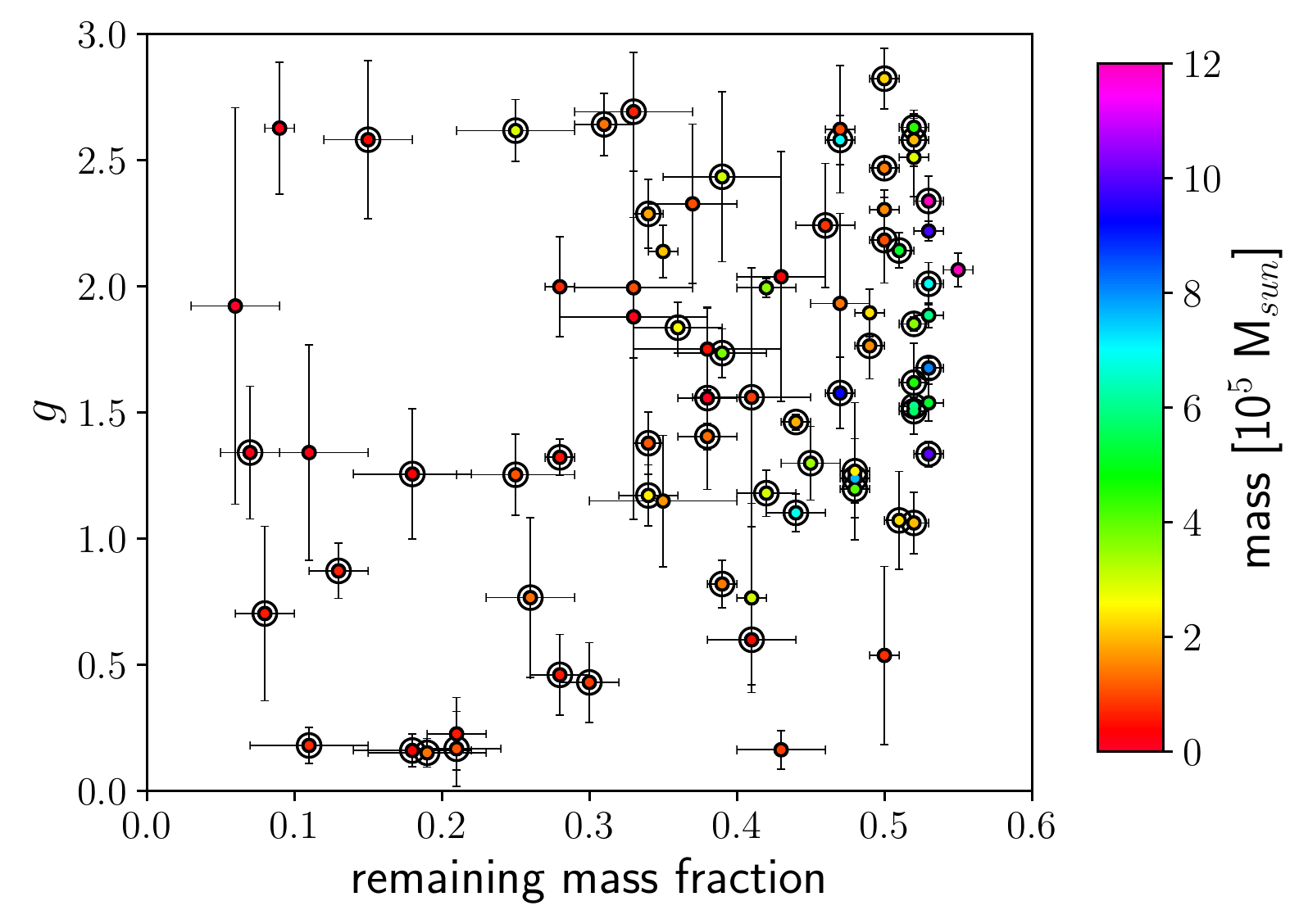}
\caption{Comparison between the remaining mass fraction $\mu$ from \citet{Balbinot18} and the  {\sc limepy} truncation parameter $g$. Lower values of $\mu$ indicate a larger fraction of the cluster has been lost, consistent with a more evolved cluster. Colours indicate the cluster mass from \citet[][2010 version]{Harris96}. GCs with large ($>$0.1 percent) fractions of PEs are shown as points with an additional circle around them. \label{GC_mu_comp}}
\end{figure}

Furthermore, Figure~\ref{GC_mu_comp} shows the {\sc limepy} truncation parameter $g$ as a function of the cluster remaining mass fraction of \citet{Balbinot18}, which is an indication of how evolved the cluster is. Simulations by \citet{Zocchi16} indicate that the cluster  truncation changes over time, with $g$ being smaller for more evolved clusters. Clusters start of with high $g$, which decreases to King-like values as they fill their Roche volume. This is indeed what we see in Figure~\ref{GC_mu_comp}, with more unevolved clusters showing Wilson-like profiles and evolved cluster with $\mu<$0.3 displaying King-like $g$. The 3 GCs with high g at low $\mu$ are pal$\_$1, NGC~6366 and ic~1276, which suffer from high background contamination or poor sampling in {\it Gaia}, which may affect the recovered $g$. A more thorough study of these GCs with {\it Gaia} DR3 would be beneficial to obtain a more accurate inner profile shape.

Figure~\ref{GC_mass_comp} shows a comparison between the integrated cluster mass from \citet[][2010 version]{Harris96} and the ratio $r_{\mathrm{h}}$/$r_{\mathrm{J}}$ from the {\sc limepy} models. A clear correlation is visible, with only little dependence on $f_{\mathrm{PE}}$, indicating that the $r_{\mathrm{h}}$/$r_{\mathrm{J}}$ fraction is driven primarily by mass. We see that cluster with lower mass are more Roche filling than the high mass clusters, or alternatively that massive clusters are under-filling their tidal radius. This could be linked to the effects of 2-body relaxation, with which larger masses have a longer relaxation time, which leads to a lower Roche lobe filling.

\section{Conclusions and discussion}\label{conclusions}
In this work, we have utilised data from {\it Gaia} DR2 to study the number density profile of GCs from the sample of \citet[][2010 version]{Harris96}. The proper motion selected samples of GC members are combined with literature data from \citet{Trager95} and \citet{Miocchi13} to obtain a full sampling of the density profile (see Section~\ref{profiles}). This is the first time that GC profiles are investigated using data covering both the inner regions and outskirts simultaneously. 

We have fit the combined density profiles using a variety of single-mass models, including often-used King and Wilson models, as well as the recently introduced {\sc limepy} models. Finally, we also utilise the recently developed {\sc spes} models (see Section~\ref{model_fits}), which include a prescription for the presence of PE stars, essential for reproducing the outskirts of GCs. 

The individual cluster fits in Appendix~\ref{GC_numdens_all} show that the King and Wilson model fits of \citet{McLaughlin05} are not sufficient to explain the density profile in the outskirts of GCs. The {\sc limepy} and {\sc spes} models fare better at reproducing the full density profile of our sample of GCs, with the {\sc spes} models in particular providing a better fit to low mass clusters like NGC~1261 (see also Figure~\ref{GC_chi2}). It is clear that including PEs in mass models is crucial for fully modelling GCs with a high Roche filling factor \citep[see also][]{2019MNRAS.483.1400H}. 
In section~\ref{results} we have compared the structural parameters of the different model fits to look for correlations with environmental parameters. Comparison of recovered tidal radii (Fig.~\ref{GC_RJ_comp}) makes it clear the fraction of PEs has a strong influence on the GC tidal radius, with fractions of 0.1 percent (by mass) leading to significantly smaller tidal radii. Comparison of best-fit parameters with environmental parameters also reveals correlations between some parameters, some of which are known and some of which are new (see Section~\ref{strucpars}). 

\begin{figure}
\centering
\includegraphics[angle=0, width=0.49\textwidth]{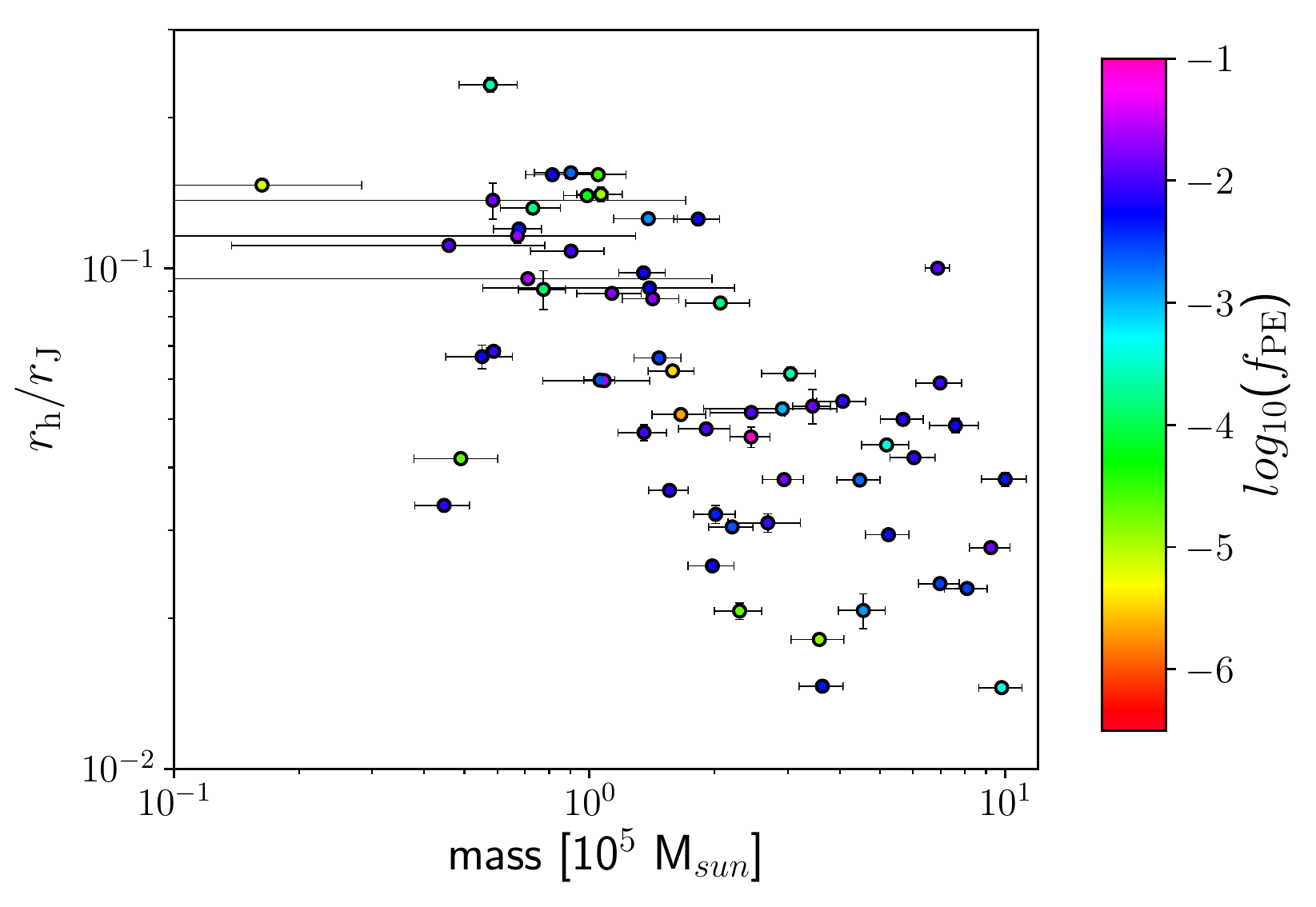}
\caption{Comparison between the mass of the clusters from \citet[][2010 version]{Harris96} and the ratio $r_{\mathrm{h}}$/$r_{\mathrm{J}}$ from the {\sc limepy} models. The colour of individual points indicates the value of $f_{\mathrm{PE}}$ from {\sc spes} models. \label{GC_mass_comp}}
\end{figure}

For instance, the comparison between {\sc limepy} dimensionless potential $W_0$ and truncation parameter $g$ in Figure~\ref{limepy_SPES_corner} shows that the expected correlation is not linear but levels out on both ends. Furthermore, it is clear that the sample of GCs cannot be described well by models using a single truncation parameter, such as King ($g=1$) or Wilson ($g=2$) models. The truncation parameter itself depends on both integrated $V$-band luminosity (probing the GC mass) and position within the Galaxy. 

Figure~\ref{GC_corr_SPES} shows that the fraction of PEs in a GC depends on both environment (pericentre distance) and structure (integrated brightness). As expected, closer pericentres result in stronger tidal fields and therefore a higher PE fraction. Finally, Figure~\ref{GC_RHRJcomp} shows us that high PE fractions are found in GCs with high $r_{\mathrm{h}}$/$r_{\mathrm{J}}$, but that this is also dependant on $W_0$. 

The analysis of structural and environmental parameters shows clear effects of current location and experienced tidal field on the properties of the cluster outskirts, such as tidal radius and fraction of PEs. The correlation between truncation parameter $g$ and Galactocentric radius shows that more King-like GCs are found preferentially at smaller radii, while more Wilson-like GCs are found further out. 
Figures~\ref{GC_corr_limepy} and~\ref{GC_corr_SPES} also show that more distant GCs are typically less concentrated than those more inward, similar to results from~\citet{vandenBergh11}. Similarly, the fraction of PEs correlates with environment, with larger distance leading to a lower fraction of PEs. This can be understood by taking into account the weaker Galactic tidal field at large distance.

Strikingly, the pericentre distance correlates with both half-mass radius and $W_0$, with low pericentre distance leading to a higher $W_0$ and smaller $r_{\mathrm{h}}$. This indicates that the Galactic tidal field has an effect on both the cluster outskirts as well as further into the centre. Figure~\ref{GC_mu_comp} shows that the structural parameters are influenced by its evolutionary state, with more evolved clusters becoming progressively more King-like, as predicted by simulations \citep{Zocchi16}. We also find that the fraction $r_{\mathrm{h}}$/$r_{\mathrm{J}}$ correlates strongly with cluster mass (Figure~\ref{GC_mass_comp}) and weakly with $f_{\mathrm{PE}}$ for clusters with a PE fraction greater than 0.1\% (Figure~\ref{GC_RHRJcomp}). Clusters which are more Roche filling have a lower mass and display a slightly higher fraction of PEs.

Finally, similar to \citet{vandenBergh11} we see little correlation between metallicity and structural parameters, apart from the correlation with tidal radius that seems more driven by Galactocentric radius. This is striking, given that samples of MW GCs are typically divided between birth environment on the basis of metallicity. 

Analysis of GCs in different environments has shown that the distinct groups of systems display different properties, among the MW, LMC and Fornax clusters. In this work, we only study GCs well within the confines of the MW, with the most distant objects reaching a Galactocentric radius of $\approx$40 kpc. Therefore, we cannot study the effect of environment on structural parameters with this sample. Reaching distant external cluster with accurate proper motions is outside the reach of {\it Gaia}, although the LMC and Fornax can be probed with limited number of stars per cluster.

It is clear that the structural properties of GCs are diverse and not simply modelled using a rigid set of distribution functions. The use of a generalised lowered isothermal model such as generated by {\sc limepy} is a first important step in fully describing their structure. However, in the future it is desirable to move away from single-mass models and employ multi-mass models with realistic mass functions for both the stars and stellar remnants to describe GCs, as done by e.g. \citet{Sollima18,Gieles18}. This will fully allow us to explore the structure and dynamics of GCs both in our local sample as well as in extra-Galactic environments.

\section*{Acknowledgements}
T.d.B., M.G. and E.B. acknowledge support from the European Research Council (ERC StG-335936). M.G. acknowledges financial support
from the Royal Society (University Research Fellowship). V.H.-B. acknowledges support from the NRC-Canada Plaskett Fellowship. The authors also thank the International Space Science Institute (ISSI, Bern, CH) for welcoming the activities of Team 407 ``Globular Clusters in the Gaia Era".

This work presents results from the European Space Agency (ESA) space mission {\it Gaia}. {\it Gaia} data are being processed by the Gaia Data Processing and Analysis Consortium (DPAC). Funding for the DPAC is provided by national institutions, in particular the institutions participating in the Gaia MultiLateral Agreement (MLA). The {\it Gaia} mission website is \url{https://www.cosmos.esa.int/gaia}. The Gaia archive website is \url{https://archives.esac.esa.int/gaia}.

This paper made used of the Whole Sky Database (wsdb) created by Sergey Koposov and maintained at the Institute of Astronomy, Cambridge by Sergey Koposov, Vasily Belokurov and Wyn Evans with financial support from the Science \& Technology Facilities Council (STFC) and the European Research Council (ERC).

The Pan-STARRS1 Surveys (PS1) and the PS1 public science archive have been made possible through contributions by the Institute for Astronomy, the University of Hawaii, the Pan-STARRS Project Office, the Max-Planck Society and its participating institutes, the Max Planck Institute for Astronomy, Heidelberg and the Max Planck Institute for Extraterrestrial Physics, Garching, The Johns Hopkins University, Durham University, the University of Edinburgh, the Queen's University Belfast, the Harvard-Smithsonian Center for Astrophysics, the Las Cumbres Observatory Global Telescope Network Incorporated, the National Central University of Taiwan, the Space Telescope Science Institute, the National Aeronautics and Space Administration under Grant No. NNX08AR22G issued through the Planetary Science Division of the NASA Science Mission Directorate, the National Science Foundation Grant No. AST-1238877, the University of Maryland, Eotvos Lorand University (ELTE), the Los Alamos National Laboratory, and the Gordon and Betty Moore Foundation.

\bibliographystyle{mn2e_fixed}
\bibliography{Bibliography}

\clearpage
\begin{appendix}
\onecolumn
\section{GC number density profile fits}
\label{GC_numdens_all}
In this appendix, we present the full set of GC number density profiles, along with the best-fit dynamical models discussed in section~\ref{model_fits}. For each GC, we show the final cluster number density profile, after tying together {\it Gaia} profiles with \citet{Trager95} or \citet{Miocchi13} profiles where available. The innermost reliable radius of the {\it Gaia} profile used to connect the profiles is shown as the solid vertical line, while the dashed vertical line indicates the Jacobi radius \citep{Balbinot18}. The background level estimated from the outer regions of the {\it Gaia} data is indicated with the horizontal dashed line. Parameters used for the models are given in Table~\ref{GCpars}.

\begin{figure}
\centering
\includegraphics[angle=0, width=0.95\textwidth]{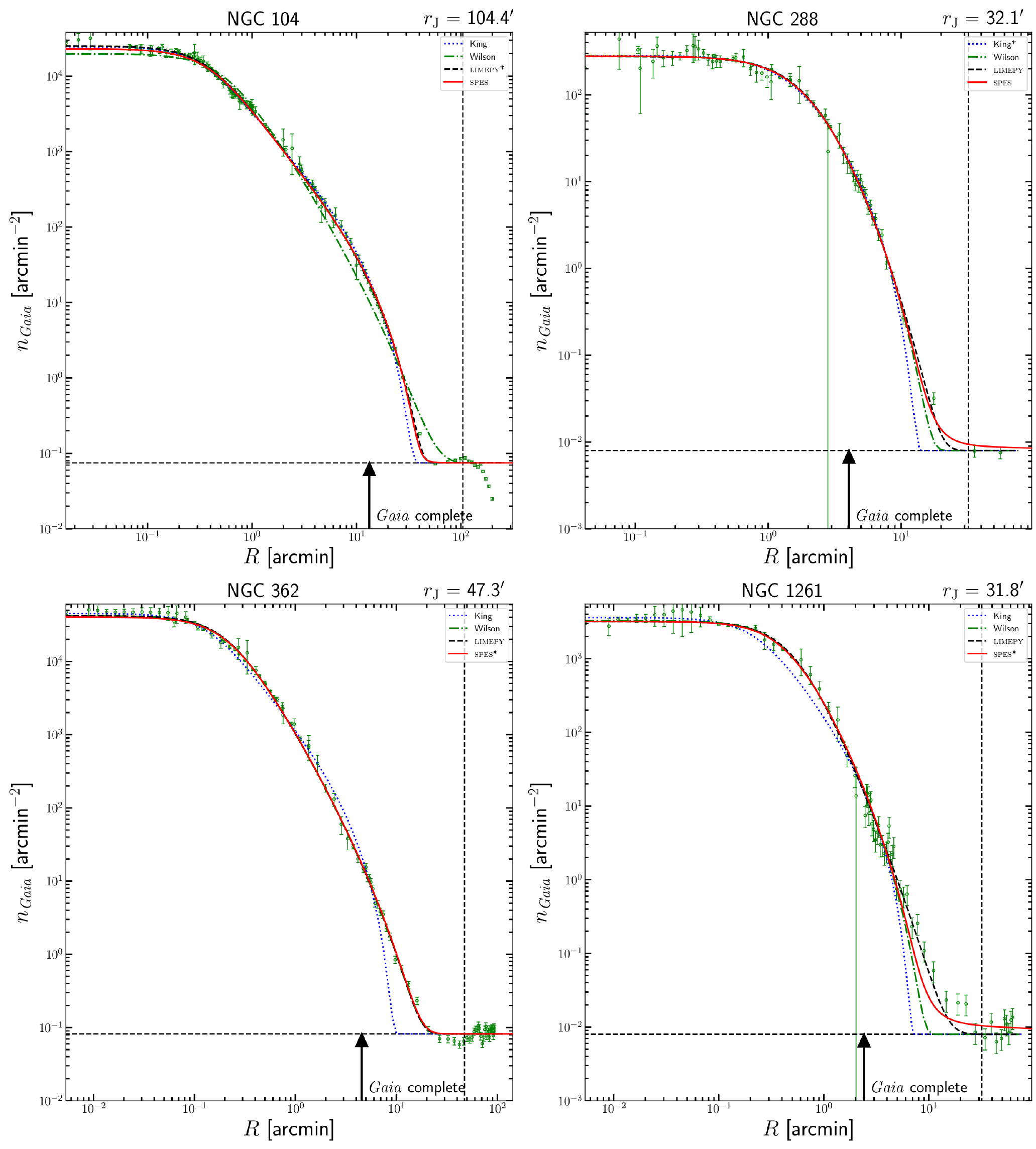}
\caption{The number density profiles of all GCs with converged fit parameters, along with best-fit dynamical models overlaid. The errorbars on individual data points are Poisson uncertainties for each radial bin. The blue and green dashed lines indicate King and Wilson models respectively, while the solid black line shows the best-fit {\sc limepy} model and the red solid line shows the {\sc spes} model fit. The model indicated by an * is the one with the lowest reduced $\chi^{2}$ value. The parameters used for the models are given in Table~\ref{GCpars}, along with the derived innermost reliable radius and tidal radius. \label{GC_dens_plots}}
\end{figure}

\begin{figure}
\centering
\includegraphics[angle=0, width=0.95\textwidth]{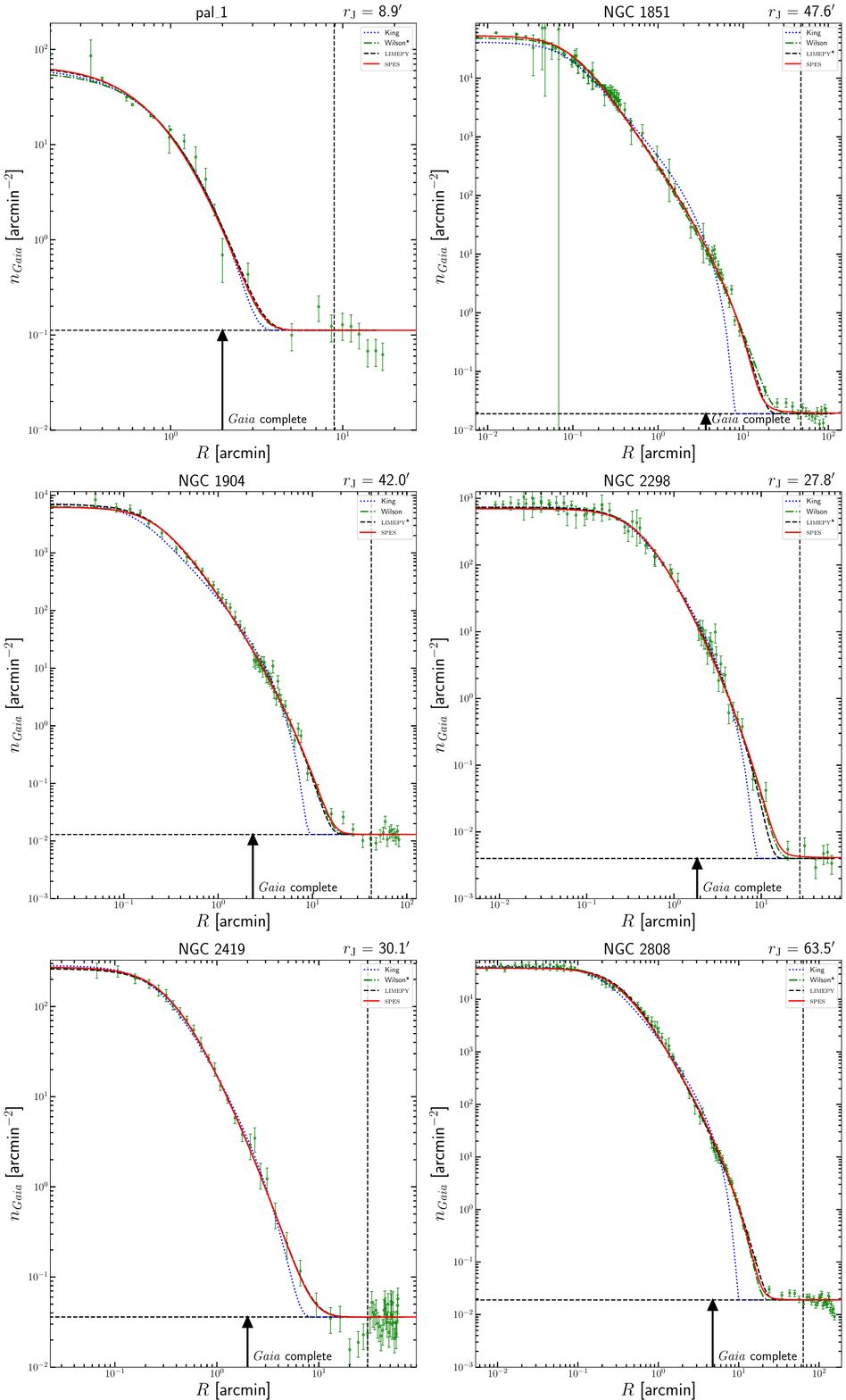}
\caption{Figure~\ref{GC_dens_plots} continued.}
\end{figure}

\begin{figure}
\centering
\includegraphics[angle=0, width=0.95\textwidth]{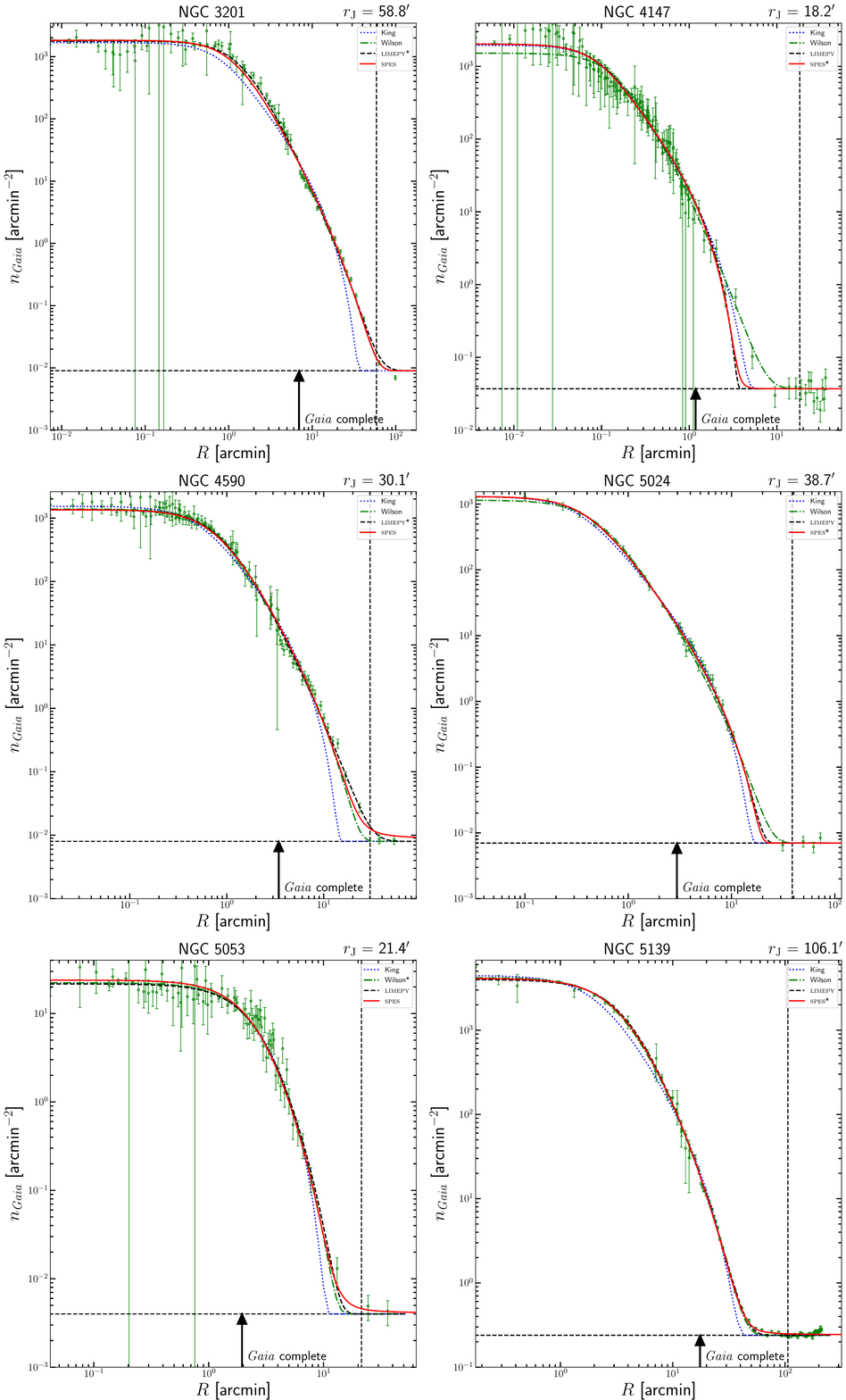}
\caption{Figure~\ref{GC_dens_plots} continued.}
\end{figure}

\begin{figure}
\centering
\includegraphics[angle=0, width=0.95\textwidth]{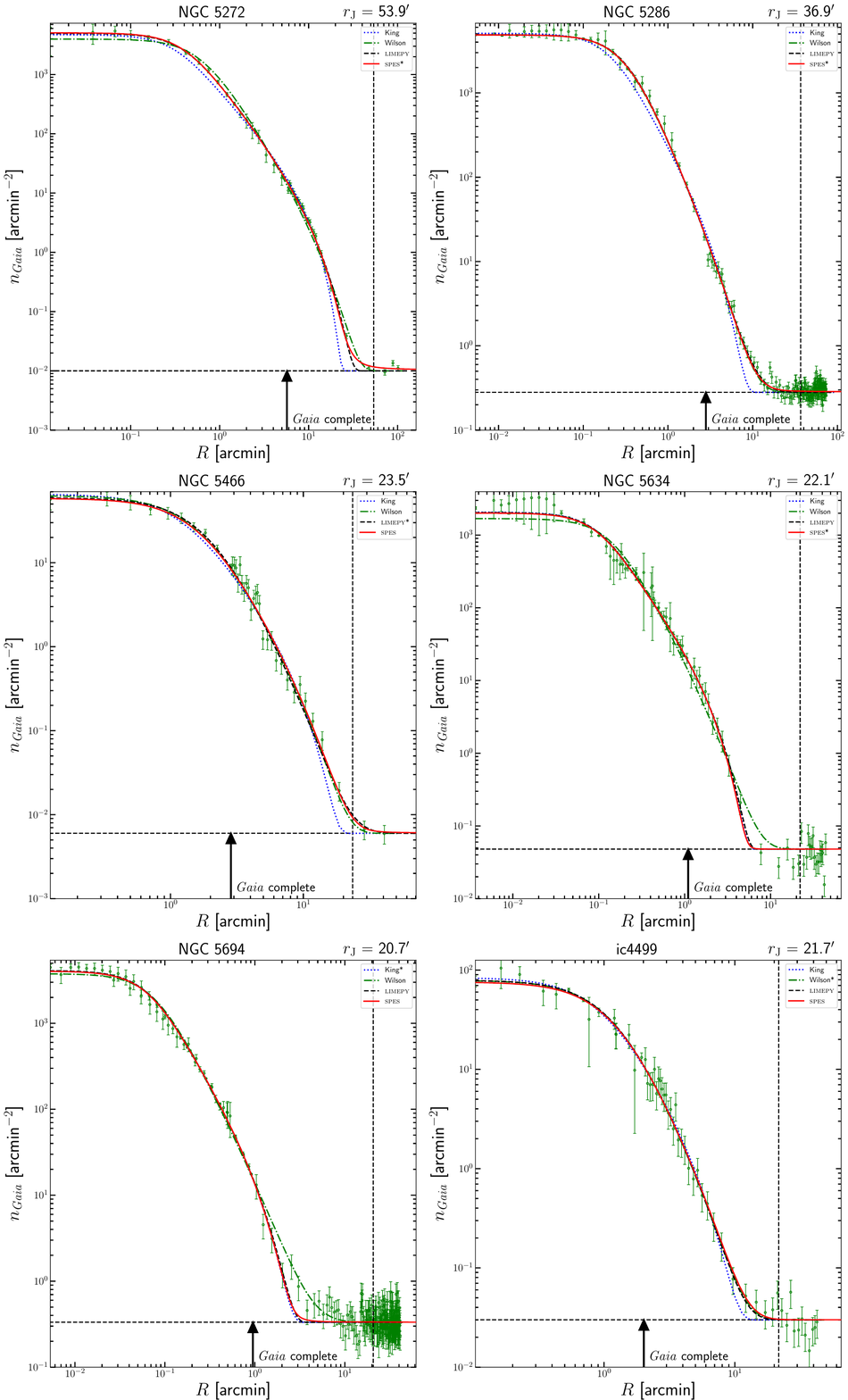}
\caption{Figure~\ref{GC_dens_plots} continued.}
\end{figure}

\begin{figure}
\centering
\includegraphics[angle=0, width=0.95\textwidth]{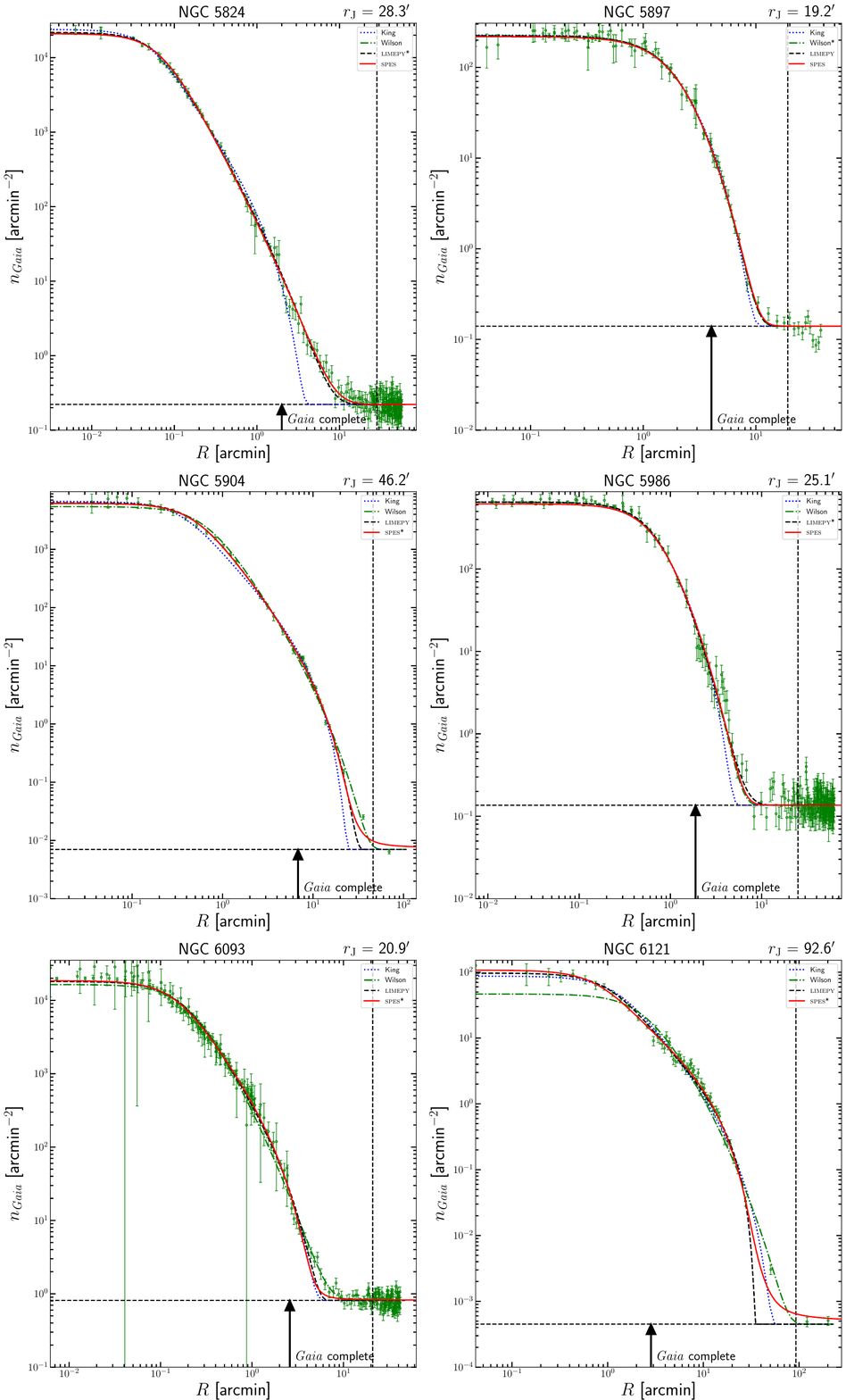}
\caption{Figure~\ref{GC_dens_plots} continued.}
\end{figure}

\begin{figure}
\centering
\includegraphics[angle=0, width=0.95\textwidth]{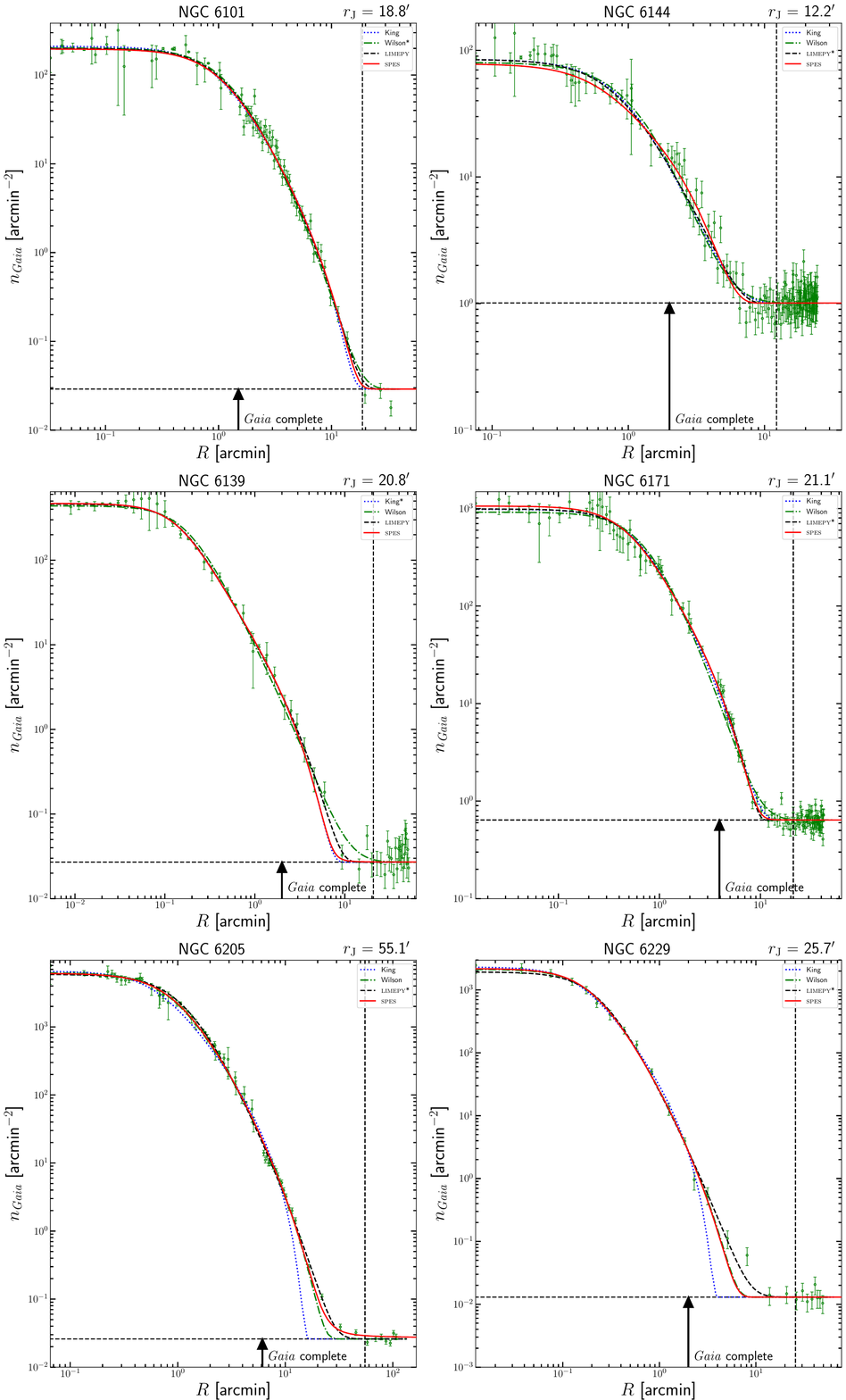}
\caption{Figure~\ref{GC_dens_plots} continued.}
\end{figure}

\begin{figure}
\centering
\includegraphics[angle=0, width=0.95\textwidth]{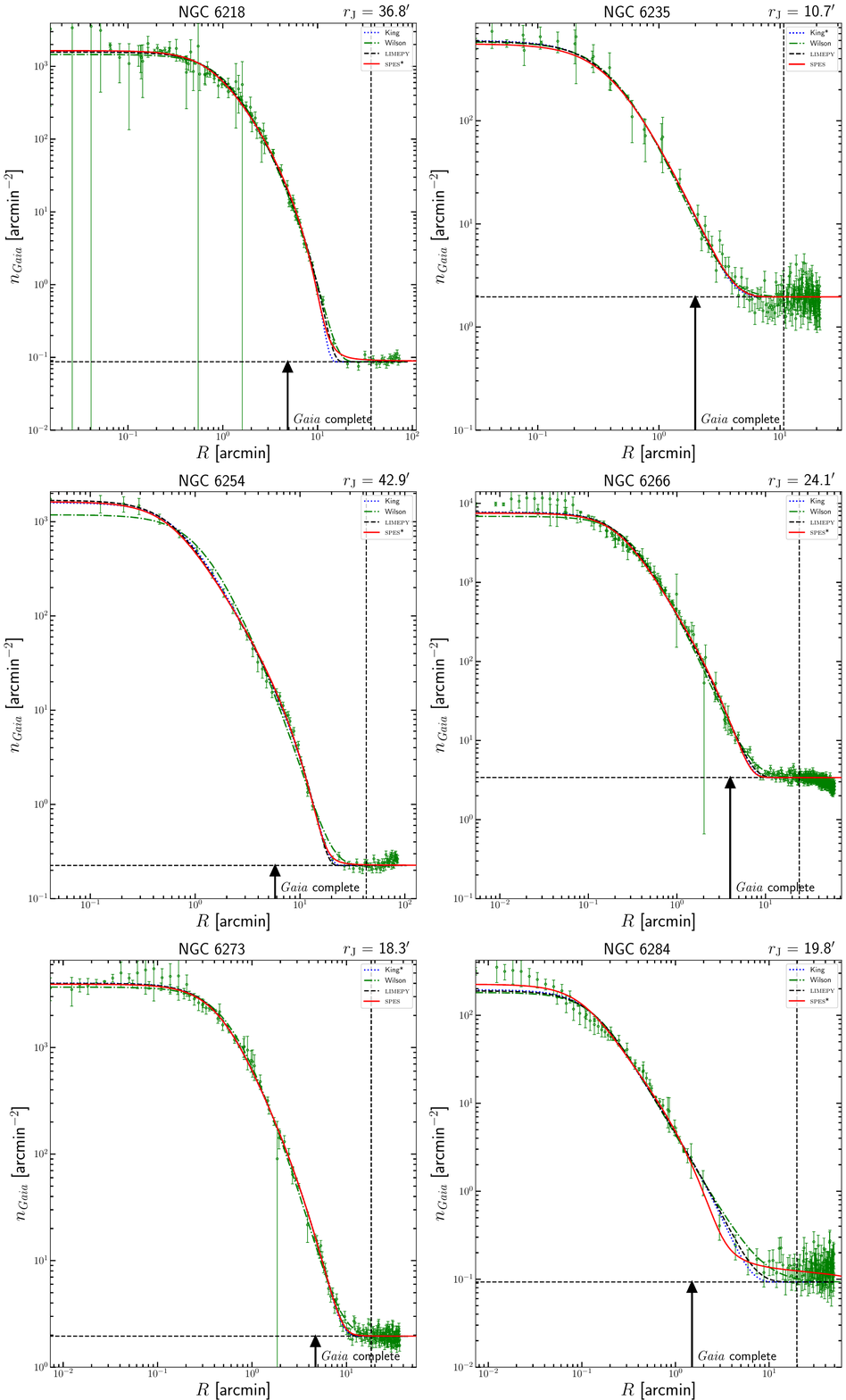}
\caption{Figure~\ref{GC_dens_plots} continued.}
\end{figure}

\begin{figure}
\centering
\includegraphics[angle=0, width=0.95\textwidth]{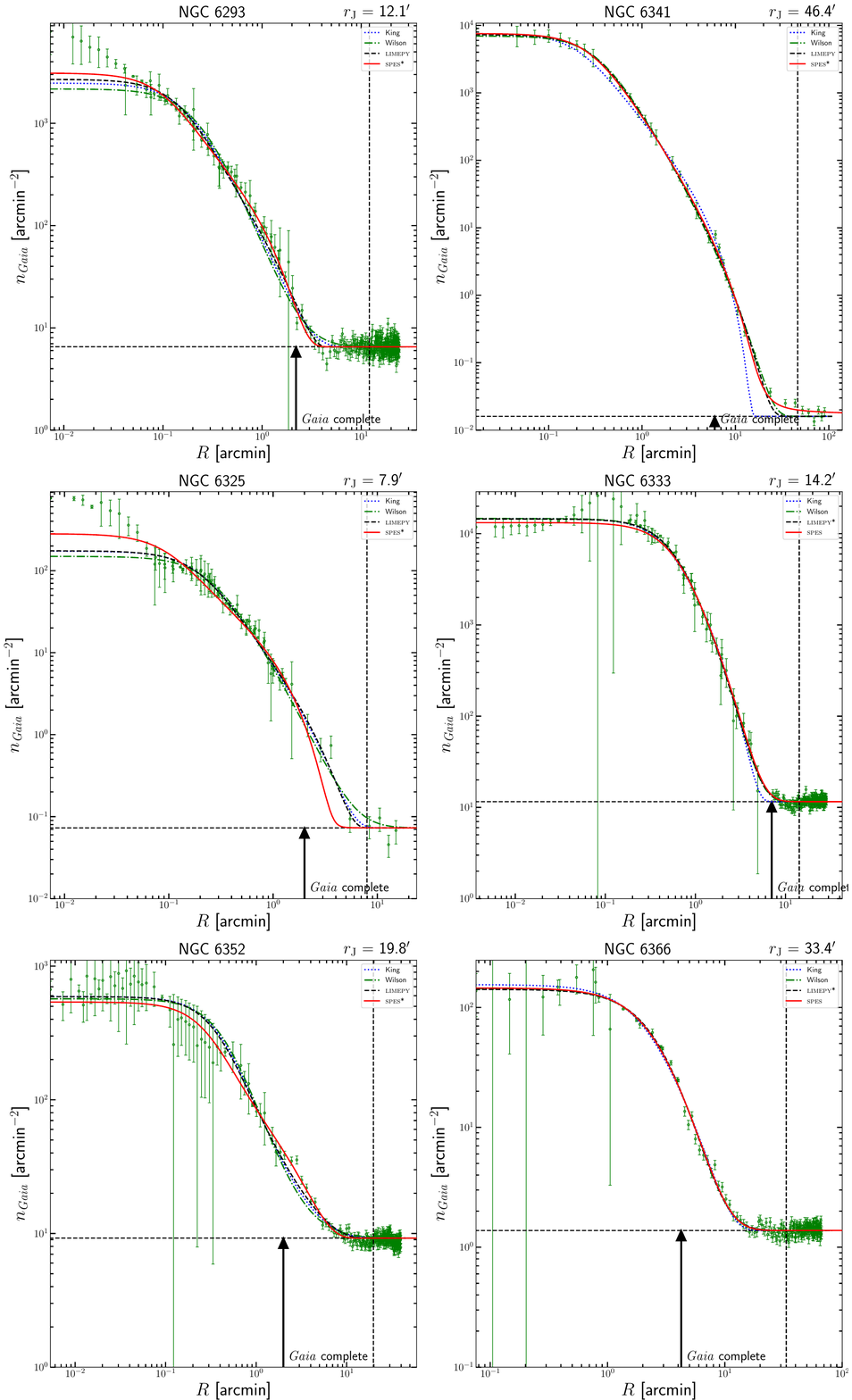}
\caption{Figure~\ref{GC_dens_plots} continued.}
\end{figure}

\begin{figure}
\centering
\includegraphics[angle=0, width=0.95\textwidth]{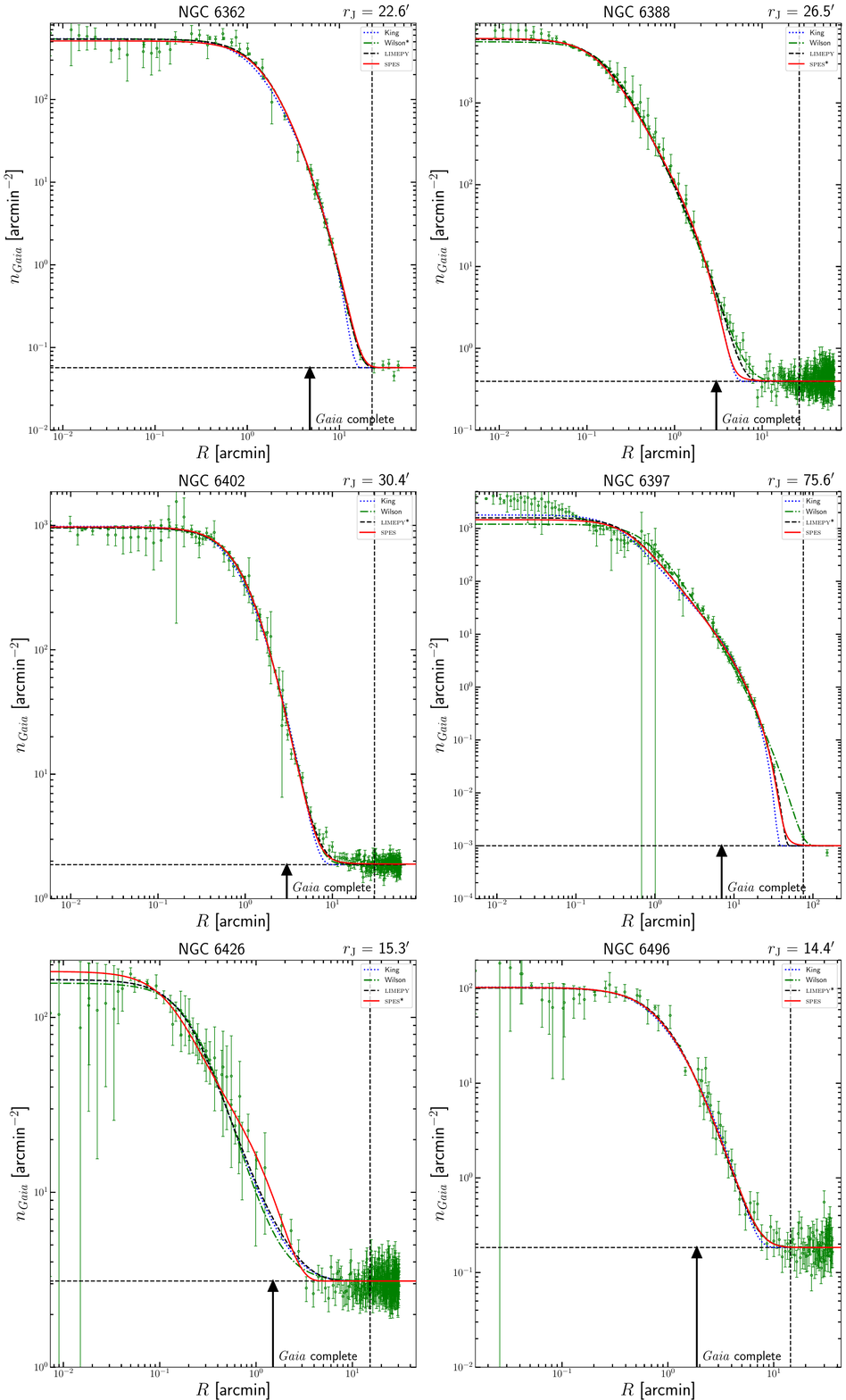}
\caption{Figure~\ref{GC_dens_plots} continued.}
\end{figure}

\begin{figure}
\centering
\includegraphics[angle=0, width=0.95\textwidth]{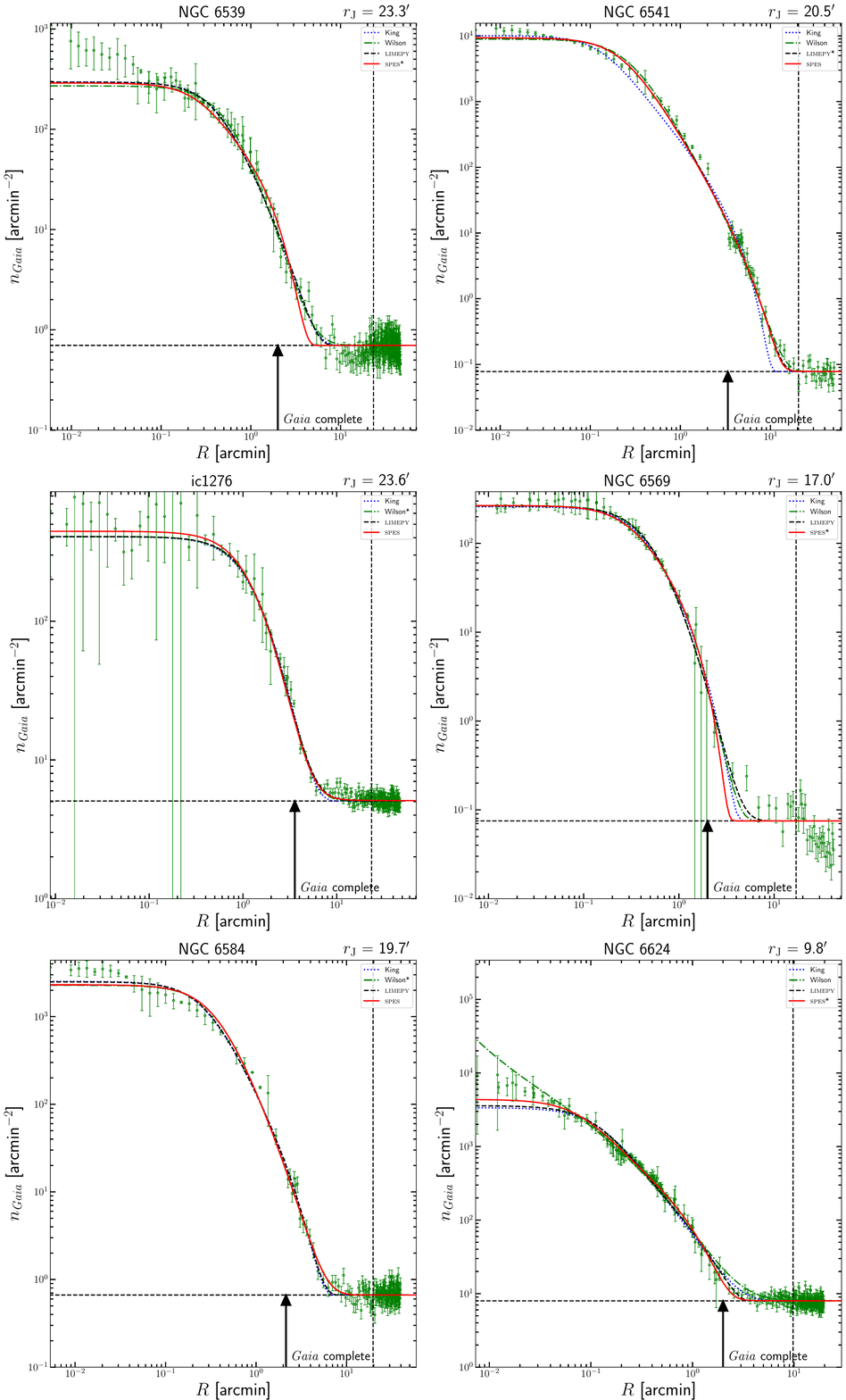}
\caption{Figure~\ref{GC_dens_plots} continued.}
\end{figure}

\begin{figure}
\centering
\includegraphics[angle=0, width=0.95\textwidth]{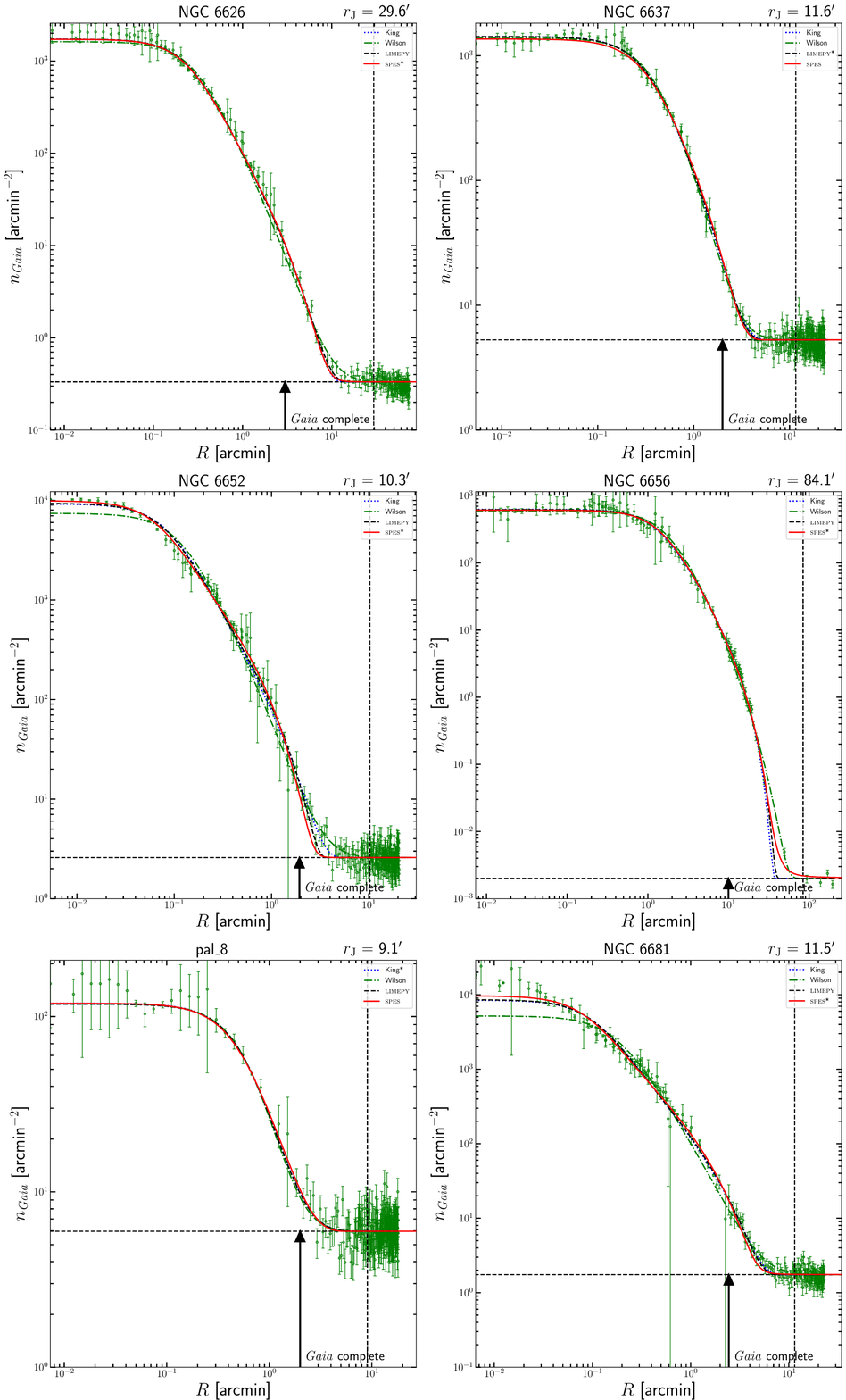}
\caption{Figure~\ref{GC_dens_plots} continued.}
\end{figure}

\begin{figure}
\centering
\includegraphics[angle=0, width=0.95\textwidth]{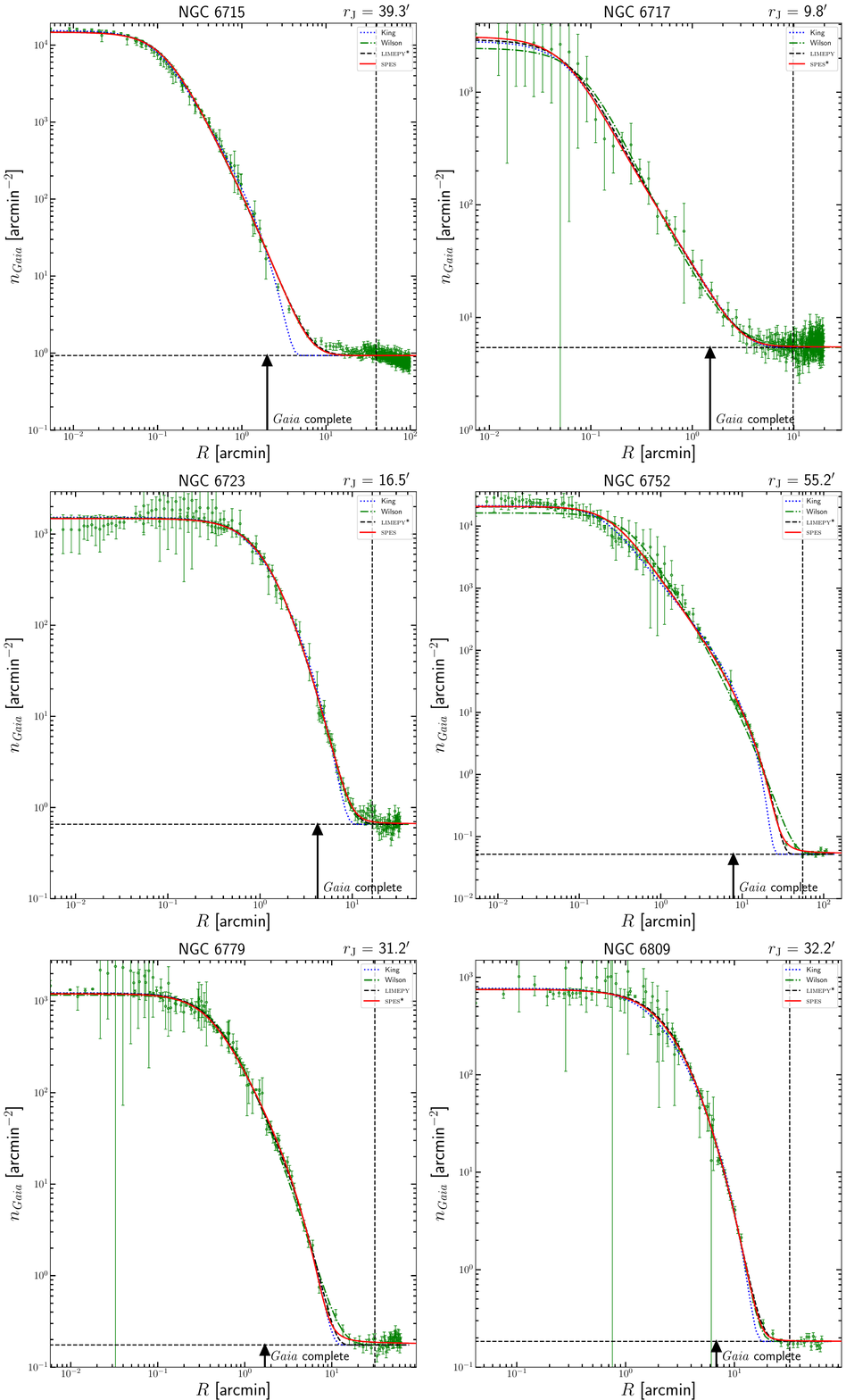}
\caption{Figure~\ref{GC_dens_plots} continued.}
\end{figure}

\begin{figure}
\centering
\includegraphics[angle=0, width=0.95\textwidth]{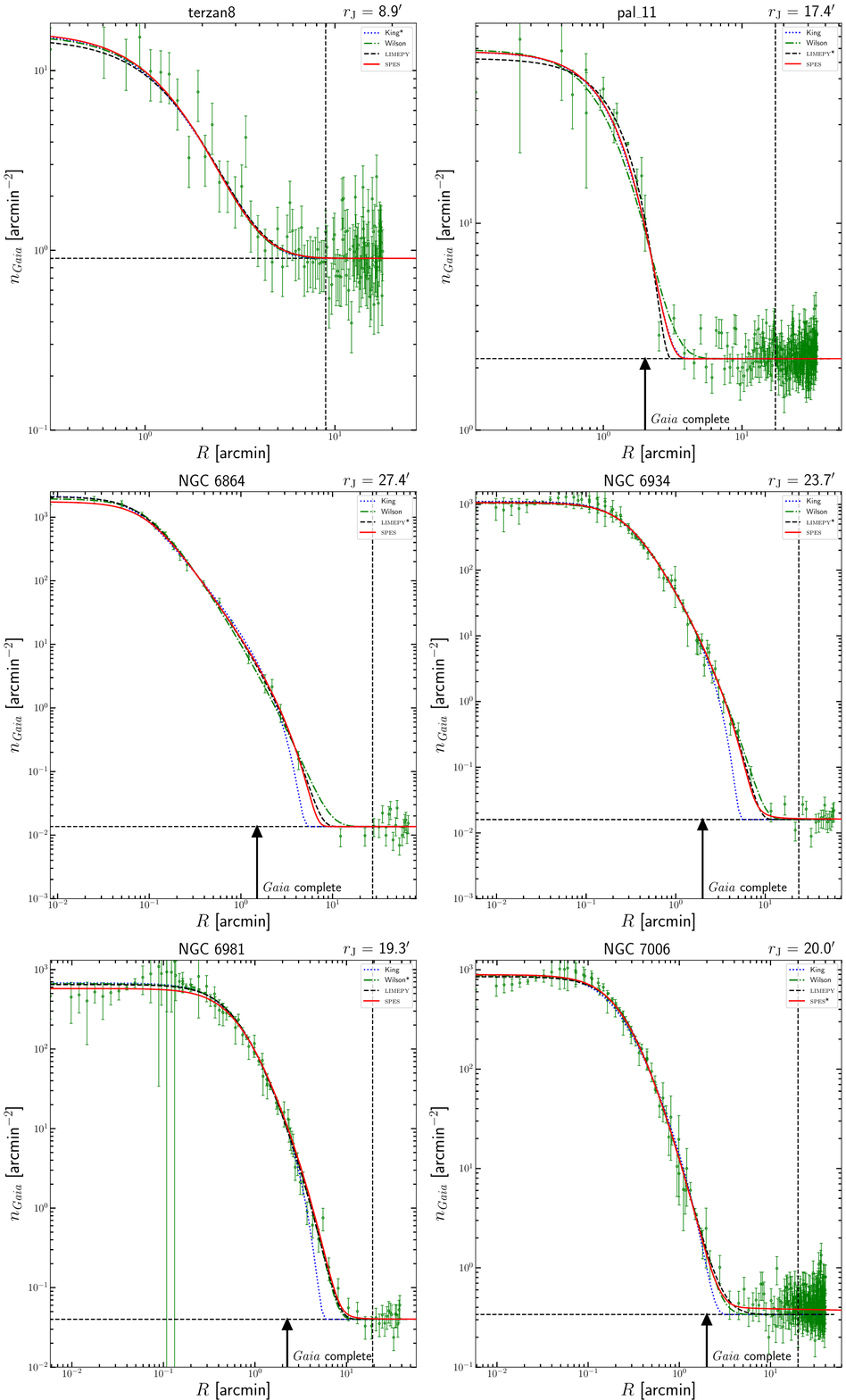}
\caption{Figure~\ref{GC_dens_plots} continued.}
\end{figure}

\begin{figure}
\centering
\includegraphics[angle=0, width=0.95\textwidth]{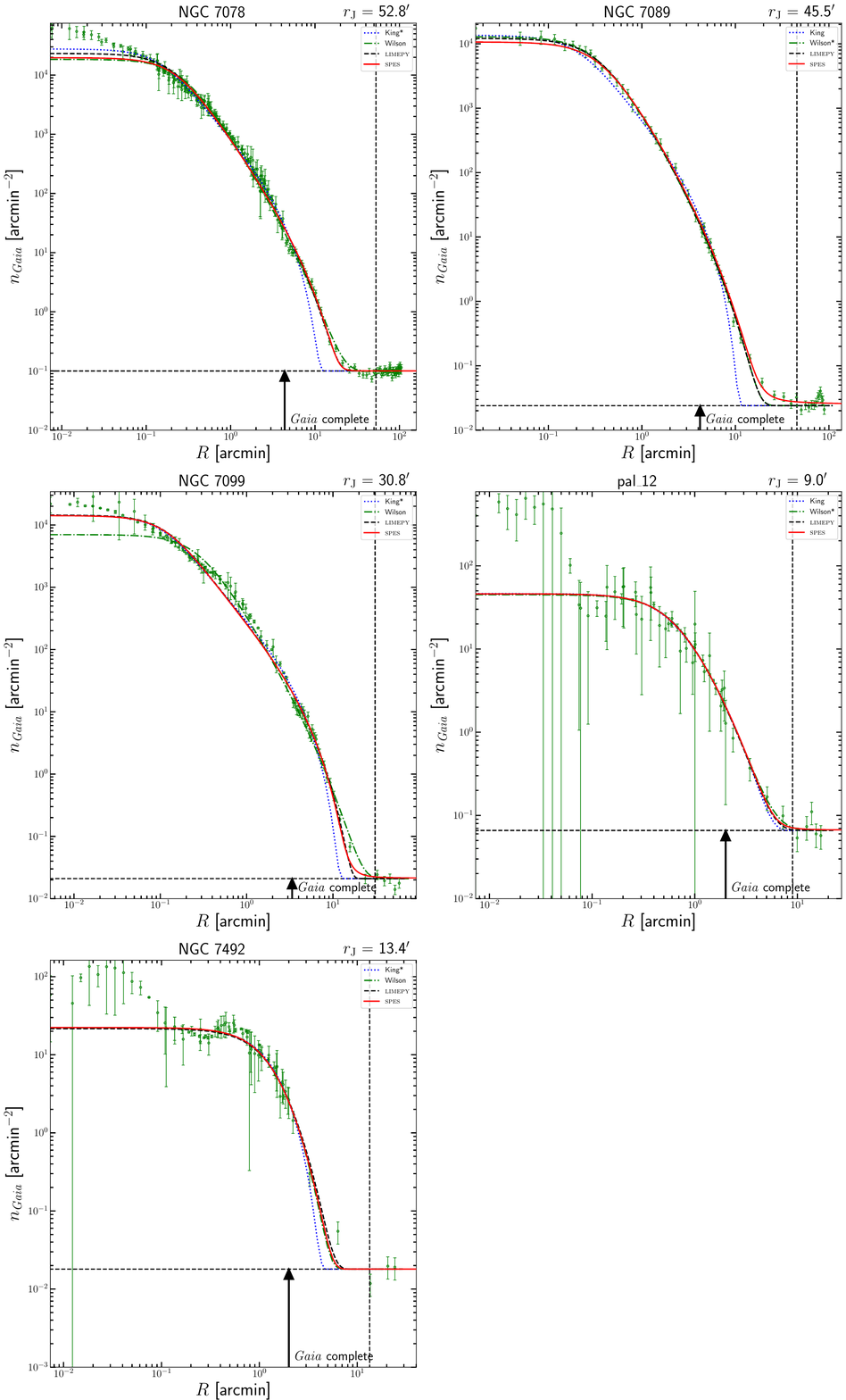}
\caption{Figure~\ref{GC_dens_plots} continued.}
\end{figure}

\end{appendix}

\clearpage
\begin{appendix}
\onecolumn
\section{GC profile fit parameters}
\label{GC_numdens_fitpars}

\begin{scriptsize}
\begin{landscape}
\begin{longtable}[c]{l|cr|cr|ccrr|cccrrc|cccc}
\caption{Best-fit parameters of {\sc limepy} and {\sc spes} models fit to 81 GCs following the procedure outlined in section~\ref{model_fits}.} \label{GCpars} \\

\hline\hline
 & \multicolumn{2}{|c|}{King} &\multicolumn{2}{|c|}{Wilson} & \multicolumn{4}{|c|}{{\textsc{limepy}}} & \multicolumn{6}{|c|}{{\textsc{spes}}} & & & \\
 id & $W$ & \multicolumn{1}{c|}{$r_{\mathrm{t}}$} & $W$ & \multicolumn{1}{c|}{$r_{\mathrm{t}}$}  & $W$ & $g$ & \multicolumn{1}{c}{$r_{\mathrm{h}}$} & \multicolumn{1}{c|}{$r_{\mathrm{t}}$} & $W$ & $\eta$ & log$_{10}$(1-B) & \multicolumn{1}{c}{$r_{\mathrm{h}}$} & \multicolumn{1}{c}{$r_{\mathrm{t}}$}  & log$_{10}$($f_{\mathrm{PE}}$) &  $r_{\mathrm{tie}}$ & BG lev & M$_{\mathrm{low}}$  \\ 
    & & \multicolumn{1}{c|}{(pc)} & & \multicolumn{1}{c|}{(pc)} & & & \multicolumn{1}{c}{(pc)} & \multicolumn{1}{c|}{(pc)} & & & & \multicolumn{1}{c}{(pc)} & \multicolumn{1}{c}{(pc)} & & {\tiny(arcmin)} & {\tiny(arcmin$^{-2}$)} & (M$_{\odot}$)   \\
\hline
\endfirsthead

\multicolumn{18}{c}{{\bfseries Table \thetable\ continued from previous page}} \\

\hline\hline
 & \multicolumn{2}{|c|}{King} &\multicolumn{2}{|c|}{Wilson} & \multicolumn{4}{|c|}{{\textsc{limepy}}} & \multicolumn{6}{|c|}{{\textsc{spes}}} & & & \\
 id & $W$ & \multicolumn{1}{c|}{$r_{\mathrm{t}}$} & $W$ & \multicolumn{1}{c|}{$r_{\mathrm{t}}$}  & $W$ & $g$ & \multicolumn{1}{c}{$r_{\mathrm{h}}$} & \multicolumn{1}{c|}{$r_{\mathrm{t}}$} & $W$ & $\eta$ & log$_{10}$(1-B) & \multicolumn{1}{c}{$r_{\mathrm{h}}$} & \multicolumn{1}{c}{$r_{\mathrm{t}}$}  & log$_{10}$($f_{\mathrm{PE}}$) &  $r_{\mathrm{tie}}$ & BG lev & M$_{\mathrm{low}}$  \\ 
    & & \multicolumn{1}{c|}{(pc)} & & \multicolumn{1}{c|}{(pc)} & & & \multicolumn{1}{c}{(pc)} & \multicolumn{1}{c|}{(pc)} & & & & \multicolumn{1}{c}{(pc)} & \multicolumn{1}{c}{(pc)} & & {\tiny(arcmin)} & {\tiny(arcmin$^{-2}$)} & (M$_{\odot}$)   \\
\hline
\endhead

\hline
\hline
\endfoot
  ngc104 & 8.58$\pm$0.02 &  52.49$\pm$0.48  & 7.05$\pm$0.03 & 200.91$\pm$6.70  & 8.30$\pm$0.08 & 1.33$\pm$0.05 &  5.18$\pm$0.16 &  72.96$\pm$4.25   &  8.12$\pm$0.10 & 0.15$\pm$0.02 & -1.82$\pm$0.18 &  5.20$\pm$0.17 &  57.33$\pm$3.24   & -2.36$\pm$0.36 & 13.28 & 0.08 & 0.60 \\ 
  ngc288 & 4.60$\pm$0.04 &  46.31$\pm$0.47  & 3.47$\pm$0.07 &  74.14$\pm$1.54  & 1.73$\pm$0.89 & 2.69$\pm$0.23 &  9.11$\pm$0.07 & 116.40$\pm$19.83  &  3.28$\pm$0.17 & 0.24$\pm$0.01 & -2.43$\pm$0.31 &  9.16$\pm$0.10 &  60.04$\pm$6.43   & -2.31$\pm$0.22 &  4.05 & 0.01 & 0.74 \\   
  ngc362 & 7.93$\pm$0.03 &  26.08$\pm$0.42  & 6.72$\pm$0.01 &  82.59$\pm$0.98  & 6.73$\pm$0.06 & 1.99$\pm$0.03 &  2.03$\pm$0.03 &  81.68$\pm$5.78   &  6.71$\pm$0.02 & 0.10$\pm$0.06 & -4.31$\pm$1.55 &  2.03$\pm$0.02 &  80.02$\pm$4.44   & -4.88$\pm$1.31 &  4.54 & 0.08 & 0.70 \\ 
 ngc1261 & 6.80$\pm$0.17 &  37.81$\pm$1.81  & 5.09$\pm$0.03 &  61.86$\pm$1.22  & 3.63$\pm$0.41 & 2.82$\pm$0.12 &  4.33$\pm$0.07 & 220.34$\pm$62.94  &  4.99$\pm$0.10 & 0.23$\pm$0.01 & -2.59$\pm$0.22 &  4.45$\pm$0.04 &  51.51$\pm$4.51   & -2.59$\pm$0.15 &  2.41 & 0.01 & 0.81 \\ 
  pal\_1 & 3.36$\pm$0.67 &  12.97$\pm$1.13  & 1.44$\pm$0.80 &  18.67$\pm$1.83  & 2.42$\pm$1.80 & 1.92$\pm$0.78 &  3.21$\pm$0.22 &  19.67$\pm$10.55  &  2.17$\pm$1.07 & 0.18$\pm$0.13 & -7.60$\pm$1.83 &  3.25$\pm$0.18 &  19.59$\pm$3.66   & -4.48$\pm$1.75 &  2.00 & 0.11 & 0.76 \\ 
 ngc1851 & 8.37$\pm$0.07 &  31.92$\pm$0.64  & 7.28$\pm$0.01 & 161.34$\pm$2.94  & 7.64$\pm$0.06 & 1.85$\pm$0.02 &  2.51$\pm$0.05 & 109.26$\pm$7.53   &  7.46$\pm$0.05 & 0.13$\pm$0.01 & -2.64$\pm$0.13 &  2.43$\pm$0.06 &  62.13$\pm$5.11   & -2.33$\pm$0.13 &  3.64 & 0.02 & 0.78 \\   
 ngc1904 & 7.93$\pm$0.11 &  38.32$\pm$1.01  & 6.56$\pm$0.03 & 112.90$\pm$2.82  & 6.79$\pm$0.20 & 1.89$\pm$0.09 &  3.20$\pm$0.10 &  97.22$\pm$13.76  &  6.57$\pm$0.07 & 0.13$\pm$0.08 & -3.94$\pm$1.69 &  3.17$\pm$0.12 & 108.73$\pm$10.38  & -4.70$\pm$1.40 &  2.33 & 0.01 & 0.74 \\ 
 ngc2298 & 6.76$\pm$0.04 &  30.05$\pm$0.62  & 6.08$\pm$0.04 &  82.20$\pm$3.04  & 6.35$\pm$0.17 & 1.75$\pm$0.16 &  3.49$\pm$0.08 &  60.05$\pm$11.80  &  6.08$\pm$0.08 & 0.19$\pm$0.11 & -3.56$\pm$1.74 &  3.38$\pm$0.06 &  80.41$\pm$6.89   & -4.69$\pm$1.81 &  1.84 & 0.01 & 0.71 \\ 
 ngc2419 & 6.82$\pm$0.10 &  {\tiny209.46$\pm$10.30} & 6.24$\pm$0.08 & 671.82$\pm$50.93 & 6.15$\pm$0.31 & 2.04$\pm$0.18 & 24.50$\pm$1.13 & 704.85$\pm$236.03 &  6.22$\pm$0.16 & 0.17$\pm$0.13 & -4.36$\pm$1.53 & 24.81$\pm$1.00 & 644.64$\pm$107.45 & -5.46$\pm$1.56 &  2.00 & 0.04 & 0.77 \\ 
 ngc2808 & 7.44$\pm$0.04 &  28.91$\pm$0.43  & 6.32$\pm$0.01 &  74.66$\pm$0.84  & 5.89$\pm$0.08 & 2.22$\pm$0.03 &  2.48$\pm$0.02 & 100.06$\pm$6.57   &  6.28$\pm$0.03 & 0.13$\pm$0.01 & -3.43$\pm$0.28 &  2.56$\pm$0.02 &  65.42$\pm$4.09   & -3.35$\pm$0.31 &  4.75 & 0.02 & 0.73 \\ 
 ngc3201 & 7.45$\pm$0.08 &  57.10$\pm$1.97  & 6.42$\pm$0.02 & 163.67$\pm$3.88  & 5.75$\pm$0.19 & 2.30$\pm$0.07 &  4.98$\pm$0.08 & 249.11$\pm$34.88  &  6.42$\pm$0.05 & 0.16$\pm$0.12 & -5.42$\pm$1.30 &  5.20$\pm$0.09 & 161.62$\pm$8.25   & -5.45$\pm$1.18 &  6.95 & 0.01 & 0.60 \\ 
 ngc4147 & 7.63$\pm$0.03 &  32.56$\pm$0.63  & 6.71$\pm$0.05 & 127.58$\pm$8.21  & 7.60$\pm$0.08 & 0.60$\pm$0.17 &  3.25$\pm$0.10 &  21.93$\pm$3.58   &  7.90$\pm$0.09 & 0.23$\pm$0.05 & -0.97$\pm$0.20 &  3.22$\pm$0.08 &  21.60$\pm$1.65   & -2.07$\pm$0.40 &  1.18 & 0.03 & 0.78 \\ 
 ngc4590 & 6.79$\pm$0.04 &  50.77$\pm$1.15  & 5.84$\pm$0.03 & 122.02$\pm$2.90  & 5.17$\pm$0.08 & 2.46$\pm$0.04 &  5.74$\pm$0.05 & 295.80$\pm$42.45  &  5.74$\pm$0.06 & 0.22$\pm$0.01 & -2.63$\pm$0.17 &  5.86$\pm$0.06 &  94.03$\pm$7.56   & -2.53$\pm$0.12 &  3.44 & 0.01 & 0.72 \\ 
 ngc5024 & 7.53$\pm$0.05 &  92.82$\pm$1.84  & 6.34$\pm$0.03 & 242.61$\pm$5.56  & 7.04$\pm$0.10 & 1.53$\pm$0.07 &  8.92$\pm$0.21 & 145.10$\pm$10.60  &  6.81$\pm$0.10 & 0.09$\pm$0.04 & -2.49$\pm$0.30 &  8.83$\pm$0.22 & 118.76$\pm$7.27   & -3.43$\pm$0.52 &  2.96 & 0.01 & 0.77 \\   
 ngc5053 & 2.96$\pm$0.19 &  59.38$\pm$2.03  & 1.58$\pm$0.53 &  91.51$\pm$7.37  & 1.07$\pm$0.71 & 2.24$\pm$0.24 & 16.11$\pm$0.29 & 106.30$\pm$19.69  &  1.50$\pm$0.59 & 0.20$\pm$0.05 & -2.99$\pm$0.67 & 15.95$\pm$0.39 &  77.32$\pm$16.51  & -2.20$\pm$0.66 &  1.95 & 0.01 & 0.77 \\ 
 ngc5139 & 6.25$\pm$0.02 &  70.25$\pm$0.56  & 4.82$\pm$0.01 & 114.30$\pm$0.85  & 3.97$\pm$0.26 & 2.33$\pm$0.09 &  9.34$\pm$0.07 & 137.30$\pm$8.62   &  4.57$\pm$0.07 & 0.25$\pm$0.01 & -2.83$\pm$0.27 &  9.36$\pm$0.06 &  97.07$\pm$7.23   & -2.67$\pm$0.17 & 17.51 & 0.24 & 0.60 \\ 
 ngc5272 & 8.10$\pm$0.07 &  77.55$\pm$1.06  & 6.48$\pm$0.02 & 197.84$\pm$5.01  & 7.46$\pm$0.08 & 1.53$\pm$0.04 &  6.71$\pm$0.14 & 121.27$\pm$4.54   &  7.22$\pm$0.09 & 0.20$\pm$0.01 & -1.75$\pm$0.07 &  6.64$\pm$0.13 &  79.92$\pm$2.97   & -2.06$\pm$0.06 &  5.71 & 0.01 & 0.72 \\ 
 ngc5286 & 7.52$\pm$0.04 &  37.78$\pm$0.79  & 6.53$\pm$0.02 & 123.14$\pm$3.24  & 6.33$\pm$0.12 & 2.14$\pm$0.07 &  3.52$\pm$0.05 & 172.84$\pm$33.08  &  6.42$\pm$0.05 & 0.20$\pm$0.01 & -2.80$\pm$0.33 &  3.57$\pm$0.04 &  86.73$\pm$14.06  & -2.32$\pm$0.19 &  2.80 & 0.28 & 0.73 \\   
 ngc5466 & 6.01$\pm$0.16 & 103.67$\pm$5.27  & 5.03$\pm$0.10 & 197.77$\pm$9.20  & 3.78$\pm$0.73 & 2.62$\pm$0.25 & 14.45$\pm$0.37 & 363.35$\pm$123.16 &  5.00$\pm$0.22 & 0.24$\pm$0.12 & -2.98$\pm$1.83 & 14.68$\pm$0.40 & 189.98$\pm$23.10  & -4.21$\pm$1.97 &  2.85 & 0.01 & 0.76 \\ 
 ngc5634 & 7.88$\pm$0.04 &  49.30$\pm$0.94  & 6.82$\pm$0.05 & 192.08$\pm$11.31 & 7.86$\pm$0.08 & 1.06$\pm$0.12 &  5.17$\pm$0.17 &  52.80$\pm$7.25   &  7.88$\pm$0.12 & 0.17$\pm$0.04 & -1.45$\pm$0.24 &  5.11$\pm$0.21 &  43.99$\pm$4.21   & -2.38$\pm$0.42 &  1.09 & 0.05 & 0.76 \\ 
 ngc5694 & 7.54$\pm$0.02 &  36.49$\pm$0.25  & 7.28$\pm$0.03 & 344.18$\pm$21.19 & 7.55$\pm$0.05 & 1.07$\pm$0.19 &  3.95$\pm$0.11 &  39.80$\pm$9.69   &  7.47$\pm$0.10 & 0.26$\pm$0.05 & -1.13$\pm$0.29 &  3.88$\pm$0.05 &  30.37$\pm$2.44   & -1.81$\pm$0.52 &  0.95 & 0.34 & 0.76 \\   
  ic4499 & 5.71$\pm$0.10 &  78.08$\pm$2.77  & 4.90$\pm$0.12 & 154.70$\pm$9.10  & 4.93$\pm$0.54 & 1.93$\pm$0.35 & 12.07$\pm$0.31 & 142.30$\pm$45.01  &  4.88$\pm$0.28 & 0.19$\pm$0.09 & -2.78$\pm$1.56 & 12.09$\pm$0.29 & 137.66$\pm$28.96  & -3.79$\pm$1.79 &  2.01 & 0.03 & 0.76 \\ 
 ngc5824 & 8.17$\pm$0.02 &  41.55$\pm$0.49  & 7.46$\pm$0.02 & 381.98$\pm$11.92 & 7.63$\pm$0.07 & 1.88$\pm$0.04 &  4.61$\pm$0.14 & 230.01$\pm$43.68  &  7.45$\pm$0.03 & 0.07$\pm$0.02 & -4.72$\pm$0.76 &  4.56$\pm$0.11 & 348.56$\pm$58.23  & -3.76$\pm$0.75 &  2.00 & 0.22 & 0.76 \\   
 ngc5897 & 3.79$\pm$0.04 &  41.94$\pm$0.45  & 2.34$\pm$0.05 &  62.22$\pm$0.79  & 2.36$\pm$0.68 & 1.99$\pm$0.27 &  9.72$\pm$0.09 &  62.44$\pm$8.28   &  2.32$\pm$0.16 & 0.19$\pm$0.10 & -3.48$\pm$1.74 &  9.69$\pm$0.08 &  60.95$\pm$2.75   & -4.57$\pm$1.63 &  4.03 & 0.14 & 0.73 \\ 
 ngc5904 & 7.93$\pm$0.05 &  56.56$\pm$0.91  & 6.40$\pm$0.03 & 143.70$\pm$3.69  & 7.23$\pm$0.17 & 1.50$\pm$0.09 &  5.06$\pm$0.15 &  82.84$\pm$7.04   &  6.98$\pm$0.15 & 0.21$\pm$0.01 & -1.70$\pm$0.09 &  5.01$\pm$0.14 &  57.36$\pm$2.69   & -2.07$\pm$0.07 &  6.82 & 0.01 & 0.68 \\ 
 ngc5986 & 4.75$\pm$0.09 &  17.62$\pm$0.48  & 4.14$\pm$0.07 &  33.77$\pm$1.02  & 3.01$\pm$0.42 & 2.61$\pm$0.12 &  3.42$\pm$0.05 &  57.49$\pm$7.16   &  4.06$\pm$0.18 & 0.20$\pm$0.04 & -2.68$\pm$0.54 &  3.46$\pm$0.04 &  29.69$\pm$3.17   & -2.97$\pm$0.70 &  1.88 & 0.14 & 0.72 \\ 
 ngc6093 & 7.13$\pm$0.01 &  18.40$\pm$0.11  & 6.33$\pm$0.02 &  54.29$\pm$1.14  & 7.07$\pm$0.05 & 1.18$\pm$0.09 &  2.04$\pm$0.02 &  21.32$\pm$1.71   &  7.26$\pm$0.12 & 0.28$\pm$0.02 & -0.94$\pm$0.11 &  2.11$\pm$0.02 &  14.25$\pm$0.76   & -1.77$\pm$0.10 &  2.59 & 0.81 & 0.70 \\   
 ngc6121 & 7.29$\pm$0.10 &  37.50$\pm$0.88  & 5.80$\pm$0.11 &  83.58$\pm$3.96  & 7.64$\pm$0.13 & 0.18$\pm$0.07 &  4.47$\pm$0.09 &  22.75$\pm$0.80   &  9.09$\pm$0.23 & 0.34$\pm$0.01 & -0.22$\pm$0.09 &  4.77$\pm$0.10 &  22.70$\pm$0.66   & -1.58$\pm$0.03 &  2.83 & 0.00 & 0.50 \\ 
 ngc6101 & 6.28$\pm$0.04 &  91.06$\pm$1.72  & 5.34$\pm$0.05 & 181.73$\pm$5.63  & 5.85$\pm$0.63 & 1.56$\pm$0.51 & 12.01$\pm$0.42 & 129.88$\pm$67.62  &  5.84$\pm$0.24 & 0.17$\pm$0.05 & -1.72$\pm$0.33 & 12.11$\pm$0.27 &  95.13$\pm$8.75   & -2.65$\pm$0.42 &  3.29 & 0.03 & 0.76 \\ 
 ngc6144 & 5.61$\pm$0.12 &  35.76$\pm$1.56  & 4.69$\pm$0.22 &  66.49$\pm$6.31  & 5.78$\pm$0.20 & 0.22$\pm$0.14 &  5.92$\pm$0.22 &  25.12$\pm$2.28   &  9.24$\pm$0.68 & 0.13$\pm$0.04 & -0.15$\pm$0.17 &  6.20$\pm$0.20 &  23.85$\pm$1.31   & -3.62$\pm$0.64 &  2.00 & 1.01 & 0.70 \\ 
 ngc6139 & 7.92$\pm$0.04 &  29.31$\pm$0.76  & 7.26$\pm$0.05 & 214.83$\pm$19.61 & 7.92$\pm$0.09 & 1.30$\pm$0.14 &  3.24$\pm$0.18 &  44.81$\pm$10.06  &  7.89$\pm$0.16 & 0.24$\pm$0.06 & -1.20$\pm$0.33 &  3.02$\pm$0.23 &  24.69$\pm$6.54   & -1.85$\pm$0.41 &  2.00 & 0.03 & 0.76 \\ 
 ngc6171 & 6.63$\pm$0.02 &  29.54$\pm$0.33  & 5.83$\pm$0.05 &  71.03$\pm$2.61  & 6.76$\pm$0.05 & 0.43$\pm$0.15 &  3.88$\pm$0.06 &  21.39$\pm$1.77   &  7.39$\pm$0.14 & 0.27$\pm$0.04 & -0.73$\pm$0.14 &  3.95$\pm$0.06 &  21.97$\pm$0.98   & -1.96$\pm$0.22 &  3.92 & 0.64 & 0.65 \\ 
 ngc6205 & 6.55$\pm$0.06 &  35.69$\pm$0.69  & 5.49$\pm$0.03 &  71.99$\pm$1.14  & 4.11$\pm$0.18 & 2.63$\pm$0.06 &  4.18$\pm$0.03 & 138.92$\pm$14.71  &  5.34$\pm$0.06 & 0.21$\pm$0.01 & -2.67$\pm$0.17 &  4.27$\pm$0.04 &  56.90$\pm$3.74   & -2.68$\pm$0.13 &  6.12 & 0.03 & 0.67 \\ 
 ngc6229 & 6.59$\pm$0.09 &  36.49$\pm$0.87  & 5.86$\pm$0.08 &  92.57$\pm$5.32  & 5.08$\pm$0.42 & 2.51$\pm$0.15 &  4.36$\pm$0.17 & {\tiny255.37$\pm$111.96} &  5.85$\pm$0.18 & 0.19$\pm$0.13 & -7.94$\pm$1.68 &  4.40$\pm$0.16 &  89.84$\pm$10.94  & -5.35$\pm$1.65 &  2.00 & 0.01 & 0.77 \\ 
 ngc6218 & 6.00$\pm$0.02 &  29.35$\pm$0.23  & 4.90$\pm$0.03 &  52.95$\pm$0.81  & 5.73$\pm$0.13 & 1.37$\pm$0.12 &  4.24$\pm$0.04 &  36.28$\pm$2.65   &  6.20$\pm$0.22 & 0.32$\pm$0.01 & -0.78$\pm$0.13 &  4.34$\pm$0.04 &  22.94$\pm$0.94   & -1.71$\pm$0.05 &  4.83 & 0.09 & 0.64 \\ 
 ngc6235 & 6.27$\pm$0.07 &  26.27$\pm$0.93  & 5.78$\pm$0.09 &  70.73$\pm$5.49  & 6.22$\pm$0.17 & 1.25$\pm$0.25 &  3.61$\pm$0.12 &  32.18$\pm$6.79   &  6.11$\pm$0.20 & 0.18$\pm$0.05 & -1.53$\pm$0.29 &  3.60$\pm$0.12 &  26.59$\pm$3.64   & -2.56$\pm$0.48 &  1.99 & 1.96 & 0.73 \\ 
 ngc6254 & 7.06$\pm$0.02 &  38.47$\pm$0.41  & 5.75$\pm$0.07 &  81.66$\pm$2.70  & 7.27$\pm$0.14 & 0.82$\pm$0.10 &  4.53$\pm$0.12 &  34.80$\pm$1.87   &  7.51$\pm$0.23 & 0.30$\pm$0.04 & -0.82$\pm$0.18 &  4.61$\pm$0.12 &  29.50$\pm$2.09   & -1.74$\pm$0.24 &  5.78 & 0.23 & 0.60 \\ 
 ngc6266 & 7.48$\pm$0.05 &  22.87$\pm$0.40  & 6.78$\pm$0.04 & 102.01$\pm$5.82  & 7.39$\pm$0.12 & 1.24$\pm$0.15 &  2.43$\pm$0.06 &  28.81$\pm$4.93   &  7.58$\pm$0.26 & 0.21$\pm$0.05 & -1.24$\pm$0.28 &  2.48$\pm$0.08 &  19.47$\pm$2.13   & -2.20$\pm$0.47 &  4.00 & 3.40 & 0.70 \\ 
 ngc6273 & 6.80$\pm$0.01 &  35.04$\pm$0.27  & 5.97$\pm$0.03 &  88.89$\pm$2.10  & 6.75$\pm$0.05 & 1.10$\pm$0.07 &  4.22$\pm$0.04 &  37.70$\pm$2.11   &  6.66$\pm$0.09 & 0.26$\pm$0.03 & -1.15$\pm$0.15 &  4.22$\pm$0.03 &  30.51$\pm$1.89   & -1.92$\pm$0.24 &  4.70 & 1.95 & 0.68 \\ 
 ngc6284 & 8.31$\pm$0.09 &  54.13$\pm$2.30  & 8.40$\pm$0.09 &1397.22$\pm$41.25 & 8.37$\pm$0.20 & 1.26$\pm$0.27 &  6.68$\pm$1.14 &  83.73$\pm$44.86  &  8.41$\pm$0.20 & 0.49$\pm$0.01 & -0.12$\pm$0.06 &  3.79$\pm$0.18 &  16.28$\pm$0.93   & -1.05$\pm$0.02 &  1.50 & 0.09 & 0.76 \\ 
 ngc6293 & 7.29$\pm$0.07 &  21.19$\pm$0.57  & 6.67$\pm$0.08 &  90.04$\pm$9.38  & 7.51$\pm$0.12 & 0.15$\pm$0.05 &  2.44$\pm$0.08 &  12.30$\pm$0.56   &  9.81$\pm$0.15 & 0.24$\pm$0.04 & -0.16$\pm$0.16 &  2.55$\pm$0.05 &  11.60$\pm$0.42   & -2.26$\pm$0.32 &  2.20 & 6.53 & 0.71 \\ 
 ngc6341 & 7.47$\pm$0.12 &  23.29$\pm$1.40  & 6.80$\pm$0.13 & 107.11$\pm$19.23 & 6.99$\pm$0.20 & 1.83$\pm$0.10 &  3.16$\pm$0.12 &  93.01$\pm$13.28  &  6.97$\pm$0.16 & 0.20$\pm$0.01 & -1.90$\pm$0.15 &  3.21$\pm$0.13 &  45.36$\pm$4.00   & -2.02$\pm$0.07 &  6.00 & 0.02 & 0.67 \\   
 ngc6325 & 5.53$\pm$0.03 &  16.22$\pm$0.19  & 4.94$\pm$0.04 &  34.21$\pm$0.74  & 7.54$\pm$0.29 & 0.70$\pm$0.34 &  2.54$\pm$0.25 &  18.48$\pm$5.27   &  9.91$\pm$0.15 & 0.30$\pm$0.08 & -0.01$\pm$0.23 &  2.14$\pm$0.17 &   9.69$\pm$1.02   & -1.82$\pm$0.58 &  2.00 & 0.07 & 0.77 \\ 
 ngc6333 & 8.05$\pm$0.10 &  40.27$\pm$0.79  & 6.68$\pm$0.03 & 118.47$\pm$3.31  & 4.75$\pm$0.16 & 2.13$\pm$0.10 &  2.62$\pm$0.02 &  38.76$\pm$4.37   &  4.91$\pm$0.09 & 0.14$\pm$0.05 & -2.99$\pm$0.77 &  2.62$\pm$0.02 &  30.94$\pm$3.13   & -3.80$\pm$0.89 &  7.02 &11.56 & 0.68 \\ 
 ngc6352 & 8.13$\pm$0.07 &  50.21$\pm$1.90  & 7.76$\pm$0.10 & 741.42$\pm$116.58& 8.15$\pm$0.12 & 0.16$\pm$0.06 &  5.19$\pm$0.31 &  25.57$\pm$1.74   & 10.73$\pm$0.43 & 0.17$\pm$0.05 & -0.25$\pm$0.24 &  5.03$\pm$0.21 &  21.82$\pm$1.32   & -2.92$\pm$0.55 &  2.00 & 9.23 & 0.66 \\ 
 ngc6366 & 4.83$\pm$0.09 &  21.00$\pm$0.51  & 3.70$\pm$0.12 &  34.80$\pm$1.31  & 2.24$\pm$0.96 & 2.62$\pm$0.26 &  4.07$\pm$0.06 &  53.13$\pm$13.01  &  3.73$\pm$0.27 & 0.24$\pm$0.14 & -4.07$\pm$1.61 &  4.07$\pm$0.07 &  34.68$\pm$3.01   & -5.13$\pm$1.74 &  4.24 & 1.38 & 0.65 \\ 
 ngc6362 & 5.64$\pm$0.06 &  38.31$\pm$0.61  & 4.47$\pm$0.06 &  63.86$\pm$1.32  & 4.45$\pm$0.38 & 1.99$\pm$0.19 &  5.91$\pm$0.07 &  63.34$\pm$8.01   &  4.44$\pm$0.13 & 0.16$\pm$0.07 & -2.95$\pm$1.28 &  5.90$\pm$0.07 &  58.36$\pm$6.76   & -3.77$\pm$1.35 &  4.77 & 0.06 & 0.68 \\ 
 ngc6388 & 7.30$\pm$0.03 &  17.98$\pm$0.18  & 6.58$\pm$0.03 &  67.62$\pm$2.27  & 7.04$\pm$0.12 & 1.57$\pm$0.14 &  1.93$\pm$0.03 &  33.53$\pm$6.53   &  7.32$\pm$0.22 & 0.26$\pm$0.03 & -1.06$\pm$0.27 &  1.96$\pm$0.04 &  14.56$\pm$2.33   & -1.84$\pm$0.22 &  2.00 & 0.39 & 0.79 \\   
 ngc6402 & 5.41$\pm$0.07 &  28.40$\pm$0.72  & 4.57$\pm$0.07 &  53.35$\pm$1.65  & 3.74$\pm$0.23 & 2.58$\pm$0.09 &  4.74$\pm$0.04 & 103.90$\pm$16.50  &  4.29$\pm$0.14 & 0.30$\pm$0.01 & -2.13$\pm$0.19 &  4.70$\pm$0.05 &  37.73$\pm$4.03   & -2.03$\pm$0.14 &  3.00 & 1.88 & 0.74 \\ 
 ngc6397 & 8.65$\pm$0.10 &  32.16$\pm$0.45  & 6.75$\pm$0.03 & 103.67$\pm$3.87  & 8.06$\pm$0.18 & 1.32$\pm$0.07 &  3.06$\pm$0.15 &  43.00$\pm$3.08   &  7.90$\pm$0.23 & 0.17$\pm$0.03 & -1.72$\pm$0.23 &  3.06$\pm$0.16 &  33.61$\pm$2.45   & -2.22$\pm$0.26 &  7.00 & 0.01 & 0.50 \\ 
 ngc6426 & 7.84$\pm$0.14 &  84.37$\pm$7.36  & 7.21$\pm$0.16 & 553.74$\pm$168.71& 7.91$\pm$0.29 & 0.53$\pm$0.35 &  9.32$\pm$1.65 &  58.48$\pm$20.72  & 11.73$\pm$0.64 & 0.11$\pm$0.04 & -0.06$\pm$0.17 &  8.06$\pm$0.71 &  30.42$\pm$3.91   & -3.88$\pm$0.62 &  1.50 & 3.11 & 0.75 \\   
 ngc6496 & 5.43$\pm$0.13 &  35.15$\pm$1.54  & 4.78$\pm$0.14 &  71.00$\pm$5.08  & 4.09$\pm$0.85 & 2.32$\pm$0.31 &  5.70$\pm$0.22 &  86.35$\pm$25.43  &  4.74$\pm$0.30 & 0.23$\pm$0.13 & -3.56$\pm$1.75 &  5.83$\pm$0.19 &  67.95$\pm$10.05  & -4.89$\pm$1.84 &  1.87 & 0.18 & 0.82 \\ 
 ngc6539 & 6.40$\pm$0.12 &  23.28$\pm$1.05  & 5.52$\pm$0.13 &  51.01$\pm$4.43  & 6.47$\pm$0.25 & 0.76$\pm$0.37 &  3.12$\pm$0.13 &  20.25$\pm$4.99   &  9.85$\pm$0.94 & 0.13$\pm$0.05 & -0.15$\pm$0.20 &  3.12$\pm$0.10 &  12.49$\pm$1.06   & -3.56$\pm$0.76 &  2.00 & 0.70 & 0.82 \\ 
 ngc6541 & 8.20$\pm$0.10 &  25.98$\pm$0.51  & 6.55$\pm$0.03 &  69.03$\pm$1.94  & 7.06$\pm$0.19 & 1.73$\pm$0.09 &  2.09$\pm$0.07 &  50.09$\pm$5.82   &  6.80$\pm$0.16 & 0.14$\pm$0.03 & -2.40$\pm$0.41 &  2.04$\pm$0.05 &  35.68$\pm$5.10   & -2.64$\pm$0.66 &  3.36 & 0.08 & 0.66 \\   
  ic1276 & 4.47$\pm$0.15 &  16.63$\pm$0.50  & 3.89$\pm$0.22 &  31.41$\pm$1.93  & 2.93$\pm$0.84 & 2.58$\pm$0.31 &  3.51$\pm$0.08 &  52.86$\pm$19.89  &  3.71$\pm$0.49 & 0.32$\pm$0.07 & -2.16$\pm$1.40 &  3.48$\pm$0.07 &  23.28$\pm$8.61   & -1.92$\pm$1.11 &  3.57 & 5.06 & 0.78 \\ 
 ngc6569 & 5.21$\pm$0.14 &  15.72$\pm$0.84  & 4.20$\pm$0.24 &  25.62$\pm$2.33  & 3.64$\pm$0.91 & 2.43$\pm$0.33 &  2.58$\pm$0.11 &  39.43$\pm$14.56  &  6.98$\pm$0.44 & 0.21$\pm$0.05 & -0.49$\pm$0.28 &  2.81$\pm$0.06 &  11.59$\pm$0.75   & -2.52$\pm$0.37 &  2.00 & 0.08 & 0.81 \\ 
 ngc6584 & 6.75$\pm$0.06 &  30.82$\pm$0.66  & 5.80$\pm$0.07 &  71.48$\pm$3.40  & 6.66$\pm$0.24 & 1.15$\pm$0.26 &  3.75$\pm$0.11 &  33.95$\pm$6.27   &  5.81$\pm$0.16 & 0.21$\pm$0.14 & -5.81$\pm$1.45 &  3.54$\pm$0.07 &  71.08$\pm$7.44   & -5.68$\pm$1.57 &  2.14 & 0.67 & 0.76 \\ 
 ngc6624 & 7.75$\pm$0.03 &  17.73$\pm$0.34  &15.43$\pm$0.11 &  85.41$\pm$2.11  & 7.66$\pm$0.09 & 0.16$\pm$0.14 &  1.72$\pm$0.07 &   8.63$\pm$1.07   &  9.39$\pm$0.12 & 0.37$\pm$0.06 & -0.00$\pm$0.22 &  1.66$\pm$0.01 &   7.29$\pm$0.33   & -1.54$\pm$0.39 &  2.00 & 7.97 & 0.80 \\   
 ngc6626 & 7.70$\pm$0.02 &  22.81$\pm$0.30  & 6.98$\pm$0.03 & 115.62$\pm$6.25  & 7.65$\pm$0.07 & 1.17$\pm$0.12 &  2.38$\pm$0.06 &  26.99$\pm$3.60   &  7.65$\pm$0.13 & 0.24$\pm$0.03 & -1.20$\pm$0.20 &  2.39$\pm$0.06 &  19.26$\pm$1.80   & -1.91$\pm$0.29 &  3.00 & 0.33 & 0.69 \\   
 ngc6637 & 5.99$\pm$0.04 &  16.47$\pm$0.34  & 5.36$\pm$0.08 &  36.87$\pm$1.82  & 6.02$\pm$0.11 & 0.76$\pm$0.31 &  2.43$\pm$0.05 &  14.30$\pm$2.76   &  6.56$\pm$0.28 & 0.25$\pm$0.05 & -0.75$\pm$0.25 &  2.44$\pm$0.04 &  12.41$\pm$0.93   & -2.22$\pm$0.42 &  2.00 & 5.28 & 0.75 \\ 
 ngc6652 & 7.83$\pm$0.04 &  16.72$\pm$0.25  & 6.92$\pm$0.04 &  78.74$\pm$4.94  & 7.92$\pm$0.06 & 0.46$\pm$0.16 &  1.78$\pm$0.03 &  10.86$\pm$1.25   &  8.99$\pm$0.14 & 0.25$\pm$0.03 & -0.55$\pm$0.22 &  1.79$\pm$0.03 &   9.58$\pm$0.44   & -2.01$\pm$0.48 &  1.93 & 2.58 & 0.75 \\   
 ngc6656 & 6.74$\pm$0.01 &  36.07$\pm$0.21  & 5.67$\pm$0.03 &  74.49$\pm$1.34  & 6.57$\pm$0.05 & 1.19$\pm$0.05 &  4.31$\pm$0.03 &  40.12$\pm$1.42   &  6.36$\pm$0.06 & 0.24$\pm$0.01 & -1.34$\pm$0.06 &  4.28$\pm$0.04 &  33.03$\pm$0.83   & -2.05$\pm$0.05 & 10.00 & 0.01 & 0.50 \\ 
  pal\_8 & 5.21$\pm$0.20 &  26.13$\pm$2.19  & 4.56$\pm$0.26 &  50.01$\pm$6.80  & 5.12$\pm$0.47 & 1.34$\pm$0.42 &  4.65$\pm$0.34 &  32.50$\pm$9.03   &  6.19$\pm$0.95 & 0.16$\pm$0.07 & -1.07$\pm$0.59 &  4.66$\pm$0.33 &  21.62$\pm$4.68   & -3.03$\pm$0.71 &  2.00 & 5.96 & 0.86 \\   
 ngc6681 & 8.27$\pm$0.04 &  23.77$\pm$0.37  & 7.03$\pm$0.05 & 121.76$\pm$8.02  & 8.33$\pm$0.12 & 0.87$\pm$0.10 &  2.53$\pm$0.07 &  21.09$\pm$2.14   &  8.88$\pm$0.21 & 0.28$\pm$0.07 & -0.77$\pm$0.36 &  2.58$\pm$0.08 &  15.17$\pm$1.74   & -1.65$\pm$0.71 &  2.44 & 1.75 & 0.69 \\   
 ngc6715 & 7.56$\pm$0.07 &  41.06$\pm$1.33  & 7.06$\pm$0.03 & 255.76$\pm$12.88 & 6.99$\pm$0.10 & 2.06$\pm$0.06 &  4.54$\pm$0.16 & 345.99$\pm$137.41 &  7.04$\pm$0.06 & 0.09$\pm$0.04 & -3.98$\pm$1.35 &  4.50$\pm$0.14 & 216.86$\pm$35.60  & -3.41$\pm$0.98 &  2.00 & 0.93 & 0.80 \\   
 ngc6717 & 9.02$\pm$0.15 &  19.50$\pm$0.88  & 8.29$\pm$0.16 & 358.05$\pm$28.63 & 8.88$\pm$0.35 & 1.34$\pm$0.26 &  2.47$\pm$0.32 &  30.60$\pm$14.39  &  8.88$\pm$0.44 & 0.34$\pm$0.11 & -0.90$\pm$0.61 &  2.33$\pm$0.31 &  12.88$\pm$6.70   & -1.18$\pm$0.66 &  1.50 & 5.41 & 0.70 \\   
 ngc6723 & 5.39$\pm$0.05 &  28.69$\pm$0.37  & 4.56$\pm$0.03 &  52.76$\pm$0.79  & 4.14$\pm$0.21 & 2.28$\pm$0.13 &  4.68$\pm$0.04 &  67.66$\pm$9.50   &  4.35$\pm$0.09 & 0.27$\pm$0.01 & -2.24$\pm$0.23 &  4.69$\pm$0.03 &  39.97$\pm$4.19   & -2.21$\pm$0.17 &  4.20 & 0.65 & 0.70 \\ 
 ngc6752 & 8.87$\pm$0.04 &  32.19$\pm$0.31  & 7.00$\pm$0.02 & 119.23$\pm$3.16  & 8.20$\pm$0.08 & 1.46$\pm$0.03 &  3.03$\pm$0.06 &  52.37$\pm$2.14   &  7.97$\pm$0.09 & 0.19$\pm$0.01 & -1.74$\pm$0.06 &  3.01$\pm$0.08 &  34.72$\pm$1.27   & -1.93$\pm$0.05 &  7.78 & 0.05 & 0.55 \\ 
 ngc6779 & 6.92$\pm$0.03 &  37.86$\pm$0.47  & 6.22$\pm$0.03 & 112.33$\pm$3.75  & 6.82$\pm$0.10 & 1.25$\pm$0.16 &  4.41$\pm$0.07 &  46.46$\pm$6.73   &  6.95$\pm$0.15 & 0.34$\pm$0.01 & -0.77$\pm$0.10 &  4.46$\pm$0.08 &  27.20$\pm$1.51   & -1.56$\pm$0.04 &  1.70 & 0.18 & 0.70 \\ 
 ngc6809 & 4.64$\pm$0.05 &  29.55$\pm$0.32  & 2.96$\pm$0.04 &  42.10$\pm$0.37  & 0.82$\pm$0.59 & 2.64$\pm$0.12 &  5.79$\pm$0.03 &  56.51$\pm$3.73   &  2.90$\pm$0.10 & 0.23$\pm$0.12 & -3.15$\pm$1.49 &  5.84$\pm$0.03 &  39.80$\pm$0.99   & -2.82$\pm$1.63 &  6.83 & 0.18 & 0.60 \\   
 terzan8 & 3.86$\pm$0.59 &  69.69$\pm$9.10  & 2.38$\pm$1.11 & 104.29$\pm$23.18 & 2.89$\pm$1.62 & 1.88$\pm$0.80 & 16.49$\pm$1.46 & 107.51$\pm$68.34  &  2.81$\pm$1.28 & 0.33$\pm$0.14 & -6.98$\pm$2.12 & 16.53$\pm$1.52 & 112.97$\pm$37.24  & -6.19$\pm$1.84 &  0.00 & 0.90 & 0.77 \\   
 pal\_11 & 0.18$\pm$0.08 &  16.38$\pm$0.29  & 0.21$\pm$0.10 &  29.34$\pm$0.72  & 0.27$\pm$0.18 & 0.16$\pm$0.07 &  5.83$\pm$0.13 &  12.10$\pm$0.41   &  1.29$\pm$1.53 & 0.04$\pm$0.05 & -0.06$\pm$0.87 &  5.78$\pm$0.16 &  16.00$\pm$0.57   & -4.42$\pm$2.69 &  2.00 & 2.22 & 0.83 \\ 
 ngc6864 & 8.02$\pm$0.08 &  34.77$\pm$1.36  & 7.05$\pm$0.04 & 165.56$\pm$11.52 & 7.60$\pm$0.21 & 1.61$\pm$0.15 &  3.34$\pm$0.22 &  74.72$\pm$20.26  &  7.41$\pm$0.21 & 0.12$\pm$0.04 & -2.22$\pm$0.39 &  3.33$\pm$0.26 &  49.88$\pm$9.89   & -2.87$\pm$0.59 &  1.02 & 0.16 & 0.77 \\ 
 ngc6934 & 6.33$\pm$0.03 &  26.12$\pm$0.32  & 6.03$\pm$0.03 &  88.01$\pm$2.66  & 6.22$\pm$0.09 & 1.76$\pm$0.13 &  3.71$\pm$0.06 &  61.21$\pm$12.27  &  5.98$\pm$0.07 & 0.23$\pm$0.04 & -1.75$\pm$0.52 &  3.71$\pm$0.06 &  36.83$\pm$14.03  & -2.13$\pm$0.47 &  2.00 & 0.02 & 0.78 \\   
 ngc6981 & 5.27$\pm$0.06 &  30.55$\pm$0.61  & 4.83$\pm$0.05 &  67.95$\pm$1.59  & 4.52$\pm$0.28 & 2.18$\pm$0.17 &  5.40$\pm$0.08 &  77.95$\pm$14.00  &  4.68$\pm$0.13 & 0.23$\pm$0.04 & -2.25$\pm$0.56 &  5.43$\pm$0.05 &  52.06$\pm$9.26   & -2.60$\pm$0.54 &  2.25 & 0.04 & 0.79 \\ 
 ngc7006 & 5.87$\pm$0.09 &  40.10$\pm$1.01  & 5.27$\pm$0.04 &  89.69$\pm$2.46  & 4.52$\pm$0.23 & 2.58$\pm$0.10 &  5.92$\pm$0.08 & 238.71$\pm$67.80  &  5.15$\pm$0.12 & 0.38$\pm$0.05 & -3.06$\pm$0.51 &  5.93$\pm$0.07 &  79.97$\pm$8.63   & -2.22$\pm$0.25 &  2.00 & 0.34 & 0.77 \\ 
 ngc7078 & 8.19$\pm$0.03 &  39.57$\pm$0.40  & 6.87$\pm$0.01 & 163.72$\pm$3.08  & 7.53$\pm$0.08 & 1.67$\pm$0.04 &  3.75$\pm$0.06 &  93.65$\pm$6.12   &  7.16$\pm$0.08 & 0.14$\pm$0.02 & -2.35$\pm$0.24 &  3.63$\pm$0.06 &  68.13$\pm$5.90   & -2.51$\pm$0.34 &  4.37 & 0.10 & 0.71 \\ 
 ngc7089 & 7.60$\pm$0.09 &  40.18$\pm$0.75  & 6.32$\pm$0.02 &  98.82$\pm$1.76  & 6.26$\pm$0.18 & 2.01$\pm$0.08 &  3.47$\pm$0.08 &  98.56$\pm$10.28  &  6.19$\pm$0.08 & 0.19$\pm$0.01 & -2.48$\pm$0.44 &  3.50$\pm$0.08 &  64.86$\pm$13.64  & -2.52$\pm$0.26 &  4.19 & 0.02 & 0.74 \\ 
 ngc7099 & 8.67$\pm$0.07 &  31.35$\pm$0.33  & 6.75$\pm$0.04 & 107.86$\pm$4.27  & 8.36$\pm$0.12 & 1.40$\pm$0.05 &  3.19$\pm$0.11 &  49.46$\pm$3.58   &  8.17$\pm$0.14 & 0.18$\pm$0.02 & -1.71$\pm$0.15 &  3.18$\pm$0.11 &  34.42$\pm$2.58   & -2.00$\pm$0.14 &  3.36 & 0.02 & 0.67 \\   
 pal\_12 & 5.57$\pm$0.19 &  47.46$\pm$3.51  & 5.14$\pm$0.23 & 112.12$\pm$16.13 & 5.32$\pm$0.45 & 1.55$\pm$0.36 &  7.74$\pm$0.36 &  68.39$\pm$20.44  &  5.16$\pm$0.44 & 0.34$\pm$0.08 & -1.09$\pm$0.72 &  7.96$\pm$0.46 &  42.74$\pm$26.53  & -1.60$\pm$0.68 &  2.00 & 0.07 & 0.88 \\   
 ngc7492 & 1.58$\pm$0.54 &  37.05$\pm$2.54  & 0.49$\pm$0.36 &  60.89$\pm$2.79  & 1.24$\pm$0.86 & 2.03$\pm$0.49 & 12.36$\pm$0.53 &  70.47$\pm$29.93  &  0.95$\pm$0.59 & 0.18$\pm$0.13 & -7.94$\pm$1.26 & 12.24$\pm$0.43 &  64.14$\pm$5.86   & -4.93$\pm$1.23 &  2.00 & 0.02 & 0.76 \\ 

\end{longtable}
\end{landscape}
\end{scriptsize}

\end{appendix}

\label{lastpage}

\end{document}